\magnification=1200 \vsize=23truecm \voffset=0.5truecm
\topskip=14truemm \font\ftitle=cmb10 at 16pt \font\craw=cmti8

\font\eightrm=cmr10 at 12pt

 \font\crease=cmr10 at 18pt

\def\section#1{\bigskip\centerline{\bf #1}\bigskip}
\def\xeqno#1{\eqno(#1)}
\def\xline#1{\hbox to 0.98\hsize{#1}}

\baselineskip=16pt
\parskip=1.5\parskip
\language=1

\footline={} {\hsize= 0.5in \settabs 1 \columns

\+ \cr \+ \cr \+ \cr \+ \cr \+ \cr

\+ \cr \+ \cr \+ \cr \+ \cr \+ \cr

\+ &{\crease Contribution {\`a} la Recherche des Equations}&\cr

\+ \cr

\+ &{\crease de Mouvement en M{\'e}canique quantique}& \cr

\+ \cr

\+ &{\crease et {\`a} une int{\'e}rpr{\'e}tation d{\'e}terministe
\footnote{\sevenrm *}{\sevenrm t{\'e}rmin{\'e} en juin 2001,
pr{\'e}sent{\'e}
le 10 
octobre 2001  {\`a}l'USTHB (Alger, Alg{\'e}rie)}} & \cr }

\bigskip
\bigskip

\centerline {by \eightrm Toufik Djama\footnote{\sevenrm
**}{\sevenrm Electronic address: {djam\_touf@yahoo.fr}}}

\vfill\eject
\centerline{REPUBLIQUE ALGERIENNE DEMOCRATIQUE ET POPULAIRE}
\centerline{MINISTERE DE L'ENSEIGNEMENT SUPERIEURE ET DE LA
RECHERCHE SCIENTIFIQUE}

\centerline{UNIVERSITER HOUARI BOUMEDIENNE DES SCIENCES ET DE LA
TECHNOLOGIE}
 \centerline {FACULTE DES SCIENCES (PHYSIQUE )}
 \centerline{DEPARTEMENT DE PHYSIQUE THEORIQUE}

\bigskip
\bigskip
\bigskip
\bigskip
\bigskip
\bigskip

\centerline {\ftitle {THESE DE MAGISTER}}

\bigskip
\bigskip

\centerline{PRESENTEE PAR: \ \ DJAMA TOUFIk}

\centerline{SPECIALITE: PHYSIQUE}

\centerline{OPTION: PHYSIQUE THEORIQUE}
\bigskip
\bigskip
\bigskip
\centerline{\bf {INTITULE}} \medskip
\centerline{{\bf CONTRIBUTION
A LA RECHERCHE DES EQUATIONS }} \centerline{{\bf DE MOUVEMENT EN
MECANIQUE QUANTIQUE}} \centerline{{\bf ET A UNE INTERPRETATION
DETERMINISTE}}

\bigskip
\bigskip
\bigskip
Soutenue le 10 octobre 2001, devant le jury :
\medskip

 {\hsize= 6.5in \settabs 5 \columns
 \+ M. M. FELLAH&\ \ \ \ PROF.&PRESIDENT&USTHB&(Alger)\cr

\+ M. A. BOUDA &\ \ \ \ M.C.&DTEUR DE THESE&UAM&(Beja\"{i}a)\cr

\+ M. A. ABADA &\ \ \ \ M.C&EXAMINATEUR &ENS&(Alger)\cr

\+ Mlle. N. H. ALLAL&\ \ \ \  PROF.&EXAMINATEUR&USTHB&(Alger)\cr

\+ M. A. CHOUCHAOUI.&\ \ \ \ M.C&EXAMINATEUR &USTHB &(Alger)\cr}

\vfill\eject

  {\hsize= 2.5in \settabs 1 \columns
  \+ \cr \+ \cr \+ \cr \+ \cr \+ \cr
  \+ \cr \+ \cr \+ \cr \+ \cr \+ \cr \+ \cr
  \+  &{\it  "...Einstein en particulier a maintes fois affirm{\'e}} \cr
  \+  &{\it  qu'{\`a} ces yeux la th{\'e}orie actuelle, tout en {\'e}tant} \cr
  \+  &{\it  parfaitement exacte dans ces pr{\'e}visions, n'est pas }\cr
  \+  &{\it  une th{\'e}orie compl{\`e}te: elle ne serait que l'aspect }\cr
  \+  &{\it  statistique d'une repr{\'e}sentation plus profonde qui }\cr
  \+  &{\it  r{\'e}tablirait l'existence d'une r{\'e}alit{\'e} plus objective."}\cr
  \+ \cr
  \+  &{ \ \ \ \ \ \ \ \ \ \ \ \ \ \ \ \ \ \ \ \ \ \ \ \ \ \ \ \
  \ \ \ \ \ \ \ \ \ \ \ \ \ \ \ \ \ \ L. de Broglie }\cr}

\vfill\eject
{\ftitle Table des mati{\`e}res}

\bigskip
\bigskip
\bigskip
\bigskip

{\bf Introduction}...........................................................................................................1
\bigskip
{\bf  Chapitre 1. Equation d'Hamilton-Jacobi quantique}...................................10
\medskip
\ \ \ 1. Th{\'e}orie de Bohm............................................................................................10

\ \ \ 2. Le mod{\`e}le de la repr{\'e}sentation des trajectoires..............................................13

\ \ \ \ \ \ \ \ 2.1. L'{\'e}quation de Hamilton-Jacobi quantique ............................................13

\ \ \ \ \ \ \ \ 2.2. Le probl{\`e}me stationnaire et l'EHJQS....................................................14

 \ \ \ \ \ \ \ \ 2.3. Approche num{\'e}rique et micro-{\'e}tats.......................................................15

 \ \ \ \ \ \ \ \ 2.4. Equation de mouvement param{\'e}tris{\'e}e par le potentiel modifi{\'e} ...........16

\ \ \ \ \ \ \ \ 2.5. Solution de l'EHJQS ............................................................................17

\ \ \ \ \ \ \ \ 2.6. Les {\'e}tats li{\'e}s et la quantification de l'{\'e}nergie........................................18

\ \ \ \ \ \ \ \ 2.7. Les micro-{\'e}tats dans l'EHJQS...............................................................20

\ \ \ \ \ \ \ \ 2.8. Etude du cas de la particule libre..........................................................22

\ \ \ 3. Le postulat d'{\'e}quivalence en m{\'e}canique quantique........................................24

\ \ \ \ \ \ \ \ 3.1. Enonc{\'e} du postulat d'{\'e}quivalence..........................................................24

\ \ \ \ \ \ \ \ 3.2. EHJQS et principe d'{\'e}quivalence...........................................................26

\ \ \ \ \ \ \ \ 3.3. La r{\'e}solution de l'EHJQS......................................................................28

\ \ \ \ \ \ \ \ 3.4. D{\'e}formation de la g{\'e}om{\'e}trie et coordonn{\'e}e quantique.........................30

\ \ \ \ \ \ \ \ 3.5. L'{\'e}quation du mouvement quantique.....................................................31

\bigskip
{\bf  Chapitre 2. La loi de Newton quantique}.......................................................33
\medskip
\ \ \ 1. Le formalisme de la m{\'e}canique quantique......................................................33

\ \ \ \ \ \ \ \ 1.1. Le formalisme analytique de la m{\'e}canique classique..............................34

\ \ \ \ \ \ \ \ 1.2. Le Lagrangien du syst{\`e}me quantique.....................................................36

\ \ \ \ \ \ \ \ 1.3. Le Hamiltonien quantique......................................................................38

\ \ \ \ \ \ \ \ 1.4. D{\'e}termination de la fonction $f(x,\Gamma)$.....................................................39

\ \ \ \ \ \ \ \ 1.5. Equation de dispersion...........................................................................41

\ \ \ 2. Equation int{\'e}grale premi{\`e}re de la loi de Newton quantique...........................42

\ \ \ \ \ \ \ \ 2.1. Etablissement de l'{\'e}quation IPLNQ.......................................................42

\ \ \ \ \ \ \ \ 2.2. Etude du cas de la particule libre..........................................................45

\ \ \ \ \ \ \ \ 2.3. Comparaison avec les r{\'e}sultats de Floyd..................................................47

\ \ \ 3. Coordonn{\'e}e quantique et version quantique du th{\'e}or{\`e}me de Jacobi.................48

\ \ \ \ \ \ \ \ 3.1. Lagrangien et Hamiltonien du syst{\`e}me quantique...................................49

\ \ \ \ \ \ \ \ 3.2. Transformations canoniques et th{\'e}or{\`e}me de Jacobi.................................53

\ \ \ \ \ \ \ \ 3.3. Comparaison avec le th{\'e}or{\`e}me de Jacobi introduit par Floyd..................54

\ \ \ 4. Conclusion..........................................................................................................55

\bigskip
{\bf Chapitre 3. G{\'e}n{\'e}ralisation de l'EHJQS {\`a} trois dimensions dans

\ \ \  le cas d'un potentiel {\`a} sym{\'e}trie sph{\'e}rique}......................................................58
\medskip
\ \ \ 1. L'EHJQS {\`a} trois dimensions...............................................................................58

\ \ \ \ \ \ \ \ 1.1. Equation de Schr{\"o}dinger pour un potentiel {\`a} sym{\'e}trie sph{\'e}rique...............59

\ \ \ \ \ \ \ \ 1.2. L'EHJQS radiale........................................................................................61

\ \ \ \ \ \ \ \ 1.3. L'EHJQS angulaire suivant $\vartheta$....................................................................62

\ \ \ \ \ \ \ \ 1.4. L'EHJQS angulaire suivant $\varphi$....................................................................62

\ \ \ 2. R{\'e}solution des EHJQS {\`a} trois dimensions.........................................................63

\ \ \ \ \ \ \ \ 2.1. La forme de l'action r{\'e}duite radiale...........................................................63

\ \ \ \ \ \ \ \ 2.1. La forme de l'action r{\'e}duite angulaire suivant $\vartheta$........................................64

\ \ \ \ \ \ \ \ 2.1. La forme de l'action r{\'e}duite angulaire suivant $\varphi$.......................................66

\ \ \ 3. L'EHJQS {\'e}crite pour l'action r{\'e}duite totale .....................................................67

\bigskip
{\bf Conclusion}..............................................................................................................69

\bigskip
{\bf R{\'e}f{\'e}rences}..............................................................................................................71

\vfill\eject

\footline={}


The standard quantum mechanics as it is taught  in graduate and post-graduate classes is
based on the probabilistic interpretation of quantum phenomena. In spite of its correct
previsions, a great debate started between the probabilistic physicists (Bohr, Heisenberg, Born...)
and the deterministic physicists (Einstein, de Broglie, Rosen, Podolsky, Bohm...) about
the nature of quantum mechanics; the questions animating this debate is:
- is standard quantum mechanics a statistical aspect of a profound representation which
will restore an objective reality? The present thesis will take a part of this debate, and
show how we should construct a deterministic approach of quantum mechanics
(see the introduction). We started from the works of Floyd [18-23,31,32] and
Faraggi and Matone [24-28,36].

Floyd started from the Quantum Stationary Hamilton-Jacobi Equation (QSHJE) in
one dimension which is a third order differential equation, already established
by Bohm (1951) (Chap. 1, Secs. 1 and 2.2 ). He solved the QSHJE and given the form
of the reduced action which
contains two real independent solutions of the Schr\"odinger equation,  and
also a set of hidden variables which can be related to the integration constants (Chap. 1, Sec. 2.5).
The following step in the work of Floyd is the determination of the quantum trajectories
of the particle. For this aim, Floyd suggest to derive these trajectories from the
Jacobi's theorem (Chap. 1, Sec. 2.4). He obtained a class of trajectories for a free particle and for
a particle moving under the linear potential and linked it to the classical trajectories
when $\hbar \to 0$ (Chap. 1, Sec. 2.8).

 Faraggi and Matone, in order to obtain an equivalent of the Classical  Stationary
Hamilton-Jacobi Equation (CSHJE) in quantum mechanics, have
enunciated an \hbox{equi-\hskip-2pt plus 2pt}\break valence
postulate from which they derived the QSHJE (Chap. 1, Sec. 3.1).
They took up the Jacobi's theorem as it is used by Floyd and
established an equation equivalent to the first integral of
quantum Newton's law (Chap. 1, Sec. 3.5). They also defined a
coordinate transformation by which they write the QSHJE in the
form of the CSHJE (Chap. 1, Sec. 3.4). With Bertoldi, Faraggi and
Matone have generalized the QSHJE into N-dimensions and for the
relativistic systems (quant-ph/9809127,quant-ph/9909201).

Taking advantage on these works we established in this thesis a new approach
of quantum mechanics. We first criticize the use of Jacobi's theorem since it is used
in classical mechanics for a first order differential equation, while the QSHJE is a
third order one. Then, we proposed an analytic approach of quantum mechanics. It is
consisting on a quantum Lagrangian [38] from which, using both the least
action principle and the QSHJE, we derive a new relation between the conjugate
momentum and the speed of the particle (Chap. 2, Sec. 1.5). Also we derived the
First Integral of the Quantum Newton's Law (FIQNL) (Chap. 2, Sec. 2.1). As an application,
we have choose the free particle case (Chap. 2, Sec.2.2 ).  It is useful to note
that the FIQNL is obtained in three context:

- a Lagrangian formalism (Chap. 2, Sec. 1.2).
- an  Hamiltonian formalism (Chap. 2, Sec. 1.3).
- a quantum version of the Jacobi's theorem (Chap. 2, Sec. 3).
Furthermore, We have attempted to generalize the QSHJE into 3-D for symmetrical
potentials (in this thesis we take the spherical symmetry potential case).
In fact, we have established three components of the QSHJE in 3-D. [39].

It is useful to indicate that this thesis is followed by three articles. The first article is
published in PLA [37] and has as subject the present work. The second one  (quant-ph/0108022),
submitted to PLA, is a description of the quantum trajectories deduced
from our present equations. The last one is a reply to a Floyd' comment on our approach
(also submitted to PLA)  .
This thesis is also the embryo of a generalization to a relativistic quantum systems
exposed in  two articles:

-The Quantum Relativistic Newton's Law and Photon Trajectories
\hbox{quant-\hskip-2pt plus 2pt}\break (ph/0111121).

-Nodes in the Relativistic Quantum Trajectories and
  Photon's Trajectories (quant-ph/0201003).

To conclude, our approach, in the framework of its generalization
into three dimensions and for the relativistic quantum cases, can
have a considerable impact for the Physic. First, it will restore
the objective reality of nature, which will have important
philosophical implications. Secondly, it present a hope to fund a
quantum theory of the gravitation which is indispensable to
achieve the unification of the forth fundamental interactions.


\vfill\eject
\footline={}

\bigskip
\bigskip
\bigskip
\bigskip

\centerline{\ftitle INTRODUCTION}

\bigskip
\bigskip
\bigskip

\bigskip
\bigskip
\bigskip

\bigskip
\bigskip
\bigskip

Dans l'{\'e}tude et l'interpr{\'e}tation des ph{\'e}nom{\`e}nes de la nature,
 la physique th{\'e}orique --- jusqu'{\`a} l'apparition de la m{\'e}canique
ondulatoire en 1923 --- se basculait entre deux sortes d'images tr{\`e}s
oppos{\'e}es. La premi{\`e}re est celle du ``grain'',
qui consistait {\`a} imaginer de petits objets se d{\'e}pla{\c c}ant
dans l'espace et ob{\'e}issant dans leur mouvement {\`a} des
lois dynamiques. La deuxi{\`e}me image, dite des
``champs'', pr{\'e}sentait les grandeurs physiques comme {\'e}tant
r{\'e}pandues contin{\^u}ment dans l'espace et variant au cours
du temps. L'image du grain {\'e}tait
la triomphante dans la dynamique du point et du syst{\`e}me
des points  mat{\'e}riels, les repr{\'e}sentations atomiques
et mol{\'e}culaires, et dans la conception de
l'{\'e}lectron. Par contre l'image des champs s'affirmait
dans la dynamique des milieux continus, en hydrodynamique,
dans la th{\'e}orie ondulatoire de la lumi{\`e}re
et dans la th{\'e}orie \hbox{{\'e}lectro-\hskip-5pt plus 5pt}\break
magn{\'e}tique. Cependant,
dans plusieurs domaines,
une coexistence entre les deux images {\'e}tait n{\'e}cessaire.
Pour expliquer l'interaction des points mat{\'e}riels,
la m{\'e}canique avait d{\^u} admettre que ces points
sont les foyers de champs de forces, tel que le champ de gravitation
et le champ {\'e}l{\'e}ctrostatique. Et, Lorentz avait introduit
la discontinuit{\'e} en {\'e}lectricit{\'e} en
avan{\c c}ant l'hypoth{\`e}se de l'{\'e}lectron dans la th{\'e}orie
{\'e}lectromagn{\'e}tique de Maxwell
qui est une th{\'e}orie de champs. Toutefois, cette coexistence
entre les deux images ne refl{\'e}tait aucune synth{\`e}se apparente,
c'{\'e}tait seulement la nature du probl{\`e}me physique lui-m{\^e}me
qui imposait le choix de l'une que l'autre. D'ailleurs,
en m{\'e}canique classique, Hamilton puis Jacobi ont
remarqu{\'e} une curieuse analogie entre une certaine
mani{\`e}re de grouper les mouvements d'un point
mat{\'e}riel dans un champs de forces, et la propagation de la
lumi{\`e}re dans un milieu r{\'e}fringent {\`a} l'approximation de
l'optique g{\'e}om{\'e}trique [1,2]. Mais, cette analogie n'{\'e}tait
que formelle dans le cadre de la m{\'e}canique classique.

En optique, l'opposition entre grain et champ {\'e}tait pr{\'e}sente
depuis le 18{\`e}me si{\`e}cle. En effet, Newton pensait d{\'e}j{\`a}
que les faisceaux lumineux se composaient de corpuscules.
En parall{\`e}le, Huyghens optait pour une r{\'e}alit{\'e} ondulatoire
de la lumi{\`e}re, comme le montraient --- depuis le 17{\`e}me si{\`e}cle
d{\'e}j{\`a} --- les exp{\'e}riences de diffraction [3]. Au d{\'e}but
du 19{\`e}me si{\`e}cle, Young r{\'e}alisa sa c{\'e}l{\`e}bre
exp{\'e}rience des fentes [3] par laquelle il remarqua le ph{\'e}nom{\`e}ne
des interf{\'e}rences. En se basant sur les travaux de Huyghens, Fresnel
expliqua ce ph{\'e}nom{\`e}ne {\`a} l'aide d'un artifice math{\'e}matique [3]
qui trancha --- du moins jusqu'au d{\'e}but du 20{\`e}me si{\`e}cle ---
pour la nature ondulatoire de la lumi{\`e}re.

Apr{\`e}s moins d'un si{\`e}cle des travaux de Fresnel,
la physique th{\'e}orique se heurtait {\`a} de nouveaux probl{\`e}mes
d'explications de quelques ph{\'e}nom{\`e}nes. Parmi ces ph{\'e}nom{\`e}nes,
on trouve le spectre du rayonnement noir, l'effet
photo{\'e}lectrique,... En 1900, Max Planck avan{\c c}a son hypoth{\`e}se
des quanta [4]. Cette hypoth{\`e}se consistait dans le fait que
les corpuscules et notamment les {\'e}lectrons, ne pouvaient avoir
n'importe lequel des \hbox{mouve-\hskip-5pt plus 5pt}\break ments que la dynamique classique
leur pr{\'e}voyait. Seuls {\'e}taient permis certains
de ces mouvements satisfaisant {\`a} des conditions de quanta d'{\'e}nergie.
L'hypoth{\`e}se de Planck permit d'expliquer le spectre du rayonnement
noir et de retrouver les lois empiriques connues dans la
thermodynamique
de ce rayonnement. Par la suite, en se basant sur l'hypoth{\`e}se de Planck,
Einstein avan{\c c}a en 1905 sa th{\'e}orie des ``quanta de lumi{\`e}re'' [4]
qui donne un aspect corpusculaire (grains) au rayonnement lumineux.
Cette nouvelle th{\'e}orie donna une
explication de l'effet
photo{\'e}lectrique qui {\'e}tait un ph{\'e}nom{\`e}ne assez {\'e}trange pour
 les physiciens de l'{\'e}poque. En effet, en recevant de
la radiation, une plaque m{\'e}tallique lib{\`e}re des {\'e}lectrons.
Elle ne lib{\`e}re ces {\'e}lectrons que pour des fr{\'e}quences
sup{\'e}rieures {\`a} une limite $\nu_0$; et quelle que soit l'intensit{\'e}
du rayonnement, cette plaque ne lib{\`e}re aucun {\'e}lectron si
la fr{\'e}quence des rayons est inf{\'e}rieure {\`a} la fr{\'e}quence
critique. En fait, en amplifiant l'intensit{\'e}, nous augmentons
seulement le nombre d'{\'e}lectrons lib{\'e}r{\'e}s, mais leurs
{\'e}nergies individuelles ne s'accroient qu'en augmentant
la fr{\'e}quence. La th{\'e}orie des quanta de lumi{\`e}re suppose
que l'{\'e}change d'{\'e}nergie entre la lumi{\`e}re et l'{\'e}lectron
ne se fait que par des quanta d'{\'e}nergie appel{\'e}s photons,
{\'e}gales {\`a} $h\nu$, o{\`u}
$$
h=6.62 \ 10^{-34}Joule \times seconde
$$
\headline={
\vbox{
    \xline{   {\bf  \folio}}}}
est la constante de Planck. Apr{\`e}s quelques ann{\'e}es,
l'effet Compton renfor{\c c}a
cette \hbox{hypo-\hskip-5pt plus 5pt}\break th{\`e}se en
mettant en {\'e}vidence la n{\'e}cessit{\'e}  d'attribution au
photon d'une impulsion dont la direction co{\"\i}ncide avec celle de la
propagation de la lumi{\`e}re. D{\'e}sormais, la physique va prendre
une nouvelle voie qui sera triomphante mais pleine de difficult{\'e}s.

Par ailleurs, on savait depuis 1910, d'apr{\`e}s les exp{\'e}riences
de Rutherford, que l'atome est compos{\'e} d'un noyau central charg{\'e}
positivement et d'{\'e}lectrons n{\'e}gatifs qui lui gravitent autour [5].
D'apr{\`e}s la th{\'e}orie {\'e}lectrodynamique,
les {\'e}lectrons devraient rayonner des ondes {\'e}lectromagn{\'e}tiques,
verraient leur {\'e}nergie diminuer progressivement et tomberaient
ainsi sur le noyau, ce qui contredirait la constatation de M. Rutherford.
Donc, l'approche classique est incapable
d'expliquer la structure plan{\'e}taire de l'atome.
D'un autre cot{\'e}, les spectres discontinus d'absorption et d'{\'e}mission
des atomes de l'hydrog{\`e}ne et du sodium n'{\'e}taient pas
interpr{\'e}tables dans le cadre d'une th{\'e}orie classique [5].
Tout ceci amena Bohr {\`a} {\'e}noncer, en 1913, son postulat de la
 quantification du moment cin{\'e}tique de l'{\'e}lectron dans le noyau [5].
 En fait, Bohr donna au moment cin{\'e}tique des valeurs
discr{\`e}tes {\'e}gales {\`a} $n\hbar$, ou $n$ est un nombre entier, et $\hbar$ est
 le rapport entre la constante de Planck et $2\pi$. Avec ce postulat,
 Bohr r{\'e}ussit {\`a} reproduire
les valeurs quantifi{\'e}es  de l'{\'e}nergie de l'{\'e}lectron
dans l'atome de l'hydrog{\`e}ne, et {\`a} expliquer ainsi la structure
plan{\'e}taire de l'atome. D{\`e}s lors, la quantification et les
nombres entiers devenaient indispensables pour approfondir notre
connaissance de la nature. Comme il est difficile, dans
un point de vue grain et en utilisant uniquement la dynamique
du point mat{\'e}riel, de comprendre comment
des nombres entiers s'introduisent pour d{\'e}crire un ph{\'e}nom{\`e}ne
tel que la quantification de l'{\'e}nergie de
 l'{\'e}lectron, la physique se trouvait astreinte {\`a} doubler l'aspect
 corpusculaire du grain d'un aspect ondulatoire o{\`u} les nombres
entiers interviendraient naturellement, comme il est le cas pour
les ph{\'e}nom{\`e}nes d'interf{\'e}rences et de r{\'e}sonance. Ainsi,
na{\^\i}tra une v{\'e}ritable synth{\`e}se entre grain et champ.

Effectivement, l'ensemble des physiciens du d{\'e}but du 20{\`e}me
si{\`e}cle venait d'amorcer une nouvelle th{\'e}orie, c'est celle
de la m{\'e}canique ondulatoire. D{\`e}s 1923,
Louis de Broglie associa {\`a} tout corpuscule de quantit{\'e}
de mouvement $P$ une onde dont la longueur est $\lambda=h/P$ [2,6].
Schr{\"o}dinger, en 1926, en reprenant les id{\'e}es de
de Broglie, {\'e}crit le premier pour l'{\'e}lectron, mais seulement
dans le cas non relativiste et sans tenir compte du spin, l'{\'e}quation
de propagation d'une onde pour laquelle il associa la fonction
d'onde $\psi$ [6]. Il est ainsi arriv{\'e} {\`a} reproduire exactement
les {\'e}tats quantifi{\'e}s d'un syst{\`e}me atomique, d{\'e}j{\`a}
connus depuis les travaux de Bohr. On notera aussi que des
exp{\'e}riences r{\'e}alis{\'e}es sur les {\'e}lectrons mettaient
en {\'e}vidence leur diffraction (Geiger et
Davisson 1927, M. B{\"o}rsch 1940 ...) [7], ainsi l'aspect ondulatoire
des corpuscules est bel et bien confirm{\'e}.

Schr{\"o}dinger et de Broglie tous deux, pensaient que l'onde $\psi$
{\'e}tait une onde physique. De Broglie consid{\'e}rait, dans sa
th{\'e}orie de l'onde pilote [8], que la particule {\'e}tait
localis{\'e}e d'une mani{\`e}re permanente et que son mouvement {\'e}tait
guid{\'e} par l'onde.\hbox{ Quant-{\`a}\hskip-1pt plus 2pt}\break
Schr{\"o}dinger, il pensait qu'il n'y avait
plus de corpuscules localis{\'e}s, et que la particule {\'e}tait une onde [9].
 Peu apr{\`e}s, en repr{\'e}sentant l'{\'e}tat d'un corpuscule par une
onde $\psi$, MM. Bohr et Heisenberg se sont aper{\c c}us , d{\`e}s 1927,
d'une impossibilit{\'e} de conna{\^\i}tre simultan{\'e}ment
la position et l'{\'e}tat de mouvement d'un corpuscule, ce qui
amena Heisenberg {\`a} {\'e}tablir les relations d'incertitudes
qui portent son nom [6,10]
$$
\triangle X\triangle P_x \geq \hbar,
$$
$$
\triangle Y\triangle P_y \geq \hbar,
$$
$$
\triangle Z\triangle P_z \geq \hbar.
$$
En fait, les physiciens de l'{\'e}poque essayaient de r{\'e}soudre
le probl{\`e}me de la non localisation de la particule par
une approximation analogue {\`a} celle de
l'optique g{\'e}om{\'e}trique qui consiste {\`a} consid{\'e}rer
 le corpuscule, dans son mouvement, comme parcourant l'un
des rayons de $\psi$ [2,6]. M{\^e}me dans cette approximation, les choses sont
d{\'e}j{\`a} compliqu{\'e}es, car il y a une infinit{\'e} de rayons,
ce qui implique une infinit{\'e} de mouvements possibles.
C'est de ce fait qu'intervient la notion de groupe d'onde pour
expliquer le mouvement du corpuscule.
Cependant, dans le cas des ondes non planes, cette
approximation n'est plus valable, et c'est  le cas dans la plupart
des probl{\`e}mes de m{\'e}canique quantique. Ainsi, la notion de
trajectoire c'est vue peu {\`a} peu
d{\'e}laiss{\'e}e, et la nouvelle th{\'e}orie a {\'e}t{\'e} amen{\'e}e
{\`a} prendre une forme probabiliste.
C'est ainsi que Born introduit (1927) la normalisation de
la fonction d'onde $\psi$, et avan{\c c}a le concept d'onde
de probabilit{\'e} qui le mena {\`a} consid{\'e}rer que le
carr{\'e} du module de la fonction d'onde $\vert \psi \vert ^2$
repr{\'e}sente la densit{\'e} de probabilit{\'e} pour que
la particule soit pr{\'e}sente dans un point de l'espace [6].
Cette approche diff{\'e}rente et tout {\`a} fait originale
conduit {\`a} un tr{\`e}s grand nombre de pr{\'e}visions exactes,
mais ne fournit aucune pr{\'e}sentation compr{\'e}hensible de
la coexistence des ondes et des corpuscules. Plus encore,
en niant la notion de trajectoire et en affirmant que les
valeurs des grandeurs physiques ne pr{\'e}existait pas avant la mesure,
la nouvelle approche contredit tout sens humain d'une r{\'e}alit{\'e}
objective de la nature. De ce fait, la mesure en m{\'e}canique quantique
cr{\'e}e la valeur des grandeurs physiques, et m{\^e}me elle cr{\'e}e un
nouvel {\'e}tat pour le syst{\`e}me. Cette mesure n'{\'e}tait dans
la th{\'e}orie classique qu'un moyen pour l'homme de d{\'e}terminer
les valeurs des grandeurs physiques et n'affectait en rien l'{\'e}tat de
ces grandeurs et leurs valeurs futures. La mesure donnait
seulement un moyen d'approfondir notre connaissance du monde physique.

Cette conception de la nature, malgr{\'e} son succ{\`e}s,
affrontait de s{\'e}rieuses difficult{\'e}s et elle suscitait
 une forte contestation aupr{\`e}s d'un grand nombre
de physiciens, dont les p{\`e}res  fondateurs  `` Einstein,
de Broglie et Schr{\"o}dinger '' constituent l'{\'e}lite [11].
Einstein lui-m{\^e}me affirmait que
la th{\'e}orie actuelle, tout en {\'e}tant parfaitement exacte
dans ses pr{\'e}visions, n'est pas une th{\'e}orie compl{\`e}te,
elle ne serait que l'aspect statistique d'une repr{\'e}sentation
plus profonde qui r{\'e}tablirait l'existence d'une r{\'e}alit{\'e}
objective.

De l{\`a}, na{\^\i}tra par la suite le d{\'e}bat entre
d{\'e}terministes ( Einstein, \hbox {de Broglie,\hskip-5pt plus 5pt}\break
Schr{\"o}dinger, Bohm, Rosen, Podolsky...) --- qui croyaient
en une r{\'e}alit{\'e} objective
\hbox { ind{\'e}-\hskip-5pt plus 5pt}\break pendante
de notre observation --- et probabilistes, repr{\'e}sent{\'e}s par les
adh{\'e}rents de l'{\'e}cole de
Copenhague  (Bohr, Born, Heisenberg,
et la plupart des jeunes physiciens de l'{\'e}poque) --- qui
croyaient en une r{\'e}alit{\'e} probabiliste de la nature
pour laquelle notre observation joue le
r{\^o}le  primordiale en
 cr{\'e}ant la valeur de la mesure.
 Ce d{\'e}bat continu {\`a} nos jours, et aucune th{\'e}orie
nouvelle n'a pu offrir un substitut {\`a} la m{\'e}canique
quantique. Tout de m{\^e}me, les arguments des partisans du
d{\'e}terminisme   sont toujours pr{\'e}sents, et parmi
lesquels l'argument suivant [9]:  `` Soit une source qui {\'e}met
une onde sph{\'e}rique transportant une particule. Un instant
apr{\`e}s, la particule manifeste sa pr{\'e}sence dans un point
de l'onde
par un fait localis{\'e} sur un d{\'e}tecteur. Il est {\'e}videmment
certain que l'{\'e}mission de la particule par la source est la ``cause''
de son arriv{\'e}e sur le d{\'e}tecteur.
Or le lien causal entre les deux ph{\'e}nom{\`e}nes ne peut
{\^e}tre {\'e}tabli que par l'existence d'une trajectoire,
et nier celle-ci c'est renoncer {\`a} la causalit{\'e},
c'est se condamner {\`a} ne pas comprendre''. D{\'e}j{\`a},
cette objection est suffisante pour soulever le doute dans
l'interpr{\'e}tation probabiliste, mais le paradoxe E.P.R [11] reste
l'argument le plus fort pour les d{\'e}terministes. Effectivement, si
on effectue des mesures sur un syst{\`e}me de deux particules (1)
et (2) qui ont interagi dans le pass{\'e}, on montre que
la mesure sur la particule (1) et celle sur la particule (2)
sont corr{\'e}l{\'e}es. Ceci veut dire que si on mesure une
grandeur $A_1$ de (1) on pourrait conna{\^\i}tre la valeur
pr{\'e}cise $A_2$ de (2) sans l'avoir mesur{\'e}e. Cela
contredit, comme l'a fait remarquer Einstein [11],
les postulats de la m{\'e}canique quantique. Car, pour
celle-ci, la valeur de la grandeur $A_2$ ne pourrait pas
pr{\'e}exister avant la  mesure sur (2). En effet, d'apr{\`e}s
l'interpr{\'e}tation probabiliste, si on associe {\`a} la grandeur
$A$ d'un syst{\`e}me quantique un ensemble de valeurs possibles
dont chacune a une probabilit{\'e} non nulle, alors il n'est
en aucun cas possible de savoir {\`a} l'avance, et sans effectuer
une mesure directe, laquelle de ces valeurs va prendre $A$.
 Si on admet, dans le cas des deux
syst{\`e}mes corr{\'e}l{\'e}s, que la mesure sur (2) ne change
pas l'{\'e}tat de (1) mais seulement notre connaissance de (1),
alors la d{\'e}composition spectrale de $\psi$ suivant les
valeurs propres de $A_1$ ne repr{\'e}sentera que notre
ignorance de la valeur exacte de $A_1$ [11]. Dans ce cas,
les probabilit{\'e}s en m{\'e}canique quantique ne serait
qu'un moyen d'approche qui rem{\'e}dierait {\`a} cette ignorance.
De l{\`a}, Einstein s'est fait une fine conviction que la
th{\'e}orie actuelle n'est pas compl{\`e}te et qu'une nouvelle
th{\'e}orie bas{\'e}e sur des variables cach{\'e}es est
indispensable pour donner un aspect objectif
aux ph{\'e}nom{\`e}nes quantiques.
En ce sens M. de Broglie d{\'e}veloppa une th{\'e}orie
hydrodynamique dite ``de l'onde pilote'' [8] qui consid{\`e}re
un fluide fictif qui se conserve pour lequel les lignes
de courant co{\"\i}ncident avec les trajectoires des particules
du fluide. Dans cette th{\'e}orie la quantit{\'e}
$\vert \psi \vert^2$ repr{\'e}sente la densit{\'e} du fluide.
Dans ce cas, si on associe {\`a} $\psi$ un grand nombre
de corpuscules, les trajectoires des particules du fluide
co{\"\i}ncident avec celles des corpuscules et
$\vert \psi \vert^2$ repr{\'e}sentera la densit{\'e} de corpuscules.
Et lorsqu'on associe un train d'onde $\psi$ pour un seul corpuscule,
on pourra d{\'e}finir une infinit{\'e} de trajectoires
hydrodynamiques mais une seule sera effectivement d{\'e}crite.
Dans ce cas $\vert \psi \vert^2 d \tau$ repr{\'e}sente
la probabilit{\'e} de trouver le corpuscule dans le volume $d \tau$.
Cette
interpr{\'e}tation reste classique parce qu'elle donne {\`a}
la particule une trajectoire quoi qu'on ne conna{\^\i}t pas
cette trajectoire. Toutefois, la th{\'e}orie de l'onde pilote
pr{\'e}sente des difficult{\'e}s insurmontables qui ont
pouss{\'e} de Broglie {\`a} l'abondonner [8]. Un quart de si{\`e}cle
plus tard ( en 1951 ), un jeune physicien nomm{\'e} David Bohm [12]
relan{\c c}a le d{\'e}bat en reprenant la th{\'e}orie de l'onde pilote.
Il posa tout d'abord
$$
\psi =A\, exp \left (iS/ \hbar \right )\ , \xeqno{1}
$$
 o{\`u} $S$ et $A$ sont des fonctions r{\'e}elles.
En rempla{\c c}ant l'expression de $\psi$ dans l'{\'e}quation de
Schr{\"o}dinger, il {\'e}crit une {\'e}quation qu'il appela
{\'e}quation d'Hamilton-Jacobi quantique dans laquelle $S$
joue le r{\^o}le de l'action du syst{\`e}me. Cette {\'e}quation ne diff{\`e}re
de l'{\'e}quation d'Hamilton-Jacobi
classique que par un certain terme dit potentiel quantique.
Pour {\'e}tudier le mouvement du corpuscule Bohm
reprenait l'hypoth{\`e}se de de Broglie selon laquelle la vitesse $\vec{v}$
est donn{\'e}e par la relation $m \vec{v}=\vec{\nabla} S_0$.
Avec Vigier [12], il postula par la suite l'existence d'un milieu
subquantique dans lequel se propagerait l'onde $\psi$ {\`a}
laquelle il attribua une r{\'e}alit{\'e} physique,
contrairement {\`a} la l'interpr{\'e}tation probabiliste.

La th{\'e}orie de Bohm est hydrodynamique tout comme la th{\'e}orie
de l'onde pilote, c'est pour cette raison qu'elles durent affronter
les m{\^e}mes difficult{\'e}s. De plus, l'approche de Bohm n'est
plus valable lors d'une {\'e}tude portant sur
des fonctions d'onde r{\'e}elles, car dans ce cas la vitesse du syst{\`e}me sera nulle.

Apr{\`e}s ces tentatives non aboutissantes, d'autres th{\'e}ories
ont {\'e}t{\'e} avanc{\'e}es sans qu'elles soient de vrais substituts
{\`a} la m{\'e}canique quantique [13,14].
En 1964, sur la base des id{\'e}es d'Einstein et de l'exp{\'e}rience
de pens{\'e}e E.P.R.B\footnote{*}
{\sevenrm En croyant r{\'e}soudre le paradoxe E.P.R,
Bohm avan{\c c}a son exp{\'e}rience de pens{\'e}e E.P.R.B
[15]. Elle consiste {\`a} faire des mesures sur un syst{\`e}me
de deux photons corr{\'e}l{\'e}s.}, Bell avan{\c c}a une nouvelle
approche par laquelle il
pr{\'e}tend r{\'e}soudre le paradoxe E.P.R [16,17]. Il se basa sur
le fait que les valeurs des grandeurs physiques pr{\'e}existaient
avant la mesure, et postula que la fonction d'onde contenait des
param{\`e}tres cach{\'e}s qui pr{\'e}d{\'e}termineraient ces valeurs.
Ainsi, il put {\'e}tablir un ensemble d'in{\'e}galit{\'e}s qui
seraient viol{\'e}es par la m{\'e}canique quantique.
Ces in{\'e}galit{\'e}s pr{\'e}sentaient le privil{\`e}ge
d'une possibilit{\'e} d'{\^e}tre mises
 {\`a} l'{\'e}preuve par l'exp{\'e}rience et on se verra ainsi
capable de trancher pour l'une des th{\'e}ories. Effectivement,
un groupe de physiciens dirig{\'e} par Aspect a
effectu{\'e} une s{\'e}rie d'exp{\'e}riences entre 1981
et 1984 [15] par lesquelles on trancha en faveur de
la m{\'e}canique quantique actuelle. Mais, ces exp{\'e}riences
ont {\'e}t{\'e} contest{\'e}es par certains physiciens [15,17].
Il reste que la th{\'e}orie de Bell et ses mod{\`e}les {\`a}
param{\`e}tres cach{\'e}s ont {\'e}t{\'e} abandonn{\'e}s.
Ceci ne constitue en aucun cas une mise {\`a} l'{\'e}cart d'un
mod{\`e}le {\`a} param{\`e}tres cach{\'e}s {\'e}tablit suivant
une mani{\`e}re diff{\'e}rente de celle des mod{\`e}les de Bell.

A partir de la fin des ann{\'e}es 70, Floyd [18,19] reprit le mod{\`e}le
de Bohm et tenta une meilleur approche du probl{\`e}me. Il reprit
l'{\'e}quation de Hamilton-Jacobi quantique (EHJQ)
{\`a} une dimension, et {\'e}tudia le probl{\`e}me avec la
th{\'e}orie de Jacobi qu'il adapta au cas
 stationnaire (EHJQS), et par laquelle il montra que la vitesse
de la particule ne co{\"\i}ncide pas avec celle du fluide
$(m\dot{x} \not= {\partial S_0 / \partial x})$. Floyd, pour traiter le cas
des fonctions d'ondes r{\'e}elles, utilisa une forme trigonom{\'e}trique pour la
relation entre $\psi$ et $S_0$ qui diff{\`e}re de celle donn{\'e}e par l'Eq. (1).
Par la suite, il parla de fonction de Hamilton
principale, d'action r{\'e}duite et de  fonction de Hamilton
caract{\'e}ristique [18,20,21,22]. Il donna alors une solution {\`a} l'EHJQS [18,21] {\`a} une dimension et put obtenir une {\'e}quation de
trajectoire pour une particule libre [18,23]. Il montra aussi l'existence
de micro-{\'e}tats contenus dans l'EHJQS que la
fonction de Schr{\"o}dinger ne detecte pas . Floyd arriva {\`a}
la conclusion suivante [18,22]:
``  La fonction de Schr{\"o}dinger n'est pas exhaustive et
l'EHJQS est plus fondamentale que l'{\'e}quation de
Schr{\"o}dinger''.

Malgr{\'e} ces r{\'e}sultats th{\'e}oriques importants,
Floyd n'a pu justifier l'utilisation du th{\'e}or{\`e}me
de Jacobi ($t-t_0 = {\partial S_0 / \partial E}$), puisque ce dernier
est utilis{\'e} en m{\'e}canique classique pour une {\'e}quation
du premier ordre alors que l'EHJQS est du troisi{\`e}me ordre.
De plus, le fait de distinguer le cas des fonctions d'ondes r{\'e}elles de
celui des fonctions d'ondes complexes laissait sa
repr{\'e}sentation incompl{\`e}te.

Avec les m{\^e}mes concepts, mais en utilisant la g{\'e}om{\'e}trie
diff{\'e}rentielle,
Faraggi et Matone ont essay{\'e} de reprendre
et de corriger la th{\'e}orie de Bohm. Ils aboutirent
{\`a} {\'e}noncer un principe d'{\'e}quivalence qui stipule que tous
les syst{\`e}mes quantiques sont connect{\'e}s par des transformations
de coordonn{\'e}es [24,25]. Ils arriv{\`e}rent
{\`a} partir de ce principe {\`a}
 {\'e}tablir l'EHJQS sans passer par la m{\'e}thode de Bohm [24,26].
 De m{\^e}me ils r{\'e}ussirent {\`a} {\'e}luder le probl{\`e}me
de la nullit{\'e} du moment conjugu{\'e}
 en donnant une forme unifi{\'e}e de la relation entre
$\psi$ et $S$ qui substituerait aux formes donn{\'e}es par l'Eq. (1),
et trigonom{\'e}trique que propose Floyd pour les fonctions d'onde r{\'e}elles.
De plus, sans faire appel {\`a}
l'interpr{\'e}tation probabiliste de la fonction d'onde,
ils ont montr{\'e}s que
l'effet tunnel et la quantification de l'{\'e}nergie sont des
cons{\'e}quences du principe d'{\'e}quivalence [27]. Apr{\`e}s, avec Bertoldi,
Faraggi et Matone ont  g{\'e}n{\'e}ralis{\'e} l'EHJQS {\`a} N dimension [28].
Ces r{\'e}sultats ont {\'e}t{\'e} obtenus par A. Bouda [29] sans faire
appel {\`a} la g{\'e}om{\'e}trie diff{\'e}rentielle. Ce qui permettait
de mieux voir les r{\'e}alit{\'e}s qui se cachaient derri{\`e}re
les {\'e}quations. Il a avanc{\'e} une nouvelle forme de la solution de
l'EHJQS dans laquelle les constantes d'int{\'e}gration sont clairement
identifi{\'e}es. Il arriva ainsi {\`a} confirmer l'existence de micro-etats
dans l'EHJQS.

Tout en {\'e}tant en progr{\`e}s successif, la
repr{\'e}sentation des trajectoires ( trajectory
repr{\'e}sentation )
avanc{\'e}e par Floyd, reste toujours une th{\'e}orie non
achev{\'e}e. La difficult{\'e} majeure de ce formalisme est l'utilisation
du th{\'e}or{\`e}me de Jacobi par sa
 version classique\footnote{*}
{{\sevenrm En fait, on parle de version classique
 du th{\'e}or{\`e}me de Jacobi parce qu'on introduit -dans ce travail-
 une version quantique pour ce th{\'e}or{\`e}me,
qui constituerait la base de la m{\'e}canique quantique. }} [19,23,24],
comme il est d{\'e}j{\`a} mentionn{\'e} plus haut.

Dans ce travail, nous introduirons de nouvelles id{\'e}es et
exposerons une approche assez originale de la repr{\'e}sentation
des trajectoires. D'abord, dans le premier chapitre nous allons
exposer les principaux devellopements qu'a connue la m{\'e}canique quantique
depuis le d{\'e}bat entre Einstein et Bohr. Nous pr{\'e}senterons
la th{\'e}orie de Bohm, les travaux de Floyd, ainsi que ceux de
Faraggi et Matone portant sur le principe d'{\'e}quivalence.
Ensuite dans le deuxi{\`e}me chapitre, on exposera un nouveau
formalisme dans le cadre de la repr{\'e}sentation des trajectoires,
tout en avan{\c c}ant une justification de nos
choix, et une interpr{\'e}tation de nos r{\'e}sultats.
On introduira le Lagrangien d'un syst{\`e}me quantique {\`a} une
dimension duquel, on tirera les {\'e}quations de mouvement du
corpuscule, entre autres l'{\'e}quation int{\'e}grale premi{\`e}re
de la loi de Newton quantique (IPLNQ). On fera ensuite un trait sur
le formalisme Hamiltonien correspondant {\`a} notre Lagrangien.
Toujours dans ce chapitre, en s'appuyant sur le principe
d'{\'e}quivalence et sur une version quantique du
th{\'e}or{\`e}me de Jacobi,  on reproduira par une nouvelle mani{\`e}re les
{\'e}quation du mouvement dont l'IPLNQ. Enfin,
nous ferons une comparaison entre nos r{\'e}sultats et ceux de Floyd.

Dans le troisi{\`e}me chapitre, nous allons exposer une
tentative de g{\'e}n{\'e}ralisation de l'EHJQS {\`a} trois dimensions
dans le cas d'un potentiel {\`a} sym{\'e}trie sph{\'e}rique.
Enfin, nous exposerons une conclusions dans laquelle
nous essayerons d'interpr{\'e}ter physiquement nos r{\'e}sultats
et leurs implications. Nous discuterons aussi, le fondement de notre
approche, et la coexistence ondes et corpuscules.

A pr{\'e}sent, nous laisserons le lecteur d{\'e}couvrir
lui-m{\^e}me, une hardie tentative de parvenir {\`a}
une th{\'e}orie d{\'e}terministe fond{\'e}e
sur un formalisme bel et bien con{\c c}u.

\vfill\eject

\headline{}
\centerline{\ftitle CHAPITRE 1}
\bigskip
\bigskip

\centerline{\ftitle EQUATION D'HAMILTON JACOBI QUANTIQUE}

\bigskip
\bigskip
\bigskip
\bigskip
\bigskip
\bigskip
\bigskip
\bigskip
\bigskip

Dans ce chapitre, nous exposerons la th{\'e}orie li{\'e}e {\`a}
l'{\'e}tablissement de l'{\'e}quation de Hamilton Jacobi quantique.
Nous d{\'e}velopperons
les grands axes d'o{\`u} est pass{\'e}e
cette th{\'e}orie, depuis son apparition dans l'approche de Bohm
jusqu'{\`a} son {\'e}tat d'avancement actuel. Ceci sera fait
en discutant de ses points faibles et des efforts fournis par des
physiciens contemporains tel que Floyd, Faraggi et Matone,
pour construire une th{\'e}orie coh{\'e}rente.

\bigskip
\bigskip
\bigskip

\noindent
{\bf 1. THEORIE DE BOHM}
\bigskip

En reprenant les id{\'e}es de de Broglie, David Bohm
d{\'e}veloppa d{\`e}s 1951 un int{\'e}ressant mod{\`e}le
bas{\'e} sur une approche hydrodynamique de la m{\'e}canique quantique.
En particulier, Bohm g{\'e}n{\'e}ralisa l'{\'e}quation de Hamilton-Jacobi {\`a}
la m{\'e}canique quantique, et cela pour une description du mouvement
des particules dans le cadre d'une interpr{\'e}tation causale [12].
L'{\'e}quation obtenue est appel{\'e} {\'e}quation de Hamilton-Jacobi
quantique (EHJQ). Bohm prit comme point de
d{\'e}part l'{\'e}quation de Schr{\"o}dinger d{\'e}pendante du temps
$$
 - {\hbar^2 \over {2m}  } \Delta\psi + V\psi = i\hbar
 {\partial\psi \over \partial t  }\ , \xeqno{1.1}
$$
et {\'e}crit la fonction d'onde $\psi$ sous la forme suivante [12]:
$$
\psi(x,y,z,t) = A(x,y,z,t)\
\exp\left({i\over\hbar }S(x,y,z,t)\right)\ ,
    \xeqno{1.2}
$$
o{\`u} $A(x$,$y$,$z$,$t)$ et $S(x$,$y$,$z$,$t)$ sont des fonctions
r{\'e}elles. En rempla{\c c}ant l'Eq. (1.2) dans l'Eq. (1.1)
et en s{\'e}parant la partie imaginaire de la partie r{\'e}elle,
Bohm aboutit {\`a}
$$
{1\over 2m }(\vec{\nabla} S)^2- {\hbar^2\over 2m }{\Delta A\over A }+V=
-{\partial S \over \partial t  }\ , \xeqno{1.3.a}
$$
$$ \vec{\nabla  }\cdot \left(A^2 {\vec{\nabla  } S\over m} \right) =
-{\partial {A^2} \over \partial t  } \ . \xeqno{1.3.b}
$$
\noindent
Le terme proportionnel {\`a} $\hbar^2$ dans l'Eq. (1.3.a), not{\'e}
$$
 V_B = -{\hbar^2\over {2m} }{\Delta A\over A } \ ,\xeqno{1.4}
$$
est appel{\'e} potentiel quantique de Bohm.

Remarquons tout d'abord que, si on pose $\hbar=0$, le potentiel
quantique s'annule et alors l'Eq. (1.3.a) se r{\'e}duit {\`a}
l'{\'e}quation  de Hamilton-Jacobi classique (EHJC) d{\'e}crivant
le mouvement d'une particule classique. Dans ce cas, $S$ est
identifi{\'e}e {\`a} l'action et $V_B$ est alors consid{\'e}r{\'e}
comme d{\'e}crivant les effets quantiques.
La relation (1.3.b), comme le fait remarquer Bohm [12],
repr{\'e}sente l'{\'e}quation de conservation du
courant de probabilit{\'e}. Effectivement,
 si on injecte l'expression donn{\'e}e par l'Eq. (1.2)
de $\psi$ dans l'expression connue en m{\'e}canique quantique
du courant de probabilit{\'e} [30]
$$
\vec{\jmath}={\hbar\over 2mi }(\psi^*\vec{\nabla}{\psi}-
\psi\vec{\nabla}{\psi^*})\ , \xeqno{1.5}
$$
on trouve que
$$
\vec{\jmath}=A^2{\vec{\nabla} S\over m}\ , \xeqno{1.6}
$$
$\vec{\jmath}$ {\'e}tant le courant de probabilit{\'e}. Ainsi,
 l'Eq. (1.3.b) s'{\'e}crit sous la forme
$$
\vec{\nabla}\cdot\vec{\jmath}+{\partial A^2 \over \partial t^2}=0 \ .
\xeqno{1.7}
$$
Cette relation est bel et bien l'{\'e}quation de conservation
du courant de probabilit{\'e}, {\'e}tant donn{\'e}e que l'Eq. (1.2) indique
que la densit{\'e} de pr{\'e}sence est
$$
\rho (x,t)=  \vert \psi(x,t) \vert ^2 = A^2(x,t) \ . \xeqno{1.8}
$$

Ce r{\'e}sultat a {\'e}t{\'e} avanc{\'e} dans les travaux de
de Broglie [8] et de Madelung [13].
En fait, Bohm et ses pr{\'e}d{\'e}cesseurs ont remarqu{\'e}
dans l'Eq. (1.7) une forte analogie avec l'{\'e}quation de
conservation d'un fluide. Ceci les incita {\`a}
associer {\`a} l'onde $\psi$ un fluide fictif de densit{\'e}
$\rho (x,t)$ et de flux $\vec{\jmath}$, et les mena {\`a} d{\'e}finir
un champ de vitesse $\vec {V}$ repr{\'e}sentant en chaque point
la vitesse de l'{\'e}l{\'e}ment fluide d{\'e}finie par
\headline={
\vtop{
\xline{ {\craw Chapitre 1}.   \hfill {\craw 1.Th{\'e}orie de Bohm}
{\bf \ \ \folio}}
    \medskip \hrule
}
}
$$
\vec{v}=\vec{\jmath} / \rho \ . \xeqno{1.9}
$$
D'apr{\`e}s les Eqs. (1.6), (1.8) et (1.9), on a
$$
\vec{v}= {\vec {\nabla} S \over m} \ . \xeqno{1.10}
$$
Cette tr{\`e}s importante relation {\'e}tait le point d'appui
de l'interpr{\'e}tation de Bohm du fluide quantique.

Pour aborder l'interpr{\'e}tation que proposa Bohm
dans son approche, essayons \hbox{d'expo-\hskip-5pt plus 5pt}\break ser
le point de vue de la th{\'e}orie de l'onde pilote
avanc{\'e}e par de Broglie [8].
La th{\'e}orie de l'onde pilote et le mod{\`e}le de
Madelung identifient la vitesse
de l'{\'e}l{\'e}ment fluide, donn{\'e}e par l'Eq. (1.10),
{\`a} la vitesse de la particule
(tel qu'un {\'e}lectron). Dans ce cas, on peut supposer
qu'{\`a} l'onde $\psi$ est associ{\'e} un grand nombre de
particules qui d{\'e}crivent les trajectoires des
{\'e}l{\'e}ments fluides. Mais dans une telle approche,
on affronte deux difficult{\'e}s
apparentes. La premi{\`e}re consiste dans le fait que
pour une fonction d'onde r{\'e}elle, le champ de vitesse
du fluide est nul, ainsi la vitesse des particules
quantiques devrait s'annuler, ce qui est {\'e}videmment
insens{\'e}. La deuxi{\`e}me difficult{\'e} est que
dans ce mod{\`e}le, on identifie un fluide continu (champ)
{\`a} un ensemble de particules (grains) qui, quel que soit
leur grand nombre, ne peuvent d{\'e}crire exactement le
comportement d'un milieu continu. D'un autre cot{\'e},
 lors de l'{\'e}tude d'un fluide, on ne peut pas d{\'e}crire
exactement une vraie localisation $\vec {r(t)}$ de la particule
--- du fait que le fluide est un milieu continu dont le
comportement est celui d'un champ et non de grain ---,
ce qui rend impossible une interpr{\'e}tation causale de la
th{\'e}orie quantique.

Pour compl{\'e}ter ce mod{\`e}le hydrodynamique,
Bohm postula l'existence de particules au sein du fluide.
Ces particules seraient des non-homog{\'e}n{\'e}it{\'e}s extr{\^e}mement
localis{\'e}es (highly localized inhomogenity) dans le fluide,
qui se d{\'e}placeraient {\`a} la vitesse de l'{\'e}l{\'e}ment
fluide $\vec {v}(x,t) $. La nature exacte de ces non-homog{\'e}n{\'e}it{\'e}s
n'est pas facile {\`a} pr{\'e}voir. Bohm avan{\c c}a qu'elles pourraient
{\^e}tre des objets {\'e}trangers de densit{\'e} qui s'approcherait
de celle du fluide, et qui serait entra{\^\i}n{\'e}s par le milieu continu.
D'une autre fa{\c c}on, ces non-homog{\'e}n{\'e}it{\'e}s peuvent {\^e}tre
interpr{\'e}t{\'e}es
par des structures dynamiques stables existant au sein du fluide,
par exemple un vortex. De telles structures seraient dues {\`a}
des non-lin{\'e}arit{\'e}s dans les {\'e}quations qui gouvernent le
mouvement de ce fluide [12]. De plus, le comportement du fluide
r{\'e}el n'est pas enti{\`e}rement d{\'e}crit par des {\'e}quations th{\'e}oriques.
Il subsisterait contin{\^u}ment des fluctuations d{\'e}sordonn{\'e}es qui
ont diverse origines. Ces fluctuations laisseraient les non-homog{\'e}n{\'e}it{\'e}s
pr{\'e}sentes de mani{\`e}re permanente. Ainsi, la densit{\'e} $\rho $
et les vitesses des particules $ \vec{v}(x$,$t)$ ne seraient que des
valeurs moyennes.
Alors, dans le cas des fonctions d'ondes r{\'e}elles, ce n'est que
la vitesse moyenne
qui s'annule, et pas la vitesse de la particule.
Malgr{\'e} cette tentative de rem{\'e}dier aux probl{\`e}mes
qu'elle affronte, la th{\'e}orie de Bohm ne r{\'e}ussit pas
{\`a} {\'e}tudier le
mouvement d'une particule sous l'action d'un potentiel tel que
la barri{\`e}re semi-infinie [31], l'oscillateur harmonique...

\bigskip
\bigskip
\noindent
{\bf 2. LE MOD{\`E}LE DE LA REPR{\'E}SENTATION DES TRAJECTOIRES}
\medskip
\headline={
\vbox{
    \xline{{\craw Chapitre 1.} \hfill  {\craw 2.La repr{\'e}sentation des trajectoires} {\bf \ \ \folio}}
    \medskip \hrule

}
}

Comme il est d{\'e}j{\`a} pr{\'e}sent{\'e} {\`a} l'introduction,
Floyd reprit les travaux de
Bohm et tenta, {\`a} son tour, une approche du probl{\`e}me en
essayant en premier lieu de surmonter les difficult{\'e}s d{\'e}crites
au paragraphe pr{\'e}c{\`e}dent. Tout d'abord, pr{\'e}cisons que la
grande partie des travaux de Floyd est faite {\`a} une dimension
et dans le cas stationnaire. Dans
son approche, il se basa sur une description dynamique
du mouvement des particules. A chacune de ces derni{\`e}res il affecta
une trajectoire individuelle.

La tentative de Floyd va-t-{\^e}tre d'un apport
consid{\'e}rable, et aboutira
 {\`a} l'{\'e}laboration d'une nouvelle th{\'e}orie dite
--- quoiqu'elle ne soit pas encore achev{\'e}e --- la repr{\'e}sentation
des trajectoires ( trajectory representation ). Cette th{\'e}orie
fera l'objet d'{\'e}tude de cette section.

\bigskip
\noindent
{\bf 2.1. {\'E}quation de Hamilton-Jacobi quantique}
\medskip

En partant de l'{\'e}quation de Schr{\"o}dinger {\`a}
une dimension

$$
 - {\hbar^2 \over {2m}  } {\partial^2{\psi} \over \partial x^2}+ V\psi = i\hbar
 {\partial\psi \over \partial t  }\ , \xeqno{2.1}
$$
et en {\'e}crivant la fonction d'onde de la forme [19]
$$
\psi(x,t) = A(x,t)\
\exp\left({i\over\hbar }S(x,t)\right)\ ,
    \xeqno{2.2}
$$
on aboutit {\`a}
$$
{1\over 2m }{\left(\partial S \over \partial x\right)}^2-
 {\hbar^2\over 2m }{({\partial ^2 A / \partial x^2}) \over A }+V(x)=
-{\partial S \over \partial t}\ , \xeqno{2.3}
$$
$$
{1 \over 2m} \left [A\ \left( {\partial^2 A /\partial x^2} \right ) +
 2 \ {\partial A \over \partial x}\ {\partial S \over \partial x} \right ]
=-{\partial A \over \partial t}\ , \xeqno{2.4}
$$
$A(x)$ et $S(x)$ {\'e}tant des fonctions r{\'e}elles.
Les Eqs. (2.3) et (2.4) repr{\'e}sentent
respectivement l'EHJQ et
l'{\'e}quation de continuit{\'e}. La quantit{\'e}
$$
 V_B = -{\hbar^2\over {2m} }
{(\partial ^2 A / \partial x^2) \over A }\ , \xeqno{2.5}
$$
est le potentiel quantique de Bohm. Comme il est dit plus haut,
il est responsable des effets quantiques li{\'e}s
au comportement dynamique de la particule.
Vu la ressemblance avec l'EHJC, Floyd appela la quantit{\'e} $S$ qui
appara{\^\i}t dans l'EHJQ ``la fonction de Hamilton principale'',
ou encore ``action''.
\bigskip
\noindent
{\bf 2.2. Probl{\`e}mes stationnaires et {\'e}quation de  Hamilton-Jacobi

quantique stationnaire }
\medskip

En physique la plupart des probl{\`e}mes non-stationnaires
se ram{\`e}nent {\`a} des probl{\`e}mes stationnaires.
Les ph{\'e}nom{\`e}nes stationnaires sont
permanents, alors que les ph{\'e}nom{\`e}nes non stationnaires
sont transitoires.

Dans le cas stationnaire, la densit{\'e} de probabilit{\'e}
$\vert \psi \vert^2$ n'est fonction que de
la \hbox{coordon-\hskip -5pt plus 5pt}\break n{\'e}e $x$. Ainsi,
$$
{\partial A \over \partial t}=0 \ , \xeqno{2.6}
$$
$$
S(x,t)=S_0(x,E)-Et\ . \xeqno{2.7}
$$
$S_0 $ est dite la fonction de Hamilton caract{\'e}ristique,
 c'est l'action r{\'e}duite.
Dans cette situation, les relations (2.3) et (2.4) s'{\'e}crivent:
$$
{1\over 2m }{\left(\partial S_0 \over \partial x\right)}^2-
 {\hbar^2\over 2m }{({\partial ^2 A / \partial x^2}) \over A }+V(x)=E\ ,
 \xeqno{2.8}
$$
$$
A\ \left( {\partial^2 S_0 \over \partial x^2} \right) +
2\ {\partial A \over \partial x}\ {\partial S_0 \over \partial x} =0 \ .
\xeqno{2.9}
$$
De l'Eq. (2.9), on d{\'e}duit  [20,21,22]
$$
A=K \left ({\partial S_0 \over\partial x}\right )^{-1/2} \ . \xeqno{2.10}
$$
En substituant cette derni{\`e}re dans l'Eq. (2.8), on aboutit {\`a}
$$
{1\over 2m} \left({\partial S_0 \over \partial x}\right)^2-
{\hbar^2\over 4m}  \left[{3\over 2}\left(
{\partial S_0 \over\partial x}\right)
^{- 2 }\left({\partial^2 S_0 \over \partial x^2}\right)^2-
\left( {\partial S_0 \over \partial x  }\right)^{- 1 }
\left({\partial^3 S_0 \over \partial x^3  }\right) \right]+V(x)=E \ ,
 \xeqno{2.11}
$$

L'Eq. (2.11) est l'{\'e}quation de Hamilton-Jacobi Quantique Stationnaire
(EHJQS). En injectant l'Eq. (2.10) dans l'Eq. (2.2), on obtient
$$
\psi (x)=K \left ({\partial S_0 \over\partial x}\right )^{-1/2}
\exp {\left[{iS_0(x) \over \hbar} \right ]} \ . \xeqno{2.12}
$$
On remarque que le cas o{\`u} $S_0=$ cte est {\`a} exclure faute
d'ind{\'e}termination dans les Eqs. (2.10) et (2.11). Dans ce cas,
en reprenant l'Eq. (2.8), on peut {\'e}crire
$$
-{\hbar^2\over 2m }{\left( {\partial ^2 A / \partial x^2}\right) \over A }+V(x)=E\ ,
 \xeqno{2.13}
$$
ce qui est identifique {\`a} l'{\'e}quation de Schr{\"o}dinger, $A(x)$
 correspond {\`a} $ \psi(x)$ {\`a} un facteur de phase pr{\`e}s.
On peut voir que pour les {\'e}tats li{\'e}s qui sont d{\'e}crits par
 des fonctions d'ondes r{\'e}elles, le moment conjugu{\'e}
g{\'e}n{\'e}ralis{\'e} d{\'e}fini par
$$
P= {\partial  S_0 \over \partial x}\ ,\xeqno{2.14}
$$
s'annule. Dans ce cas, une approche dynamique s'av{\`e}re
impossible {\`a} faire. Pour rem{\'e}dier {\`a}  cela,
Floyd propose d'{\'e}crire la fonction d'onde sous une nouvelle
forme [19] diff{\'e}rente de celle
 donn{\'e}e par Bohm (Eq. (2.2))
$$
\psi(x)=A(x) \left [ \sigma \cos {\left ({S_0(x) \over \hbar} \right )}+
\beta \sin {\left ({S_0(x) \over \hbar} \right )} \right ]\ , \xeqno{2.15}
$$
o{\`u} $\sigma$ et $\beta$ sont des coefficients r{\'e}els. Avec
cette nouvelle forme de $\psi$, on peut s'assurer que $S_0$ n'est jamais
constant dans le cas des {\'e}tats li{\'e}s. Il est facile
de v{\'e}rifier que la forme (2.15) de la fonction
d'onde $\psi$ nous m{\`e}ne aux
relations (2.10) et (2.11) (voir Ref. [29]).
Finalement, pour l'{\'e}tude du comportement dynamique des objets
quantiques dans le cas des probl{\`e}mes stationnaires,
Floyd propose  de reprendre l'EHJQS donn{\'e}e par
l'Eq. (2.11), et cela en tenant compte de l'Eq. (2.10) et en
liant la fonction d'onde {\`a} l'action
r{\'e}duite par la relation (2.2) dans le cas des fonctions
d'ondes complexes, ou par la relation (2.15) dans le cas des fonctions
d'ondes r{\'e}elles.
\medskip
\noindent
{\bf 2.3. Approche num{\'e}rique et micro-{\'e}tats:}
\medskip
Contrairement {\`a} Bohm qui a propos{\'e} une approche hydrodynamique,
le raisonnement de Floyd consiste {\`a} consid{\'e}rer un
comportement dynamique des
particules.
Il consid{\'e}ra alors, que l'EHJQS est plus fondamentale
que l'{\'e}quation de Schr{\"o}dinger. Ceci {\'e}tait
in{\'e}vitable du fait que cette derni{\`e}re ne contient aucun
aspect dynamique. En fait l'{\'e}quation de Schr{\"o}dinger est
une {\'e}quation d'une onde.
Cette assertion poussa Floyd {\`a} {\'e}tudier l'EHJQ
donn{\'e}e par l'Eq. (2.8) en
introduisant le potentiel modifi{\'e}
$$
U(x,E)= V(x)+V_B \ ,\xeqno{2.16}
$$
o{\`u} $V_B$ est le potentiel quantique de Bohm d{\'e}fini
par (2.5). Il {\'e}crit l'Eq. (2.8) sous la forme
$$
{1 \over 2m} \left({\partial S_0 \over \partial x }\right)^2+U(x,E)=E\ .
\xeqno{2.17}
$$
En {\'e}crivant cette derni{\`e}re {\'e}quation comme
$$
{\partial S_0 \over \partial x}= \left [ 2m \left ( E-U(x,E) \right )
\right]^{1 \over 2} \ , \xeqno{2.18}
$$
et en l'injectant dans l'Eq. (2.11), on arrive {\`a}
$$
U(x,E)+{\hbar^2 \over 8m} \left[ {(\partial ^2 U / \partial  x^2) \over E-U}
\right]+
{5 \hbar^2 \over 32m} \left[{\partial U / \partial  x \over E-U} \right]^2=V(x)
\ . \xeqno{2.19}
$$
Cette {\'e}quation diff{\'e}rentielle en $U$ est du second ordre
par rapport {\`a} $x$. En la r{\'e}solvant, on trouve la forme du
potentiel modifi{\'e} $U(x)$
qui nous donne une description  du comportement de
la particule. A l'aide
d'une approche num{\'e}rique, Floyd d{\'e}montra  en appliquant l'Eq. (2.19)
{\`a} l'oscillateur harmonique la possibilit{\'e} d'existence
d'{\'e}tats de mouvement appel{\'e}es micro-{\'e}tats, que la fonction d'onde
de Schr{\"o}dinger ne d{\'e}tectait pas.

\bigskip
\noindent
{\bf 2.4 {\'E}quation de mouvement
param{\'e}tris{\'e}e par le potentiel modifi{\'e}}
\medskip
Bohm, dans ces travaux, fait correspondre la vitesse de l'{\'e}l{\'e}ment
fluide {\`a} celle du corpuscule
$$
\dot{x}={({\partial S_0 / \partial x}) \over m} \ , \xeqno{2.20}
$$
d'o{\`u} l'on tire
$$
\ddot{x}=-{1 \over m}{\partial U \over \partial x}\ . \xeqno{2.21}
$$
Mais, Floyd assume que l'on doit d{\'e}river l'{\'e}quation de
mouvement {\`a} partir du th{\'e}or{\`e}me de Jacobi
$$
t-t_0= {\partial S_0 \over  \partial E} \ . \xeqno{2.22}
$$
Ainsi, en utilisant la relation (2.18), on peut {\'e}crire
$$
S_0(x,E)=\int^{x} \sqrt{2m \left ( E-U(x,E) \right )}\ dx \ . \xeqno{2.23}
$$
En rempla{\c c}ant dans l'Eq. (2.22) et en d{\'e}rivant par rapport {\`a} $t$,
on aboutit {\`a} [19]
$$
\dot{x}={{\partial S_0 / \partial x} \over m(1-{\partial U / \partial E} )}\ .
\xeqno{2.24}
$$
Cette derni{\`e}re {\'e}quation donne la relation entre
le moment conjugu{\'e} g{\'e}n{\'e}ralis{\'e}
$\partial S_0 / \partial x$ et la vitesse $\dot{x}$. Cette relation
diff{\`e}re de celle de Bohm par le facteur $1-{\partial U / \partial E} $.
Floyd explique ceci par le fait que l'{\'e}nergie n'est plus
une constante s{\'e}parable du potentiel modifi{\'e}, puisque $U$
d{\'e}pend de $E$. Dans ce cas, on trouve que le moment conjugu{\'e}
$\partial S_0 / \partial x$ n'est pas {\'e}gale {\`a}
la quantit{\'e} de mouvement $m \dot{x}$.
Maintenant, en reprenant la relation (2.24) on peut {\'e}crire
$$
\ddot{x}=-{\partial U / \partial x \over m(1-{\partial U / \partial E})^2}+
{2(E-U) \over   m(1-{\partial U / \partial E})^3}
{\partial  ^{2} U \over \partial x \partial E} \ . \xeqno{2.25}
$$
L'{\'e}quation diff{\'e}rentielle (2.19) {\'e}tant du second ordre, apr{\`e}s r{\'e}solution,
le potentiel $U$ d{\'e}pendra de $(x,E)$ et de deux param{\`e}tres suppl{\'e}mentaires
$(a,b)$ jouant le r{\^o}le de constantes d'int{\'e}gration. Ainsi, d'apr{\`e}s (2.19)
il y aurait plusieurs valeurs possibles pour $\ddot{x}$, et
donc diff{\'e}rents mouvements seraient autoris{\'e}s pour la particule.
Comme il est d{\'e}j{\`a} mentionn{\'e} dans le
paragraphe 2.3, la pr{\'e}sence de micro-{\'e}tats dans l'EHJQS
pr{\'e}sente plusieurs \hbox{possibili-\hskip -5pt plus 5pt}\break
t{\'e}s pour le mouvement du corpuscule,
et ces possibilit{\'e}s sont exprim{\'e}es par la relation (2.25).
\bigskip
\noindent
{\bf 2.5. Solution de l'EHJQS:}
\medskip

Maintenant, reprenons l'EHJQS donn{\'e}e par l'Eq. (2.11).
En r{\'e}solvant cette {\'e}quation, Floyd {\'e}crit le moment conjugu{\'e}
sous la forme [21,22,23]
$$
{\partial S_0 \over \partial x }= {\sqrt{2m} \over a\, \phi ^2+b\, \theta ^2
+c\, \phi \theta }\ ,\xeqno{2.26}·
$$
o{\`u} $\theta$ et $\phi$ sont deux solutions r{\'e}elles ind{\'e}pendantes de
l'{\'e}quation de Schr{\"o}dinger
$$
 - {\hbar^2\over 2m }\ {\partial ^2 \psi \over\partial x^2}+ V(x) \psi= E\psi
\ ,
$$
$a$, $b$, $c$ sont des constantes r{\'e}elles telles que $a$, $b>0$ et
 $ab-{c^2/4}>0$.
Les solutions $\theta$ et $\phi$ sont normalis{\'e}es de fa{\c c}on que leur Wronskien
$$
{\cal W}=\theta_x \phi -\theta \phi_x\ .
$$
soit
$$
{\cal W} =\pm{\sqrt{2m} \over \hbar \sqrt{ab-{c^2/4}}}\ ,
$$
Ce Wronskien est une constante puisque $\theta$ et $\phi$
 sont solutions d'une {\'e}quation
\hbox{diff{\'e}rentiel-\hskip-5pt plus 5pt}\break le lin{\'e}aire
et homog{\`e}ne du second ordre, dont le coefficient de la premi{\`e}re
d{\'e}riv{\'e}e est nul. Le fait que les constantes $a$ et $b$ sont positives est
d{\^u} {\`a} la r{\'e}alit{\'e} de
l'action r{\'e}duite $S_0$ [21,22,23].

A partir de l'Eq. (2.26) on peut facilement montrer que la forme
de l'action r{\'e}duite est [21]
$$
S_0=\hbar
\arctan \left \{ {b\, (\theta / \phi) +c/2 \over \sqrt{ab-c^2/4}} \right \}+K
\ . \xeqno{2.27}
$$
\noindent
$K$ est une constante additive qu'on peut mettre {\'e}gale {\`a} $0$.

Cette forme de l'action r{\'e}duite sera d'un apport pr{\'e}cieux pour
la repr{\'e}sentation des trajectoires.
Elle d{\'e}pend de l'{\'e}nergie $E$,
et des constantes $a$, $b$ et $c$. L'EHJQS {\'e}tant du second ordre
par rapport {\`a} $(\partial S_0/\partial x)$, sa solution (2.26)
ne devrait d{\'e}pendre que de deux constantes d'int{\'e}gration.
En cons{\'e}quence, parmi les trois param{\`e}tres $a$, $b$ et $c$,
seulement deux sont ind{\'e}pendantes.

\bigskip
\noindent
{\bf2.6. Les {\'e}tats li{\'e}s et la quantification de l'{\'e}nergie:}
\medskip

Dans cette partie nous allons introduire la variable de l'action.
Elle est d{\'e}finie par la quantit{\'e} [21]
$$
{\cal J}=\oint P dx \ . \xeqno{2.28}
$$
\noindent
En fait, cette quantit{\'e} ressemble {\`a} l'action {\`a}
laquelle est impos{\'e}e la condition de
\hbox{quantifica-\hskip -5pt plus 5pt}\break tion Bohr-Sommerfeld. Cependant,  dans notre cas ${\cal J}$
repr{\'e}sente l'int{\'e}grale
du moment conjugu{\'e} g{\'e}n{\'e}ralis{\'e}
$\partial S_0 / \partial x$,
et non plus l'int{\'e}grale de la quantit{\'e} de mouvement
classique $m \dot{x}$. En utilisant la relation (2.26), on peut {\'e}crire
$$
{\cal J}= 2 \int_{-\infty}^{\infty}
 {\sqrt{2m}\, dx \over a\, \phi ^2+b\, \theta ^2+c\, \phi \theta }\ .
$$
En multipliant et en divisant par le Wronskien dans l'argument de
l'int{\'e}grale on arrive {\`a}
$$
{\cal J}= 2 \int_{-\infty}^{\infty} {\sqrt{2m} \over {\cal W}}
{{d \left( {\theta / \phi} \right) \over dx}\ dx \over
 a+b\, \left( {\theta / \phi} \right)^2+c\, \left( {\theta / \phi} \right)}\ ,
$$
\noindent
ce qui se ram{\`e}ne {\`a}
$$
{\cal J}= 2 \hbar \int_{-\infty}^{\infty} \left( ab-{c^2 / 4}\right)^{1 \over 2}
{{d \left( {\theta / \phi} \right) \over dx}\ dx \over
 a+b\, \left( {\theta / \phi} \right)^2+c\, \left( {\theta / \phi} \right)}\ .
\xeqno{2.29}
$$
Cette relation constitue la d{\'e}finition explicite de la variable de
l'action ${\cal J}$ dans la  \hbox{repr{\'e}sen-\hskip -5pt plus 5pt}\break tation des trajectoires.
Si on veut calculer l'int{\'e}grale de l'Eq. (2.29), alors il faudrait
faire un changement de variables de $ x $ vers $ ( {\theta / \phi})$.
Choisissons $ \phi$ comme {\'e}tant la solution physique du
 probl{\`e}me des {\'e}tats li{\'e}s, et $\theta$ comme {\'e}tant
la solution math{\'e}matique associ{\'e}e {\`a} $\phi$.
Notons $N-1$ le nombre de noeuds pour lesquels la fonction $\phi$
 s'annule [33] ( $\phi$ poss{\`e}de deux autres
points, {\`a} $\pm \infty$,  pour lesquels elle s'annule mais qui
 ne sont pas des noeuds). Cependant, $\theta$ poss{\`e}de $N$ noeuds
({\`a} $\pm\infty$\ , $\theta$ est divergente).
 Consid{\'e}rons d'abord le cas o{\`u} {\`a} la limite $x \to -\infty$, $\phi$
et $\theta$ sont de signes diff{\'e}rents. On a alors
$$
\lim_{x \to -\infty} ( {\theta / \phi} ) =-\infty\ .
$$
Comme les nombres des noeuds des fonctions $\theta$ et $\phi$ sont de parit{\'e}s
diff{\'e}rentes, alors {\`a} la limite $x \to +\infty$, seulement celle dont
le nombre de noeuds est impaire changera de signe. De ce fait
$$
\lim_{x \to +\infty} ( {\theta / \phi} ) =+\infty\ .
$$
De plus,  $({\theta / \phi})$ poss{\`e}de des singularit{\'e}s
aux $N-1$ noeuds $x_n$ de la fonction $\phi$:
$$
\lim_{x \to x_{n}^{+}} ( {\theta / \phi} )=-\infty\ ,
$$
$$
\lim_{x \to x_{n}^{-}} ( {\theta / \phi} ) =+\infty \ ,
$$
En cons{\'e}quence lors du changement de variables de $x$ {\`a}
 $({\theta / \phi})$, et pour $x$ passant de \  $-\infty$
{\`a} \  $+\infty$, la quantit{\'e} $( {\theta / \phi} )$ passe
$N$ fois de \  $-\infty$ {\`a} \  $+\infty$.
Dans ce cas, l'Eq. (2.29) se ram{\`e}nera {\`a}
$$
{\cal J}= 2N \hbar \int_{-\infty}^{\infty}
\left( ab-{c^2 / 4} \right)^{1 \over 2}
{d \left( {\theta / \phi} \right) \  \over
 a+b\, \left( {\theta / \phi} \right)^2+c\, \left( {\theta / \phi} \right)}\ ,
$$
\noindent
ce qui nous donne
$$
{\cal J}= 2N \hbar
\left[ \arctan \left \{ {b\, (\theta / \phi) +c/2 \over \sqrt{ab-c^2/4}}
\right \}\right]_{\theta / \phi=-\infty}^{\theta / \phi=+\infty}\ ,
$$
Enfin, on aura
$$
{\cal J}= 2N \hbar \pi=Nh\ . \xeqno{2.30}
$$
Donc, dans le cas des {\'e}tats li{\'e}s,
la variable de l'action est quantifi{\'e}e,
et ceci quelles que soient $a$, $b$ et $c$. De ce fait, on sait d{\`e}s lors  que
la quantification de la variable de l'action nous conduit directement {\`a}
la quantification de l'{\'e}nergie donn{\'e}e par l'{\'e}quation
de Schr{\"o}dinger.
On remarque aussi que cette quantification est diff{\'e}rente de
celle d{\'e}duite de la m{\'e}thode WKB [34] qui donne la valeur
${\cal J}_{WKB}=(N+{1 \over 2})h $. Nous savons en fait, que
la m{\'e}thode WKB
n'est qu'une m{\'e}thode d'approximation.

\bigskip
\noindent
{\bf 2.7. Les micro-{\'e}tats dans l'EHJQS: }
\medskip

Nous avons consid{\'e}r{\'e} dans le paragraphe 2.3 les
micro-{\'e}tats d{\'e}tect{\'e}s par l'approche du potentiel modifi{\'e}.
Maintenant, nous parlerons de micros-{\'e}tats d{\'e}duits directement
de la solution g{\'e}n{\'e}rale de l'EHJQS. Floyd [22]
fait remarquer que les micro-{\'e}tats
existent pour les {\'e}tats li{\'e}s et que pour les
{\'e}tats non li{\'e}s il n'y aurait pas de micro-{\'e}tats.

Consid{\'e}rons d'abord le cas des {\'e}tats li{\'e}s. En portant les Eqs. (2.10),
(2.26) et (2.27) dans l'Eq. (2.15), on peut {\'e}crire
$$
\psi=(a\ \phi^2+b\ \theta^2 +c\ \theta \phi )
\left \{
\sigma \cos \left[ \arctan \left( {b\ ({\theta / \phi})+{c/2} \over
\sqrt{ab-c^2/4}} \right) \right]+ \right.
$$
$$
\left.\beta \sin \left[ \arctan \left( {b\ ({\theta / \phi})+{c/2} \over
\sqrt{ab-c^2/4}} \right) \right] \right \}
$$
ce qui donne
$$
\psi= \sigma \ \sqrt{a-{c^2 \over 4b}} \ \phi+
\beta \ \sqrt{b} \ \theta+ {c \ \beta \over \sqrt{2b}} \ \phi\ ,
$$
$$
\ \ \ \ \ \ \ \ =\left( \sigma \ \sqrt{a-{c^2 \over 4b}}+ {c \ \beta \over \sqrt{2b}} \
\right) \ \phi+
\beta \ \sqrt{b} \ \theta \ . \     \xeqno{2.31}
$$

On sait que la solution physique de
l'{\'e}quation de Schr{\"o}dinger est la combinaison lin{\'e}aire de deux solutions
math{\'e}matiques $\psi= \alpha \phi^{'} +\beta \theta^{'} $.
Maintenant, en choisissant la paire des fonctions ($\phi$,$\theta$)
qui figurent dans l'expression de l'action r{\'e}duite, de mani{\`e}re que
l'une d'elles (par exemple $\phi$) correspond {\`a} la solution
physique $\psi$, donc $\psi=\phi$, on peut voir que
$$
\beta \ \sqrt{b}=0\ \ \ ; \ \ \
\sigma \ \sqrt{a-{c^2 \over 4b}}+ {c \ \beta \over \sqrt{2b}}=1 \ .
$$
La constante $b$ {\'e}tant strictement positive, alors
$$
\beta =0\ \ \ \ \ \ ; \ \ \ \ \
\sigma \ \sqrt{a-{c^2 \over 4b}}=1 \ . \xeqno{2.32}
$$
Cette derni{\`e}re {\'e}quation ne permet pas de fixer compl{\`e}tement les param{\`e}tres
$a$, $b$, et $c$. Ainsi, pour une m{\^e}me fonction d'onde $\phi$ qui d{\'e}crit
un {\'e}tat physique, nous avons une multitude de choix de $(a,b,c)$
et par cons{\'e}quent, une multitude de trajectoires possibles {\'e}tant donn{\'e} que
le g{\'e}n{\'e}rateur du mouvement $S_0$ est une fonction d{\'e}pendante
de ces param{\`e}tres (Eq. (2.27)). Ces {\'e}tats de mouvement possibles,
correspondant {\`a} la m{\^e}me fonction d'onde d{\'e}finissent les micro-{\'e}tats.

Dans le cas des {\'e}tats non li{\'e}s, la fonction d'onde est complexe, la
relation (2.12) repr{\'e}sentera alors, la liaison entre $\psi$ et $S_0$
$$
\psi (x)=K \left ({\partial S_0 \over\partial x}\right )^{-1/2}
\exp {\left[{iS_0(x) \over \hbar} \right ]}\ .
$$
\noindent
Dans ce cas, on a
$$
\psi=K^{'}(a\ \phi^2+b\ \theta^2 +c\ \theta \phi )
\left \{
 \cos \left[ \arctan \left( {b\ ({\theta / \phi})+{c/2} \over
\sqrt{ab-c^2/4}} \right) \right]+ \right.\ \ \ \ \ \ \ \ \ \ \ \ \ \ \ \ \ \ \ \ \ \ \ \ \
$$
$$
\ \ \ \ \ \ \ \ \ \ \ \ \ \ \ \ \ \ \ \ \ \ \ \ \ \ \ \ \ \ \ \ \ \ \ \ \ \ \ \ \ \ \ \ \ \
\left. i \sin \left[ \arctan \left( {b\ ({\theta / \phi})+{c/2} \over
\sqrt{ab-c^2/4}} \right) \right] \right \}\ ,
$$
Remarquons tout d'abord que le fait que remplacer les constantes $a$, $b$ et
$c$ par les constantes $a^{'}=a/K^{'}$, $b^{'}=b/K^{'}$ et \ $c^{'}=c/K^{'}$
ne change en rien la nature de la fonction d'onde $\psi$,
ou m{\^e}me de l'action $S_0$. Donc, on peut {\'e}crire
$$
\psi=(a^{'}\ \phi^2+b^{'}\ \theta^2 +c^{'}\ \theta \phi )
\left \{
 \cos \left[ \arctan \left( {b^{'}\ ({\theta / \phi})+{c^{'}/2} \over
\sqrt{a^{'}b^{'}-c^{'2}/4}} \right) \right]+ \right. \ \ \ \ \ \ \ \ \ \ \ \ \ \ \ \ \ \ \ \ \ \ \ \ \
$$
$$
\ \ \ \ \ \ \ \ \ \ \ \ \ \ \ \ \ \ \ \ \ \ \ \ \ \ \ \ \ \ \ \ \ \ \ \ \ \ \ \ \ \ \ \ \ \  \left. i \sin \left[ \arctan \left( {b^{'}\ ({\theta / \phi})+{c^{'}/2} \over
\sqrt{a^{'}b^{'}-c^{'2}/4}} \right) \right] \right \}\ ,
$$
$$
=  \left(  \sqrt{a^{'}-{c^{'2} \over 4b^{'}}}
+ {i \over 2}{c^{'} \over \sqrt{b^{'}}} \
\right) \ \phi+
 i \sqrt{b^{'}} \ \theta \ ,  \xeqno{2.33}
$$
\noindent
Comme la solution physique s'{\'e}crit sous la forme de la combinaison
lin{\'e}aire $\psi= \alpha \phi +\beta \theta$
($\alpha$ et $\beta$ sont des nombres complexes), alors
$$
\alpha=\left(  \sqrt{a^{'}-{c^{'2} \over 4b^{'}}}
+ {i \over 2}{c^{'} \over \sqrt{b^{'}}} \right) \ \ \ \ ; \ \ \ \
\beta=i \ \sqrt{b^{'}} \ . \xeqno{2.34}
$$

Du fait que $\alpha$ et $\beta$
sont des constantes d{\'e}termin{\'e}es {\`a} partir des
conditions initiales, il est clair que les deux
relation de (2.34) d{\'e}terminent de mani{\`e}re univoque
les valeurs des deux constantes ind{\'e}pendantes de l'ensemble
($a^{'}$,$b^{'}$,$c^{'}$),
la troisi{\`e}me {\'e}tant d{\'e}termin{\'e}e {\`a} partir
du Wronskien.
Donc, {\`a} la fonction d'onde $\psi$ correspond une
seule trajectoire possible et les micro-{\'e}tats
n'existent pas pour les {\'e}tats non li{\'e}s.
Tous ces r{\'e}sultats concernant les micro-{\'e}tats ont
{\'e}t{\'e} retrouv{\'e}s par la suite par
Bouda [29] en utilisant une forme unique pour
les fonctions d'onde r{\'e}elles et complexes
$$
\psi_E=\left( {\partial S_0 \over \partial x} \right)^{-{1 \over 2}}
\left[ \alpha \exp \left( {{i \over \kappa} S_0}\right)+
\beta \exp \left( {-{i \over \kappa} S_0} \right) \right]\ .
$$
Cette forme a {\'e}t{\'e} d{\'e}j{\`a} introduite
par Faraggi et Matone dans leur
approche par la g{\'e}om{\'e}trie diff{\'e}rentielle (voir Sec.3).

\bigskip
\noindent
{\bf 2.8. {\'E}tude du cas de la particule libre}
\medskip

 Floyd proposa dans l'un de ses articles [23] d'{\'e}tudier la
particule libre. Il choisit alors, la paire de  solutions
ind{\'e}pendantes de l'{\'e}quation de Schr{\"o}dinger suivante:
$$
\phi=\left[E(ab-{c^2 /4}) \right]^{-{1 \over 2}}
\cos \left[ (2mE)^{1 \over 2}{x \over \hbar}\right]\ , \xeqno{2.35}
$$
$$
\theta=\left[E(ab-{c^2 /4}) \right]^{-{1 \over 2}}
\sin \left[ (2mE)^{1 \over 2}{x \over \hbar}\right]\ . \xeqno{2.36}
$$
Dans ce cas, l'action r{\'e}duite (Eq. (2.27)) peut {\^e}tre
{\'e}crite sous la forme
$$
S_0=\hbar
\arctan \left \{ {b\, \tan[(2mE)^{1 \over 2} x/ \hbar] +c/2
\over
(ab-c^2/4)^{1 \over 2}} \right \} \ . \xeqno{2.37}
$$

Pour $a=b$ et $c=0$, on a $S_0=(2mE)^{1 \over 2} x$. Cette derni{`e}re
relation co{\"\i}ncide avec l'action r{\'e}duite classique.
Pour {\'e}tudier le comportement de
l'action r{\'e}duite {\`a} la limite classique, dans un cas o{\`u}
$a\not=b$ et $c\not=0$,  Floyd utilisa la r{\`e}gle de l'Hospital
$$
\lim_{\hbar \to 0}S_0=
{2(ab-c^2/4)^{1 \over 2} \sqrt{2mE}\ x
\over
a+b+\left[(a-b)^2+c^2 \right]^{1 \over 2} \cos \left \{
2(2mE)^{1 \over 2}{x / \hbar}+\cot^{-1}[(b-a)/c]
\right\}}\ .  \xeqno{2.38}
$$

On remarque bien que le facteur \ $({1 / \hbar})$\  pr{\'e}sent dans
l'argument du cosinus induit, {\`a} la limite classique, une certaine
ind{\'e}termination dans l'expression de l'action r{\'e}duite [23].

\noindent
Floyd essaya par la suite d'{\'e}tablir les {\'e}quations de mouvement.
Il {\'e}crit tout d'abord \hbox{l'expres-\hskip-5pt plus 5pt}\break sion du moment conjugu{\'e}
$$
{\partial S_0 \over \partial x}=
{2(ab-c^2/4)^{1 \over 2} \sqrt{2mE}\ x
\over
a+b+\left[(a-b)^2+c^2 \right]^{1 \over 2} \cos \left \{
2(2mE)^{1 \over 2}{x / \hbar}+\cot^{-1}[(b-a)/c]
\right\}}\ .  \xeqno{2.39}
$$

De m{\^e}me que pour l'action r{\'e}duite,
lors du passage {\`a} la limite classique
une \hbox{ind{\'e}termi-\hskip-5pt plus 5pt}\break nation
r{\'e}side dans l'expression du moment conjugu{\'e}.
Pour retrouver les r{\'e}sultats de la m{\'e}canique classique, Floyd
propose de moyenner la valeur du moment conjugu{\'e} sur un cycle du cosinus
qui appara{\^\i}t dans l'expression de ${\partial S_0 / \partial x}$,
et cela tout en utilisant les
tables standard d'int{\'e}grales [35]. Il trouve
$$
<\lim_{\hbar \to 0}{\partial S_0 \over \partial x}>=
\lim_{\hbar \to 0}{(2mE)^{1 \over 2} \over \hbar \pi}
\int_{-\hbar \pi \over (8mE)^{1 / 2}}^{\hbar \pi \over (8mE)^{1 / 2}}
\left( \partial S_0 \over \partial x \right)(E\ ,a\ ,b\ ,c\ ,x+x^{'}) dx^{'}
$$
$$
={2(2mE)^{1 \over 2}(ab-c^2/4)^{1 \over 2} \over (a+b)
[1-{(a-b)^2-c^2 \over (a+b)^2}]^{1 \over 2}}
 \ \ \ \ \ \ \ \ \ \ \ \ \ \ \ \ \ \ \ \ \ \ \ \ \
$$
$$
=(2mE)^{1 \over 2}\; . \ \ \ \ \ \ \ \ \ \ \ \ \ \ \ \ \ \ \ \ \ \ \ \ \ \ \ \ \ \ \ \ \ \ \ \ \ \ \ \ \ \ \  \xeqno{2.40}
$$
On voie clairement que la moyenne
du moment conjugu{\'e} correspond au moment classique,
et que les micro-etats sp{\'e}cifi{\'e}s par les constantes
$a$, $b$ et $c$ disparaissent.
Pour {\'e}tablir l'{\'e}quation de mouvement, Floyd utilise le theoreme
de Jacobi sous sa version classique
$$
{\partial S_0 \over \partial E}=t-t_0\ ,
$$
{\`a} partir duquel il {\'e}crit
$$
t-t_0={(ab-c^2/4)^{1 \over 2} \sqrt{2m/E}\ x
\over
a+b+\left[(a-b)^2+c^2 \right]^{1 \over 2} \cos \left \{
2(2mE)^{1 \over 2}{x / \hbar}+\cot^{-1}[(b-a)/c]
\right\}} \ . \xeqno{2.41}
$$

Tout comme pour le moment conjugu{\'e}, Floyd {\'e}valua la moyenne de $t-t_0$
sur un cycle du cosinus et trouva
$$
<\lim_{\hbar \to 0}(t-t_0)>_{moy}=
{{(2m/E)^{1 \over 2}x}(ab-c^2/4)^{1 \over 2} \over (a+b)
[1-{(a-b)^2-c^2 \over (a+b)^2}]^{1 \over 2}}=
\left( {m \over 2E} \right)^{1 \over 2} x
\ . \xeqno{2.42}
$$
Ce qui correspond au r{\'e}sultat de la m{\'e}canique classique.

Il est clair que le fait de moyenner sur le cycle du cosinus a permis {\`a}
Floyd de retrouver les r{\'e}sultats de la m{\'e}canique classique,
bien que l'ind{\'e}termination r{\'e}siduelle subsiste toujours.
Floyd expliqua cette ind{\'e}termination par le fait qu'en
m{\'e}canique \hbox{classi-\hskip-5pt plus 5pt}\break que,
on utilise une seule constante d'int{\'e}gration $(E)$, alors qu'en
m{\'e}canique quantique on a plus d'une constante
d'int{\'e}gration. De plus, il a fait la remarque que $\hbar$
est une constante qui ne doit pas {\^e}tre trait{\'e}e comme
une variable, donc une approche du type $\hbar \to 0$ n'est pas
justifi{\'e}e.

Il est {\`a} noter que pour la particule libre la relation
$$
\lim_{\hbar \to 0}S_0=2E(t-t_0)\ . \xeqno{2.43}
$$
est v{\'e}rifi{\'e}e. Cette relation peut {\^e}tre directement d{\'e}duite
des relations (2.38) et (2.41). De m{\^e}me, on peut {\'e}crire
la relation suivante
$$
\lim_{\hbar \to 0} S =E(t-t_0)=\int_{t_0}^{t} \;
L^{classique}\; dt=S^{classique}\ . \xeqno{2.44}
$$
\headline={
\vbox{
    \xline{{\craw Chapitre 1.} \hfill  {\craw 3. Postulat d'{\'e}quivalence
    en m{\'e}canique quantique} {\bf \ \ \folio}}
    \medskip \hrule

}
}
\bigskip
\bigskip
\noindent
{\bf 3. POSTULAT D'EQUIVALENCE EN M{\'E}CANIQUE QUANTIQUE}
\bigskip

Dans le cadre de la g{\'e}om{\'e}trie diff{\'e}rentielle, Faraggi
et Matone [24,25,26] ont montr{\'e}
r{\'e}cemment qu'{\`a} partir d'un postulat d'{\'e}quivalence,
on peut reproduire la m{\'e}canique
\hbox{quanti-\hskip-5pt plus 5pt}\break que. En fait, ils
ont pu construire l'EHJQS
d'un syst{\`e}me non relativiste {\`a} une dimension. Ils ont montr{\'e}
que cette d{\'e}rni{\`e}re se ram{\`e}ne {\`a}
l'{\'e}quation de Hamilton Jacobi classique en rempla{\c c}ant
la d{\'e}riv{\'e}  $\partial_x$ par la d{\'e}riv{\'e}
$\partial_{\hat{x}} $, o{\`u} $\hat{x} $ est la coordonn{\'e}e
r{\'e}sultant de la transformation de $x$ d{\'e}finie par
$d\hat{x}={dx \over \sqrt{ 1-{\cal B}^2}}$,
${\cal B}^2$ {\'e}tant proportionnel au potentiel
\hbox{quanti-\hskip-5pt plus 5pt}\break que.

\bigskip
\noindent
{\bf 3.1 {\'E}nonc{\'e} du postulat d'{\'e}quivalence}
\medskip
Avant d'arriver {\`a} {\'e}noncer le postulat d'{\'e}quivalence,
Faraggi et Matone
 \hbox{cherchaient {\`a} \hskip-5pt plus 5pt}\break construire  certaines classes de transformations canoniques qui
permetteraient de relier plusieurs syst{\`e}mes quantiques,
et parmi lesquelles il existerait une transformation qui nous m{\'e}nerait {\`a}
l'EHJQS [24]. Mais, le fait que les {\'e}quations canoniques de
la m{\'e}canique quantique
{\'e}taient inconnues rendait difficile la construction de telles
\hbox{transforma-\hskip-5pt plus 5pt}\break tions. Au cours de leur recherche,
Faraggi et Matone ont remarqu{\'e} qu'en m{\'e}canique classique, deux syst{\`e}mes de
particules $A$ et $B$ d'actions $S_0^{cl}(x)$ et $\tilde{S}_0^{cl}(\tilde{x})$
sont {\'e}quivalent sous une transformation de coordonn{\'e}es d{\'e}finie par
$$
\tilde{S}_0^{cl}(\tilde{x})=S_0^{cl}(x) \ .
$$

Cette remarque n'est plus valable dans le cas o{\`u} l'une des
particules est au repos (action constante).
Donc, en d{\'e}crivant les syst{\`e}mes
physiques par l'EHJCS, on ne peut pas avoir une {\'e}quivalence entre deux
syst{\`e}mes lorsque l'un d'eux est au repos. Pour cette raison,
les deux physiciens {\'e}tait curieux de voir quel type d'{\'e}quation
{\'e}manerait lors d'une transformation de $x \to \tilde{x}$
telle que
$$
\tilde{S_0} ( \tilde{x})=S_0[x(\tilde{x})]\ , \xeqno{3.1}
$$
par laquelle les deux syst{\`e}mes, d{\'e}crits suivant $x$ et
$\tilde{x}$, sont {\'e}quivalents m{\^e}me si l'un d'eux est au repos.
Ainsi, une telle transformation r{\'e}duirait tout syst{\`e}me
quantique {\`a} un syst{\`e}me libre d'{\'e}nergie nulle.
Cette question incita les deux physiciens {\`a}
formuler le postulat d'{\'e}quivalence suivant [24,25,26]:

\bigskip
{\it pour toute paire $ W^a $, $ W^b $, il existe une
transformation de coordonn{\'e}es

$$
x \to \tilde{x}= \tilde{S_0}^{-1} \circ S_0(x) \ , \xeqno{3.2}
$$

pour laquelle on a
$$ W^a(x) \to \tilde {W}^a(\tilde{x})=W^b(\tilde{x}) \ ,
$$
}
\medskip
\noindent
avec $ W(x)=V(x)-E $. $ W^a$ et $ W^b$  constituent deux
vecteurs de l'espace ${\cal H}$ des formes des quantit{\'e}s $W(x)$.

Le principe pr{\'e}c{\`e}dent peut {\^e}tre interpr{\'e}t{\'e} comme suit:
Si on peut passer d'un syst{\`e}me d'action r{\'e}duite $S_0(x)$
{\`a} un autre syst{\`e}me d'action r{\'e}duite
$\tilde {S}_0(\tilde{x})$ via une transformation
$x \to \tilde{x} $, telle que $\tilde {S}_0(\tilde{x})=S_0(x)$,
alors les deux syst{\`e}mes sont dit {\'e}quivalents
s'il existe dans l'espace ${\cal H}$ un vecteur $ W^b$
qui serait le vecteur transform{\'e} de $W^a$. En fait, si le syst{\`e}me
$S_0$ est d{\'e}crit par une
{\'e}quation caract{\'e}ristique qui contient $W^a$ et $S_0(x)$,
alors, le syst{\`e}me $\tilde {S}_0$  sera d{\'e}crit par une
{\'e}quation caract{\'e}ristique de forme apparent{\'e}e {\`a} celle
du syst{\`e}me  $S_0$, mais qui sera {\'e}crite pour
$\tilde {S}_0(\tilde{x})$ et $ W^b(\tilde {x})$.

D'apr{\`e}s ce principe, on peut toujours trouver une transformation
de coordonn{\'e}es qui nous fait passer d'un syst{\`e}me
dont le potentiel et l'{\'e}nergie sont $V(x)$ et $E$ {\`a} un syst{\`e}me
dont le potentiel et l'{\'e}nergie sont nuls.

Maintenant, essayons d'appliquer ce principe dans le cas d'un
syst{\`e}me classique qui est d{\'e}crit par l'{\'e}quation de
Hamilton-Jacobi classique

$$
{1\over 2m} \left({\partial S_{0}^{cl} \over \partial x}\right)^2 +W(x)=0\ .
\xeqno{3.3}
$$
En appliquant alors la transformation d{\'e}finie par l'Eq. (3.2),
on peut {\'e}crire
$$
{1\over 2m} \left({\partial \tilde{S}_{0}^{cl} \over \partial \tilde{x}}\right)^2+
\tilde {W}(\tilde{x})=0 \ ,\xeqno{3.4}
$$
avec
$$
\tilde {W}(\tilde{x})=\left({\partial x  \over \partial \tilde{x}}\right)^2 \,W(x)
 \ . \xeqno{3.5}
$$
D'apr{\`e}s cette derni{\`e}re {\'e}quation, on voit que,
transformer un vecteur $W(x)$ en un vecteur
$\tilde{W}(\tilde{x})=0$ n'est possible que
si le vecteur $W(x)$ est lui-m{\^e}me nul. Ceci contredit le principe
d'{\'e}quivalence pour lequel il serait possible de ramener tout vecteur
$W(x)$ vers le vecteur nul de l'espace $\cal{H}$ des formes des $W$.
Donc, en m{\'e}canique classique le principe d'{\'e}quivalence
n'est  pas v{\'e}rifi{\'e}.
De ce fait, Faraggi et Matone {\'e}taient contraints de chercher
une {\'e}quation diff{\'e}rente de l'EHJCS et pour laquelle
le principe d'{\'e}quivalence est v{\'e}rifi{\'e}.
%
%
\bigskip
\noindent
{\bf 3.2 EHJQS et principe d'{\'e}quivalence}
\medskip

Dans le but de d{\'e}crire les syst{\`e}mes quantiques, Faraggi et Matone
d{\'e}riv{\`e}rent l'{\'e}quation v{\'e}rifiant le principe d'{\'e}quivalence tout
en respectant les propri{\'e}t{\'e}s suivantes [26]:

1)- Lors de la transformation de coordonn{\'e}es $x \to \tilde{x}$
tel que $\tilde{S}_0(\tilde{x})=S_0(x)$, l'{\'e}quation doit garder
la m{\^e}me forme g{\'e}n{\'e}rale par rapport aux nouvelles grandeurs
$\tilde{W}_0(\tilde{x})$ et $\tilde{S}_0(\tilde{x})$.

2)- Lors d'un passage {\`a} la limite classique, notre {\'e}quation
devrait se ramener {\`a} l'{\'e}quation de Hamilton Jacobi classique.

3)- Tous les {\'e}tats $W \in {\cal H}$ sont {\'e}quivalent sous
 la transformation $x \to \tilde{x}$.

La premi{\`e}re propri{\'e}t{\'e} est li{\'e}e {\`a} l'invariance, lors
des transformations de coordonn{\'e}es, des {\'e}quations
qui d{\'e}crivent les syst{\`e}mes physiques.
La deuxi{\`e}me est due {\`a} l'existence
de la m{\'e}canique classique. La troisi{\`e}me
propri{\'e}t{\'e} sera d{\'e}cisive
pour la construction de l'{\'e}quation diff{\'e}rentielle
{\`a} laquelle ob{\'e}it $S_0$.
En ce sens, Faraggi et Matone propos{\`e}rent de chercher la
nouvelle {\'e}quation sous la forme
$$
{1 \over 2m} \left( {\partial S_0 \over \partial x}\right)^2+
W(x)+Q(x)=0\ , \xeqno{3.6}
$$
\bigskip
\noindent
o{\`u} $Q(x)$ est une fonction qui d{\'e}pend de $x$ et qui s'annule
lors du passage {\`a} la limite classique.

Introduisons maintenant, l'identit{\'e} de base suivante [26,36]

\bigskip
$$
\left( {\partial S_0 \over \partial x}\right)^2=
{\kappa^2 \over 2} \left( \left \{ e^{{2i \over \kappa}S_0},x \right \}-
\left \{ S_0,x \right \} \right)\ . \xeqno{3.7}
$$
La quantit{\'e} $\left \{T(q),q \right \}$ d{\'e}finie par
$$
\left \{ T,q\right \}=\left[{3\over 2}\left(
{\partial T \over\partial q}\right)
^{- 2 }\left({\partial^2 T \over \partial q^2}\right)^2-
\left( {\partial T \over \partial q}\right)^{- 1 }
\left({\partial^3 T \over \partial q^3  }\right) \right]
$$
repr{\'e}sente le Schwarzien de la fonction
$T(q)$, alors que $\kappa$ est une constante qui
a la dimension d'une action. Cette identit{\'e} est
math{\'e}matiquement v{\'e}rifi{\'e}e, quelle que soit $\kappa$.
A partir des Eqs.(3.6) et (3.7), on a
$$
W(x)={\kappa^2 \over 4m} \left(\left \{ S_0,x \right \}-
\left \{ e^{{2i \over \kappa}S_0},x \right \} \right)-Q(x)\ . \xeqno{3.8}
$$
Par la suite, Faraggi et Matone d{\'e}montr{\`e}rent [25] que la quantit{\'e}
$$
Q(x)={\kappa^2 \over 4m}\left \{ S_0,x \right \}\ , \xeqno{3.9}
$$
constitue l'unique solution pour $Q(x)$.
Ainsi, {\`a} partir des Eqs. (3.8) et (3.9), on arrive {\`a}
$$
W(x)=-{\kappa^2 \over 4m}\left \{ e^{{2i \over \kappa}S_0},x \right \}
\ . \xeqno{3.10}
$$
Si on remplace l'Eq. (3.9) dans l'Eq. (3.6), on a alors
$$
{1 \over 2m}\left( {\partial S_0 \over \partial x}\right)^2+V(x)-E+
{\kappa^2 \over 4m}\left \{ S_0,x \right \}=0 \ . \xeqno{3.11}
$$
Ces deux derni{\`e}res {\'e}quations sont {\'e}quivalentes.

Notons que, puisqu'en m{\'e}canique classique on ne trouve pas une constante
qui a la dimension d'une action, alors $\kappa$ repr{\'e}sentera
le param{\`e}tre naturel par lequel on arrive {\`a} retrouver
la m{\'e}canique classique.
Effectivement, si $\kappa \to 0$, l'Eq. (3.11) se r{\'e}duira {\`a} l'EHJCS.

\bigskip
\noindent
{\bf 3.3 La r{\'e}solution de l'EHJQS}
\medskip

Nous allons maintenant donner la solution de l'EHJQS introduite par Faraggi
et Matone. Tout d'abord, remarquons que les relations [24,26]
$$
{\partial \over \partial x}(h^{'{1 \over 2}}h^{'-{1 \over 2}})=0
\ , \xeqno{3.12}
$$
et
$$
{\partial \over \partial x}\left[h^{'-1 }
{\partial \over \partial x}(h^{'{1 \over 2}}h^{'-{1 \over 2}}h) \right]=0
\ , \xeqno{3.13}
$$
\noindent
sont v{\'e}rifi{\'e}es quelle que soit la fonction $h(x)$,
$h^{'}(x)$ {\'e}tant la d{\'e}riv{\'e}e de $h$
par rapport {\`a} $x$. De plus, on a
$$
{\partial^2  \over \partial x^2}+{1 \over 2} \left \{ h,x \right \}=
h^{'{1 \over 2}}{\partial \over \partial x}
\left( h^{'-1 }{\partial \over \partial x}(h^{'{1 \over 2}})\right)
\ . \xeqno{3.14}
$$
\noindent
{\`A} partir des Eqs. (3.12), (3.13) et (3.14) on voit que l'{\'e}quation
$$
{\partial^2 \psi \over \partial x^2}+
{1 \over 2} \left \{ h,x \right \} \psi=0
 \xeqno{3.15}
$$
\noindent
a une solution de la forme
$$
\psi=h^{'-{1 \over 2}}(ah+b)\ . \xeqno{3.16}
$$
Si on choisit la fonction $h(x)$ telle que
$$
h(x)=e^{{2i \over \kappa}S_0}\ , \xeqno{3.17}
$$
l'Eq. (3.15) correspondra {\`a} l'{\'e}quation de Schr{\"o}dinger
en identifiant $\kappa$ avec $\hbar$.
\hbox{Effective-\hskip-5pt plus 5pt}\break ment, puisque $h(x)=e^{{2i \over \kappa}S_0}$, alors
$$
\left \{ h(x),x \right \}= \left \{ e^{{2i \over \kappa}S_0},x \right \}\ ,
$$
\noindent
et d'apr{\`e}s l'Eq. (3.10), on a
$$
\left \{ h \ ,x \right \}=-{4m \over \kappa^2}W(x)=
-{4m \over \kappa^2} [V(x)-E]
 \ . \xeqno{3.18}
$$
\noindent
En rempla{\c c}ant la relation (3.18) dans l'Eq. (3.15), on obtient
$$
-{\kappa^2 \over 2m} {d^2\psi \over dx^2}+ V(x)\psi =E\psi
\ . \xeqno{3.19}
$$
Donc, l'EHJQS, telle quelle est donn{\'e}e par l'Eq. (3.10),
g{\'e}n{\`e}re l'{\'e}quation de
Schr{\"o}dinger.
D'apr{\`e}s l'Eq. (3.16), la solution de l'EHJQS est
reli{\'e}e aux deux solutions ind{\'e}pendantes $\theta$
et $\phi$ de l'{\'e}quation de Schr{\"o}dinger par
$$
\theta=-\left( {\partial S_0 \over \partial x} \right)^{-{1 \over 2}}
\kappa^2 e^{-{i \over \kappa}S_0}
\left(ae^{2{i \over \kappa}S_0}+b \right)
\ , \xeqno{3.20.a}
$$
$$
\phi=-\left( {\partial S_0 \over \partial x} \right)^{-{1 \over 2}}
\kappa^2 e^{-{i \over \kappa}S_0}
\left(ce^{2{i \over \kappa}S_0}+d \right)
\ . \xeqno{3.20.b}
$$
Ces deux {\'e}quations nous m{\`e}nent {\`a} la relation suivante
$$
e^{2{i \over \kappa}S_0}=
{-d \theta +b \phi \over c\theta -a \phi} \ . \xeqno{3.21}
$$

\noindent
Enfin, la solution que donne Faraggi et Matone est
$$
S_0={\kappa \over 2i} \ln \left[ {-d \theta +b \phi \over c\theta -a \phi}
\right]\ . \xeqno{3.22}
$$

En fait, cette {\'e}quation ne devrait contenir que trois constantes
d'int{\'e}gration \hbox{ind{\'e}-\hskip-5pt plus 5pt}\break pendantes,
car l'EHJQS est du troisi{\`e}me ordre.
C'est pour cela que Faraggi et Matone {\'e}crivent plut{\^o}t
$$
e^{2{i \over \kappa}S_0}=e^{i \omega} {{\theta / \phi}+i \bar{\ell}
 \over {\theta / \phi}-i \bar{\ell}}\ , \xeqno{3.23}
$$
\noindent
o{\`u} $\omega \in \Re$ et $\ell \in \Im$ sont les
constantes d'int{\'e}gration. $\Re$ et $\Im$ sont
respectivement le corps des nombres r{\'e}els et celui des
nombres complexes. Cette derni{\`e}re {\'e}quation
est {\'e}quivalente {\`a} celle de Floyd (Eq. (2.27)).
L'Eq. (3.23) est {\'e}quivalente {\`a}
$$
S_0=\kappa \arctan \left[
{ \theta +\mu \phi \over \nu \theta + \phi} \right]
\ , \xeqno{3.24}
$$
qui est donn{\'e}e par A. Bouda dans la Ref. [29]
sans avoir eu recours {\`a} la g{\'e}om{\'e}trie
diff{\'e}rentielle. Remarquons aussi que d'apr{\`e}s les
Eqs. (3.20), la solution g{\'e}n{\'e}rale de
l'{\'e}quation de Schr{\"o}dinger peut s'{\'e}crire sous la forme
$$
\psi_E=\left( {\partial S_0 \over \partial x} \right)^{-{1 \over 2}}
\left[ \alpha \exp \left( {{i \over \kappa} S_0} \right)+
\beta \exp \left( {-{i \over \kappa} S_0} \right) \right]\ , \xeqno{3.25}
$$
o{\`u} $\alpha$ et $\beta$ sont des constantes complexes,
et $S_0$ l'action r{\'e}duite li{\'e}e {\`a} l'EHJQS.
Il est tout {\`a} fait clair que la forme de la liaison entre
$\psi_E$ et $S_0$ donn{\'e}e par Faraggi et Matone (Eq. (3.25))
diff{\`e}re de celle donn{\'e}e par Bohm (Eq. (2.2)). Donc,
l'action r{\'e}duite d{\'e}finie par Faraggi et Matone diff{\`e}re de celle
de Bohm.

La forme donn{\'e}e par l'Eq. (3.25) a {\'e}t{\'e}
obtenue par A. Bouda dans
la Ref. [29]. En fait, il a d{\'e}montr{\'e} que cette relation est valable
aussi bien pour les fonctions d'ondes complexes que
pour les fonctions d'ondes r{\'e}elles.
Pour traiter le cas des fonctions d'ondes r{\'e}elles, il posa
 $ \vert \alpha \vert =\vert \beta \vert $.
Bouda a pu aussi {\'e}crire le courant de probabilit{\'e} (Eq. (1.6))
sous la forme
$$
J={(\vert \alpha \vert^2-\vert \beta \vert^2)\over m} A^2
{\partial S_0 \over \partial x} \ .
$$
Donc, pour la fonction d'onde r{\'e}elle, on a un courant de probabilit{\'e}
nul sans avoir pour autant {\`a} annuler le moment conjugu{\'e}
de la particule.
Dans ce cas, $S_0$ ne serait pas constant, et une approche dynamique
serait vraisemblablement possible dans le cas des {\'e}tats li{\'e}s.

\bigskip
\noindent
{\bf 3.4 D{\'e}formation de la g{\'e}om{\'e}trie et coordonn{\'e}e quantique}
\medskip

Tout comme en relativit{\'e} g{\'e}n{\'e}rale, le principe d'{\'e}quivalence en
m{\'e}canique quantique g{\'e}n{\`e}re une d{\'e}formation de la g{\'e}om{\'e}trie de l'espace.

En fait, si on veut ramener l'EHJQS {\`a} une forme qui correspondrait {\`a}
l'EHJCS, il suffit d'appliquer une transformation de
coordonn{\'e}es telle que [36]
$$
\left({\partial x \over \partial \hat{x}} \right)^2=1-
{\hbar^2 \over 2}\left({\partial S_0 \over \partial x} \right)^{-2}
\left \{ S_0,x \right \}
\ . \xeqno {3.26}
$$
Effectivement, si on remplace l'Eq. (3.26) dans l'EHJQS on aboutit {\`a},
$$
{1 \over 2m}\left({\partial \hat{S_0} \over \partial \hat{x}} \right)+
\hat{W}(\hat{x})=0\ . \xeqno {3.27}
$$
avec
$$
\hat{S_0}(\hat{x})=S_0(x)
$$
et
$$
\hat{W}(\hat{x})=W(x(\hat{x}))\ . \xeqno {3.28}
$$
\noindent
De ce fait, l'Eq. (3.27) est {\'e}quivalente {\`a}
l'EHJQS, mais {\'e}crite avec une nouvelle coordonn{\'e}e
$\hat{x}$ d{\'e}finie par (3.26).
La coordonn{\'e}e $\hat{x}$ est dite  {\it coordonn{\'e}e quantique} et
s'{\'e}crit comme suit
$$
\hat{x}=\int^{x} { dx \over \sqrt{1+2m
\left({\partial S_0 / \partial x}\right)^{-2}V_B}} \ . \xeqno{3.29}
$$
o{\`u} $V_B$ est le potentiel quantique d{\'e}fini par
$$
V_B=-{\hbar^2 \over 4m}\left \{ S_0,x \right \} \ .
$$
En utilisant l'EHJQS, on peut {\'e}crire
$$
\hat{x}=\int^{x} {(\partial S_0 / \partial x)\; dx \over \sqrt{2m\left[
E-V(x)\right]}} \ . \xeqno{3.30}
$$
On remarque bien d'apr{\`e}s les Eqs. (3.29) et (3.30) que
le potentiel quantique joue un r{\^o}le dans la d{\'e}formation
de la g{\'e}om{\'e}trie de l'espace.
Il est de m{\^e}me pour le potentiel classique.
Cette nouvelle vue de la m{\'e}canique quantique sera exploit{\'e} dans
le chapitre suivant pour {\'e}tablir les {\'e}quations de mouvement
d'une particule {\`a} une dimension.

\bigskip
\noindent
{\bf3.5 {\'E}quation du mouvement quantique:}
\medskip

Du fait de l'existence en m{\'e}canique quantique
d'une {\'e}quation de Hamilton Jacobi, un raisonnement bas{\'e} sur la
notion de trajectoire devient in{\'e}vitable. Pour cette raison,
on se demandait si l'EHJQS ne d{\'e}riverait pas d'une
{\'e}quation plus fondamentale qui d{\'e}crirait
le mouvement dynamique  de la particule de mani{\`e}re plus explicite.

C'est dans ce sens que Faraggi et Matone ont reprit [24]
les id{\'e}es de Floyd (Sec.2) [19,23], en admettant que le
th{\'e}or{\`e}me de Jacobi {\'e}crit sous la forme
$$
t-t_0= {\partial S_0 \over \partial E}\ ,
$$
\noindent
est d'autant valable pour la m{\'e}canique quantique que pour
la m{\'e}canique classique.
Dans cet {\'e}tat d'esprit, ils reprirent la forme
du moment conjugu{\'e} de Floyd
donn{\'e}e par l'Eq. (2.24) de ce chapitre
$$
P= {\partial S_0 \over \partial x}=
m\dot{x}\; (1-\partial V_B / \partial E ) \ , \xeqno{3.31}
$$
o{\`u} $P$ est le moment conjugu{\'e}.

Maintenant, si on d{\'e}rive l'EHJQS par rapport {\`a} $E$
on arrive {\`a}

$$
m{\partial V_B \over \partial E} =m-P{\partial P \over \partial E}
\ . \xeqno{3.32}
$$
\noindent
En injectant cette relation dans l'Eq. (3.31), on {\'e}crira
l'{\'e}quation
$$
\dot{x}={1 \over \partial _E P} \ , \xeqno{3.33}
$$
qui est aussi classiquement satisfaite. Particuli{\`e}rement, pour la
particule libre on a
$$
E={m \over 2} \dot{x}^2 (1-\partial _E V_B)+V_B \ . \xeqno{3.34}
$$
De la relation (3.34), Faraggi et Matone d{\'e}finirent ``le champ de
masse quantique'' ({\sl the quantum mass field}) par la quantit{\'e} suivante
$$
m_q=m(1-\partial _E V_B) \ , \xeqno{3.35}
$$
\noindent
de telle mani{\`e}re {\`a} avoir
$$
P=m_q \dot{x}\ . \xeqno{3.36}
$$

\noindent
Cette forme du moment conjugu{\'e} nous permet d'{\'e}crire l'EHJQS
en fonction de
$\dot{x}$, $\ddot{x}$ et $\dot{\ddot{x}}$.
De l'Eq. (3.36), on peut d{\'e}duire

$$
P^{'}=m_q^{'}\dot{x}+m_q {\ddot{x} \over \dot{x}}\ \ \ ; \ \ \ \
P^{''}=m_q^{''}\dot{x}+2m_q^{'} {\ddot{x} \over \dot{x}}+
m_q\left( \dot{{\ddot{x}} \over \dot{x}^2}-
{\ddot{x}^2 \over \dot{x}^3} \right) \ , \xeqno{3.37}
$$
le signe `` $'$ '' repr{\'e}sente la d{\'e}riv{\'e}e par rapport {\`a} $x$.
En se rappelant que
$$
V_B=-{\hbar^2 \over 4m} \left \{ S_0,x \right \}=
-{\hbar^2 \over 4m}
\left( {P^{''} \over P}-{3 \over 2}{P^{'2} \over P^2}
\right)\ , \xeqno{3.38}
$$
et en rempla{\c c}ant les relations (3.37) et (3.38) dans l'EHJQS,
on peut {\'e}crire l'{\'e}quation \hbox{dyna-\hskip-5pt plus 5pt}\break mique suivante

$$
{m_q^2 \over 2m}+V(x)-E+{\hbar^2 \over 4m}
\left(
{m_q^{''} \over m_q}-{3 \over 2}{m_q^{'2} \over m_q^2}-
{m_q^{'} \over m_q}{\ddot{x} \over \dot{x}^2}+
{\dot{\ddot{x}} \over \dot{x}^3}-
{5 \over 2}{\ddot{x}^2 \over \dot{x}^4}
\right)=0 \ . \xeqno{3.39}
$$

Pour Faraggi et Matone, l'Eq. (3.39) constitue l'int{\'e}grale
premi{\`e}re de la loi de Newton d{\'e}crivant le comportement
dynamique de la particule [24].

Cette {\'e}quation, bien qu'elle se ram{\`e}ne
{\`a} l'{\'e}quation de conservation
d'{\'e}nergie classique, et malgr{\'e} que sa
solution contient des constantes
d'int{\'e}gration qui jouent le r{\^o}le de variables cach{\'e}es, elle
reste non explicite. Effectivement, pour r{\'e}soudre l'Eq. (3.39),
il faut d'abord conna{\^\i}tre la quantit{\'e} $m_q$,
qui {\`a} son tour d{\'e}pend du potentiel quantique,
lequel doit {\^e}tre d{\'e}termin{\'e}
{\`a} partir de l'EHJQS. Tout ceci fait de l'Eq. (3.39) une {\'e}quation non
explicite et lourde {\`a} manipuler.

\vfill\eject

\headline={}
\centerline{\ftitle CHAPITRE 2}\bigskip
\bigskip
\bigskip

\centerline{\ftitle LA LOI DE NEWTON QUANTIQUE }
\bigskip
\bigskip
\bigskip

\bigskip
\bigskip
\bigskip

\bigskip
\bigskip
\bigskip

Le formalisme de la `` repr{\'e}sentation des trajectoires ''
a {\'e}t{\'e} fond{\'e} par Floyd en admettant que le mouvement d'une particule est
d{\'e}crit par le th{\'e}or{\`e}me de Jacobi [19]
$$
t-t_0={\partial S_0 \over \partial E}\ .
$$
En s'appuyant sur les travaux de Floyd, Faraggi et Matone
ont {\'e}labor{\'e} une {\'e}quation repr{\'e}sentant l'analogue de l'{\'e}quation de
conservation de l'{\'e}nergie [24]. Bien que ces travaux soient
int{\'e}ressants, l'utilisation du th{\'e}or{\`e}me
Jacobi reste non justifi{\'e}e,
puisqu'il est, en m{\'e}canique classique, la cons{\'e}quence
d'une transformation canonique particuli{\`e}re avec laquelle le
nouveau Hamiltonien s'annule. L'absence d'une proc{\'e}dure analogue en m{\'e}canique
quantique, menant {\`a} l'EHJQS, cr{\'e}e un doute sur la validit{\'e} de ce th{\'e}or{\`e}me
tel qu'il est {\'e}crit par Floyd. Pour cette raison,
nous avons jug{\'e} imp{\'e}ratif d'approcher le probl{\`e}me par un autre formalisme
con{\c c}u sp{\'e}cialement pour la m{\'e}canique quantique. Ainsi, on {\'e}tablira dans ce
chapitre une {\'e}quation de mouvement qui d{\'e}crira le comportement dynamique
des particules {\`a} l'{\'e}chelle quantique.

\bigskip
\bigskip
\noindent
{\bf 1. LE FORMALISME DE LA M{\'E}CANIQUE QUANTIQUE}
\bigskip

Dans cette section, on introduira le formalisme analytique qu'on a con{\c c}u
pour la m{\'e}canique quantique.
Tout d'abord, rappelons le formalisme analytique de la
m{\'e}canique classique.

\vfill\eject
\noindent
{\bf 1.1. Formalisme analytique de la m{\'e}canique classique}
\medskip

Consid{\'e}rons un syst{\`e}me physique {\`a} une dimension d{\'e}crit par
la coordonn{\'e}e $x$, et plong{\'e} dans un potentiel $V(x)$.
Le Lagrangien du syst{\`e}me est d{\'e}fini par
$$
L(x\, ,\dot{x})= {m \over 2}\ \dot{x}^2 - V \ . \xeqno{1.1}
$$
L'action $S$ du syst{\`e}me est d{\'e}finie  {\`a} partir du Lagrangien $L$
comme suit
$$
S=\int L \ dt \ . \xeqno{1.2}
$$

Pour d{\'e}duire les {\'e}quations de mouvement, Hamilton
{\'e}non{\c c}a le principe de moindre action suivant:

`` pour un mouvement r{\'e}el, le corpuscule d{\'e}crit un chemin
pour lequel l'action du syst{\`e}me est stationnaire ($\delta S=0)$ ''.

\noindent
Ce principe nous conduit {\`a} l'{\'e}quation de Lagrange
$$
{d \over dt} \left({\partial L \over \partial \dot{x}}\right)-
{\partial L \over \partial x}=0 \ . \xeqno{1.3}
$$

L'invariance du Lagrangien, lors d'une transformation au cours du temps,
nous conduit {\`a} la conservation de la quantit{\'e}
$$
{\partial L \over \partial \dot{x}}\dot{x}-L \ .
$$
Cependant, nous savons que la quantit{\'e} qui se conserve
lors d'une transformation au cours du temps est
l'{\'e}nergie totale du syst{\`e}me. Donc, on a
$$
E={\partial L \over \partial \dot{x}}\dot{x}-L \ . \xeqno{1.4}
$$
L'invariance du Lagrangien, lors d'un d{\'e}placement virtuel
$\delta x$, nous conduit {\`a} la \hbox{conserva-\hskip-2pt plus 2pt}\break
tion de la quantit{\'e}
$$
p={\partial L \over \partial \dot{x}} \ , \xeqno{1.5}
$$
qui co{\"\i}ncide avec la quantit{\'e} de mouvement du syst{\`e}me.
A partir des Eqs. (1.4) et (1.5), on peut d{\'e}finir la
fonction de Hamilton (Hamiltonien) par la quantit{\'e}

\headline={
\vbox{
    \xline{{\craw  Chapitre 2.} \hfill {\craw 1.Le formalisme de la m{\'e}canique
    quantique} {\bf \ \ \folio}}
    \medskip \hrule
}
}
\noindent

$$
H=p\, \dot{x} -L \ , \xeqno{1.6}
$$
qui correspond {\`a} l'{\'e}nergie E du syst{\`e}me dans le cas
stationnaire.

Injectons l'Eq. (1.1) dans l'Eq. (1.5), et rempla{\c c}ons la relation obtenue
dans l'Eq. (1.6). On obtient alors
$$
H={p^2 \over 2m}+V(x) \ . \xeqno{1.7}
$$
Dans cette derni{\`e}re {\'e}quation, $H$ n'est fonction que de $p$ et $x$.
Maintenant, tout en tenant compte des Eqs. (1.3) et (1.5),
diff{\'e}rencions les deux membres de l'Eq. (1.6). On obtient
alors la relation
$$
\left({\partial H \over \partial x}+\dot{p}\right)\, dx+
\left({\partial H \over \partial p}-\dot{x}\right) \, dp =0 \ ,
$$
qui nous m{\`e}ne aux {\'e}quations canoniques

%
%
$$
{\partial H \over \partial p}=\dot{x} \ ,
$$
$$
\xeqno{1.8}
$$
$$
\ \  {\partial H \over \partial x}=-\dot{p} \ ,
$$
lesquelles d{\'e}crivent le mouvement du corpuscule.

Maintenant, envisageons une transformation de la coordonn{\'e}e
$
x \to Q \ ,
$
accompagn{\'e}e de la transformation
du moment g{\'e}n{\'e}ralis{\'e}e
$
p \to \Phi \ .
$
Ces transformations sont dites \hbox{canoni-\hskip-2pt plus 2pt}\break
ques s'il existe une fonction
${\cal K}(Q\ ,\Phi)$ telle que les {\'e}quations canoniques par rapport
aux nouvelles variables soient

$$
{\partial {\cal K} \over \partial \Phi}=\dot{Q} \ ,
$$
$$
\xeqno{1.9}
$$
$$
\ \ \  {\partial {\cal K} \over \partial Q}=-\dot{\Phi} \ .
$$
${\cal K}$ joue le r{\^o}le du nouveau Hamiltonien du syst{\`e}me.
Pour toute transformation canonique,
il existe une fonction g{\'e}n{\'e}ratrice
$F(x\ ,\Phi,t)$ qui v{\'e}rifie les relations
$$
{\partial F \over \partial x}=p \ ,
$$
$$
\ \ \  {\partial F \over \partial \Phi}=-Q \ ,   \xeqno{1.10}
$$
$$
\ \ \ \ \ \ \  {\partial F \over \partial t}={\cal K}-H \ .
$$

Dans la th{\'e}orie de Jacobi, une transformation particuli{\`e}re nous
m{\`e}ne {\`a} un syst{\`e}me avec un Hamiltonien nul. Dans ce cas,
la fonction $F$ correspond {\`a}
l'action $S$ du syst{\`e}me. En rempla{\c c}ant les relations (1.10) dans
l'{\'e}quation (1.7), on obtient l'EHJC
$$
{1 \over 2m}\left({\partial S \over \partial x} \right)^2+V(x)=
{\partial S \over \partial t}\ . \xeqno{1.11}
$$
Dans le cas stationnaire, l'action du syst{\`e}me s'{\'e}crit comme suit
$$
S=S_0(x \ , E)-Et \ ,\xeqno{1.12}
$$
$S_0$ {\'e}tant l'action r{\'e}duite. Ceci revient
{\`a} {\'e}crire l'Eq. (1.11) sous la forme
$$
{1 \over 2m}\left({\partial S_0 \over \partial x } \right)^2+V(x)=E \ . \xeqno{1.13}
$$
Les {\'e}quations de mouvement se r{\'e}duisent {\`a}
$$
{\partial S_0 \over \partial x }=p \ , \xeqno{1.14.a}
$$
et
$$
\ \ \ \ \ \ {\partial S_0 \over \partial E\ }= t-t_0 \ . \xeqno{1.14.b}
$$
Cette derni{\`e}re {\'e}quation constitue le th{\'e}or{\`e}me de Jacobi.
Ce dernier est valable pour le cas d'une {\'e}quation
de Hamilton-Jacobi de premier ordre.
 Il ne serait donc pas correct de l'appliquer
--- du moins par sa version classique --- en m{\'e}canique
quantique, puisque l'EHJQS est du troisi{\`e}me ordre.
C'est pour cette raison qu'il est n{\'e}cessaire de construire
un nouveau formalisme pour la m{\'e}canique quantique.

\bigskip
\noindent
{\bf 1.2. Lagrangien du syst{\`e}me quantique et {\'e}quation

de mouvement de premi{\`e}re esp{\`e}ce }
\medskip
\noindent
{\it 1.2.1. Motivation:}

Par comparaison avec l'action r{\'e}duite classique, on remarque que
l'expression de $S_0$ (Eq. (2.27)) en m{\'e}canique
quantique contient deux constantes
suppl{\'e}mentaires $\mu$ et $\nu$ en plus de la constante $E$
contenue implicitement dans
les fonctions d'ondes $\theta$ et $\phi$.
Ceci sugg{\`e}re que la loi fondamentale du mouvement doit
{\^e}tre une {\'e}quation diff{\'e}rentielle du quatri{\`e}me ordre.
Ce qui veut dire que le Lagrangien du syst{\`e}me doit {\^e}tre
une fonction de $x$,$\dot{x}$, $\ddot{x}$ et peut {\^e}tre de
$\dot{\ddot{x}}$ avec une d{\'e}pendance lin{\'e}aire. Cependant,
il n'est pas facile de construire,
{\`a} partir d'un tel Lagrangien, un formalisme qui reproduirait l'EHJQS.
Pour surmonter cette difficult{\'e}, on propose un Lagrangien qui d{\'e}pend de
$x$, $\dot{x}$ et de l'ensemble des variables cach{\'e}es $\Gamma$
qui est li{\'e} au constantes d'int{\'e}gration.

\medskip
\noindent
{\it 1.2.2. Hypoth{\`e}ses:}
\medskip

On pr{\'e}sentera maintenant un ensemble d'hypoth{\`e}ses
qu'on utilisera dans ce chapitre pour {\'e}tablir
les {\'e}quations du mouvement.

a)- On supposera que, comme en m{\'e}canique classique, l'action
quantique $S$ d{\'e}duite de l'EHJQS est li{\'e}e {\`a} une
quantit{\'e} $L_q$, appel{\'e}e Lagrangien quantique, par la relation
$$
S=\int L_q \ dt \ , \xeqno{1.15} \ .
$$
Comme $S$ n'est fonction que de $x$ et de $t$, alors $L_q$
ne pourra {\^e}tre qu'une fonction de $x$ et $\dot{x}$.

b)- La deuxi{\`e}me hypoth{\`e}se est que le Lagrangien quantique, dans
le cas stationnaire, s'{\'e}crit comme suit [37]
$$
L_q(x,\dot{x},\Gamma)={m\over 2}  {\dot{x}}^2 f(x,\Gamma)-V(x) \ ,
\xeqno{1.16}
$$
o{\`u} $f(x$,$\Gamma)$  est une fonction de $x$ et de l'ensemble $\Gamma$
des variables cach{\'e}es. Cette fonction doit {\^e}tre continue et
d{\'e}rivable sur $\Re$, et doit v{\'e}rifier la relation
$$
\lim_{\hbar \to 0}f(x,\Gamma)=1 \ .\xeqno{1.17}
$$
La fonction $f(x,\Gamma)$ sera d{\'e}termin{\'e}e ult{\'e}rieurement.

c)- Les {\'e}quations de mouvement d{\'e}rivent d'un principe de
moindre action, de la m{\^e}me mani{\`e}re qu'en m{\'e}canique
classique. Donc,
$$
\delta S =0 \ .\xeqno{1.18}
$$

Dans le cadre de ces trois hypoth{\`e}ses, nous allons construire
une {\'e}quation dynamique du mouvement.

\bigskip
\noindent
{\it 1.2.3. {\'E}quation de mouvement de premi{\`e}re esp{\`e}ce}
\medskip

En tenant compte des  hypoth{\`e}ses du paragraphe pr{\'e}c{\'e}dent,
on arrive {\`a} {\'e}crire
\hbox{l'{\'e}qua-\hskip-5pt plus 5pt}\break tion
de Lagrange suivante:
$$
{d \over dt} \left({\partial L_q \over \partial \dot{x}} \right)-
{\partial L_q \over \partial x}=0 \ . \xeqno{1.19}
$$
En utilisant l'Eq. (1.16), on peut {\'e}crire
$$
{\partial L_q \over \partial \dot{x}}=m \dot{x} f(x\, ,\Gamma)\ ,\ \ \ \ \
\xeqno{1.20}
$$
et
$$
{\partial L_q \over \partial x}=
{m \over 2} \dot{x}^2 {df \over dx}-
{dV \over dx} \ .\ \   \xeqno {1.21}
$$
En rempla{\c c}ant ces derni{\`e}res {\'e}quations dans l'Eq. (1.19),
on arrive {\`a}
$$
{d \over dt} \left[ m \dot{x} f(x,\Gamma) \right]-
{m\dot{x}^2 \over 2}  {df \over dx}+
{dV \over dx}=0 \ ,
$$
ce qui nous donne [37]
$$
m  f(x\ ,\Gamma)\ddot{x}+
{m\dot{x}^2 \over 2}  {df \over dx}+{dV \over dx}=0   \ .
\xeqno{1.22}
$$
On appellera l'Eq. (1.22) l'{\'e}quation dynamique de premi{\`e}re esp{\`e}ce.
Remarquons que lorsque $\hbar \to 0$,  $f \to 1$ et l'Eq. (1.22)
se ram{\`e}ne {\`a} l'{\'e}quation dynamique classique ({\'e}quation de Newton).

\bigskip
\noindent
{\bf 1.3. Hamiltonien quantique}
\medskip

En m{\'e}canique classique, on sait que si le Lagrangien
d{\'e}pend seulement de $x$ et $\dot{x}$, la quantit{\'e} qui se conserve
lors d'une transformation dans le temps est l'Hamiltonien. Il est
de m{\^e}me en m{\'e}canique quantique. Donc, on a
$$
H_q={\partial L_q \over \partial \dot{x}} \dot{x}-L_q(x,\dot{x},\Gamma) \ .
\xeqno{1.23}
$$
De plus, la quantit{\'e} qui se conserve lors d'une translation dans l'espace
est
$$
P={\partial L_q \over \partial \dot{x}}=m \dot{x} f(x\ ,\Gamma) \ ,
\xeqno{1.24}
$$
ce qui correspond {\`a} la quantit{\'e} de mouvement quantique.
L'Eq. (1.24) permet d'{\'e}crire
\hbox{l'Eq.\hskip-5pt plus 5pt}\break(1.23) sous la forme
$$
H_q=P \dot{x}-L_q(x\ ,\dot{x},\Gamma) \ .
\xeqno{1.25}
$$
De m{\^e}me, en utilisant les Eqs. (1.16) et (1.24),
l'Eq. (1.23) peut {\^e}tre {\'e}crite  sous la forme
$$
H_q={m \dot{x}^2 \over 2}  f(x\ ,\Gamma)+
V(x) \ , \xeqno{1.26}
$$
ou encore
$$
H_q={P^2 \over 2m}{1 \over f(x\ ,\Gamma)}+
V(x) \ . \xeqno{1.27}
$$
Pour des probl{\`e}mes stationnaires, $H_q$ correspond
{\`a} l'{\'e}nergie $E$ du syst{\`e}me. Ainsi, les {\'e}quations (1.26)
et (1.27) se r{\'e}duisent {\`a}
$$
E={m \over 2} \dot{x}^2 f(x\ ,\Gamma)+V(x) \ .
  \xeqno{1.28.a}
$$
$$
E= {P^2 \over 2m}{1 \over f(x\ ,\Gamma)}+V(x) \xeqno{1.28.b}
$$
Remarquons qu'en d{\'e}rivant l'Eq. (1.28.a)
par rapport {\`a} $x$, on retrouve exactement
\hbox{l'{\'e}qua-\hskip -5pt plus 5pt}\break tion
dynamique de premi{\`e}re esp{\`e}ce (1.22).

\bigskip
\noindent
{\bf 1.4. D{\'e}termination de la fonction $f(x,\Gamma)$}
\medskip

Nous savons que l'action quantique s'{\'e}crit
\medskip
$$
S=\int L_q dt \ ,
$$
donc, en diff{\'e}renciant on arrive {\`a}
$$
dS=L_qdt=(P \dot{x}-H_q) \ dt \ ,
$$
$$
\ \ =P dx-H_q dt \ .\ \ \ \ \ \ \ \ \ \  \xeqno{1.29}
$$
Comme $S$ ne d{\'e}pend que de $x$, $t$ et $\Gamma$ on a
$$
dS={\partial S \over \partial x} dx+{\partial S \over \partial t} dt
\ . \xeqno{1.30}
$$
En identifiant l'Eq. (1.29) avec l'Eq. (1.30), on peut {\'e}crire
$$
P={\partial S \over \partial x} \ , \xeqno{1.31}
$$
$$
H_q=-{\partial S \over \partial t} \ . \xeqno{1.32}
$$

\noindent
En rempla{\c c}ant les Eqs. (1.31) et (1.32) dans l'Eq. (1.27),
on obtient
$$
-{\partial S \over \partial t}={1 \over 2m }\left( {\partial S \over \partial x} \right)^2
{1 \over f(x\ ,\Gamma)}+V(x) \ , \xeqno{1.33.a}
$$
repr{\'e}sentant l'EHJQ. Dans le cas stationnaire, on a
$$
S=S_0(x,\Gamma)-Et\ , \ \ \ P={\partial S_0 \over \partial x}\ ,\ \ \
{\partial S \over \partial t}=-E\ ,
$$
et l'Eq. (1.33.a) prend la forme
$$
E={1 \over 2m }\left( {\partial S_0 \over \partial x} \right)^2
{1 \over f(x\ ,\Gamma)}+V(x) \ . \xeqno{1.33.b}
$$
L'Eq. (1.33.b) repr{\'e}sente l'EHJQS. En l'identifiant avec
l'EHJQS construite {\`a} partir de l'{\'e}quation de Schr{\"o}dinger et donn{\'e}e par l'Eq. (2.11) du Chap.1, on obtient [37]
$$
f(x\ ,\Gamma)=\left[
1-{\hbar^2 \over 2m} \left( {\partial S_0 \over \partial x} \right)^{-2}
\left \{ S_0 \ , x \right \} \right]^{-1}\ , \xeqno{1.34}
$$
$S_0$ {\'e}tant donn{\'e}e par
$$
S_0=\hbar \arctan \left \{ {\theta+\mu \phi \over \nu \theta+ \phi} \right \}
\ .
$$
En substituant cette expression de $S_0$ dans l'Eq. (1.34),
on voit clairement que $f$ est une fonction de $x$, des param{\`e}tres
$(\mu,\nu)$ et de l'{\'e}nergie $E$. Ainsi, l'ensemble des
variables cach{\'e}es $\Gamma$ correspond
{\`a} l'ensemble des constantes d'int{\'e}gration qui apparaissent
dans l'expression de $S_0$.
$$
\Gamma \equiv \{\mu,\nu,E\}
$$
En utilisant l'EHJQS, on peut {\'e}crire l'Eq. (1.34) sous la forme suivante:
$$
f(x,E,\mu,\nu,)={\left( {\partial S_0 / \partial x} \right)^{2} \over 2m(E-V(x))}
\ . \xeqno{1.35}
$$
Remarquons que lorsque $\hbar \to 0$, la fonction $f$ tend vers 1,
l'Eq. (1.35) indique que
$$
{\partial S_0 \over \partial x}= \sqrt{2m(E-V(x))}\ ,
$$
ce qui correspond bien au r{\'e}sultat de la m{\'e}canique classique.
Notons aussi, d'apr{\`e}s l'Eq. (1.35), que dans les r{\'e}gions classiquement
permises la fonction $f$ est positive. Cependant, dans les r{\'e}gions
classiquement interdites
la fonction $f$ est de signe n{\'e}gatif, ce qui fait que la partie cin{\'e}tique de
l'{\'e}nergie de la particule devient n{\'e}gative bien que
la vitesse $\dot{x}$ soit toujours r{\'e}elle. Ceci explique, dans une premi{\`e}res approche, l'effet tunnel connu en m{\'e}canique quantique.
\bigskip
\noindent
{\bf 1.5. {\'E}quation de dispersion et {\'e}quation dynamique de deuxi{\`e}me esp{\`e}ce}
\medskip
En rempla{\c c}ant l'Eq. (1.35) dans l'Eq. (1.28.a), on peut {\'e}crire
$$
E={m \dot{x}^2 \over 2}  {\left( {\partial S_0 / \partial x} \right)^{2} \over 2m(E-V(x))}+
V(x)\ .
$$
Comme il est indiqu{\'e} par les Eqs. (1.24) et (1.35), la vitesse $\dot{x}$
et le moment conjugu{\'e} ${\partial S_0 / \partial x}$ sont de m{\^e}me
signe dans les r{\'e}gions classiquement permises. Pour cette raison,
dans l'{\'e}quation pr{\'e}c{\'e}dente,
on ne tient compte que de la racine qui est en accord avec cette
constatation. Ainsi, on peut {\'e}crire [37]
$$
\dot{x}{\partial S_0 \over \partial x}=2(E-V(x)) \ . \xeqno{1.36}
$$

On remarque dans l'Eq. (1.36) une sorte de
dispersion de l'{\'e}nergie cin{\'e}tique de la particule entre la
quantit{\'e} de mouvement classique $m \dot{x}$ et le moment
conjugu{\'e} g{\'e}n{\'e}ralis{\'e} ${\partial S_0 / \partial x}$.
C'est pour cela qu'on appellera cette {\'e}quation ``la relation de
dispersion''. On note aussi que le
moment conjugu{\'e} n'est plus {\'e}gale {\`a} la quantit{\'e}
de mouvement $m \dot{x}$.

Maintenant, en rempla{\c c}ant la relation (1.24) dans l'Eq. (1.22)
tout en sachant que ${\partial S_0 / \partial x}=
{\partial L /\partial \dot{x}}$
on peut {\'e}crire la relation
$$
{1 \over 2}\; {d \over dt}\; \left({\partial S_0 \over \partial x} \right)+{1 \over 2}\;
{\ddot{x} \over \dot{x} }\; \left({\partial S_0 \over \partial x} \right)+{dV \over dx}=0
\ . \xeqno {1.37}
$$
Cette derni{\`e}re {\'e}quation est {\'e}quivalente {\`a} l'Eq. (1.22). On l'appellera {\'e}quation dynamique de deuxi{\`e}me
esp{\`e}ce.
Remarquons qu'en l'int{\'e}grant, on aboutit {\`a} l'Eq. (1.36).
\hbox{Effective-\hskip -5pt plus 5pt}\break ment, si on multiplie l'Eq. (1.37) par $\dot{x}$, on
peut {\'e}crire
$$
{1 \over 2}\;  \dot{x}\; {d \over dt}\left({\partial S_0 \over \partial x}
\right)+{1 \over 2}\;
\ddot{x}\;  {\partial S_0 \over \partial x} +{dV \over dt}=0 \ ,
$$
ou encore
$$
{1 \over 2}\left[{d \over dt}\left(\dot{x}{dS_0 \over dx}\right)\right]+
{dV \over dt}=0 \ .
$$
En int{\'e}grant cette {\'e}quation, on obtient
$$
{1 \over 2}\;\dot{x}\;{\partial S_0 \over \partial x}+V=E \ .
$$
Dans le membre de droite, la constante d'int{\'e}gration $E$ est
identifi{\'e} {\`a} l'{\'e}nergie du fait qu'{\`a} la limite classique,
$\hbar \to 0$, on doit trouver un r{\'e}sultat classique. L'{\'e}quation
pr{\'e}c{\'e}dente se ram{\`e}ne facilement {\`a} la forme (1.36), c'est
l'{\'e}quation de dispersion.

Cette derni{\`e}re {\'e}quation est l'{\'e}quation de dispersion.
C'est donc une {\'e}quation int{\'e}grale {\`a} partir de
laquelle on peut d{\'e}duire les {\'e}quations de la trajectoire
$x(t)$.

Notons que dans le cas classique,
$(\partial S_0^{cl} / \partial x)=m\dot{x}$ et l'Eq. (1.36)
se ram{\`e}ne {\`a} l'{\'e}quation de conservation de l'{\'e}nergie
$$
E={1 \over 2}m \dot{x}^2 + V(x)\ .
$$
Ainsi, l'Eq. (1.36) est une g{\'e}n{\'e}ralisation {\`a}
la m{\'e}canique quantique
de l'{\'e}quation de conservation de l'{\'e}nergie.
On verra dans les sections suivantes de ce chapitre l'importance
et le r{\^o}le que joue la relation de dispersion. En fin de compte,
l'Eq. (1.36) nous donnera une description compl{\`e}te du comportement dynamique
de la particule.

\bigskip
\bigskip
\noindent
{\bf 2. {\'E}QUATION INT{\'E}GRALE PREMI{\`E}RE
DE LA LOI DE NEWTON

QUANTIQUE (IPLNQ)}
\medskip
\headline={
\vbox{
    \xline{{\craw Chapitre 2.} \hfill {\craw 2.{\'E}quation IPLNQ}
{\bf \ \ \folio}}
    \medskip \hrule

}
}

\noindent
{\bf 2.1. {\'E}tablissement de l'{\'e}quation IPLNQ}
\medskip
\noindent
{\it 2.1.1 Premi{\`e}re m{\'e}thode}
\medskip
Du fait que
$$
{d\dot{x} \over dx}={d\dot{x} \over dt}{dt \over dx}=
{\ddot{x} \over \dot{x}}\ ,
$$
$$
{d\ddot{x} \over dx}={d\ddot{x} \over dt}{dt \over dx}=
{\dot{\ddot{x}} \over \dot{x}}\ ,
$$

En utilisant l'Eq. (1.36), on peut {\'e}crire

$$
{\partial^2S_0 \over \partial x^2}=-{2\over \dot{x}}
{dV \over dx}-{2(E-V){\ddot{x}}
\over \dot{x}^3} \ , \ \ \ \ \ \ \ \ \ \ \ \ \ \ \ \ \ \ \ \ \ \ \ \ \ \ \ \
\ \ \ \ \ \ \ \ \ \ \ \ \ \ \ \   \xeqno{2.1}
$$
et
$$
{\partial^3S_0 \over \partial x^3}=-{2\over \dot{x}}
{d^2 V\over dx^2}+{4\ddot{x}\over {\dot{x}}^3} {dV \over dx}+
{6(E-V){\ddot{x}}^2\over {\dot{x}}^5}-
{2(E-V){\dot{\ddot{x}}} \over {\dot{x}}^4} \; . \ \ \ \ \ \ \ \ \ \ \ \   \xeqno{2.2}
$$
En injectant les Eqs. (2.1) et (2.2) dans l'EHJQS (Eq. (2.11) du Chap. 1),
on obtient [37]
$$
(E-V)^4-{m{\dot{x}}^2 \over 2}(E-V)^3+{{\hbar}^2 \over 8}
{\left[{3 \over 2}
{\left({\ddot{x} \over \dot{x}}\right)}^2-{\dot{\ddot{x}} \over \dot{x}}
\right]} (E-V)^2\ \ \ \ \ \ \ \ \ \ \ \ \ \ \ \ \ \ \ \ \ \ \ \ \ \ \ \
$$
$$
\ \ \ \ \ \ \ \ \ \ \ \ \ \ -{{\hbar}^2\over 8}{\left[{\dot{x}}^2
{d^2 V\over dx^2}+{\ddot{x}}{dV \over dx}
 \right]}(E-V)-{3{\hbar}^2\over 16}{\left[\dot{x}
{dV \over dx}\right]^2}=0 \; . \xeqno{2.3}
$$
Puisqu'elle d{\'e}pend de la constante d'int{\'e}gration $E$,
cette derni{\`e}re {\'e}quation repr{\'e}sente l'{\'e}quation
Int{\'e}grale Premi{\`e}re de la Loi de Newton Quantique
(IPLNQ). Elle est une {\'e}quation diff{\'e}rentielle du
troisi{\`e}me ordre en $x$ contenant la premi{\`e}re et
la deuxi{\`e}me d{\'e}riv{\'e}es du potentiel par rapport {\`a} $x$.
Il s'en suit que la solution $x(t,E,a,b,c)$ de l'Eq. (2.3)
contient quatre constantes d'int{\'e}gration qui peuvent {\^e}tre
d{\'e}termin{\'e}es {\`a} partir des conditions
\hbox{initia-\hskip -5pt plus 5pt}\break les
$$
x(t_0)=x_0\ , \ \ \ \  \dot{x}(t_0)= \dot{x}_0\ , \ \ \ \
\ddot{x}(t_0)= \ddot{x}_0\ , \ \ \ \  \dot{\ddot{x}}(t_0)= \dot{\ddot{x}}_0\ .
$$
On peut se passer de l'une de ces conditions lorsque l'{\'e}nergie
$E$ de la particule est connue. Bien s{\^u}r, si on pose $\hbar=0$,
l'Eq. (2.3) se r{\'e}duit {\`a} la premi{\`e}re int{\'e}grale de la loi
classique de Newton $E=m\dot{x}^2/2+V(x)$. Si on r{\'e}sout (2.3)
par rapport {\`a} $(E-V)$, alors en d{\'e}rivant les racines obtenues
par rapport {\`a} $x$, on obtient la loi de Newton quantique.
Cette derni{\`e}re sera une {\'e}quation diff{\'e}rentielle
du quatri{\`e}me ordre par rapport {\`a} $x$ et contiendra la
premi{\`e}re, la deuxi{\`e}me
et la troisi{\`e}me d{\'e}riv{\'e}es de $V$ par rapport {\`a} $x$,
alors que la loi classique
$m\ddot{x}=-dV/dx$ est du deuxi{\`e}me ordre et contient seulement la
premi{\`e}re d{\'e}riv{\'e}e de $V(x)$. Notons aussi qu'en
utilisant l'Eq. (1.36) dans l'equation IPLNQ, nous obtenons l'EHJQS.

En comparant  l'{\'e}quation IPLNQ (2.3)
 avec l'{\'e}quation dynamique {\'e}tablie par Faraggi et Matone,
on trouve qu'elles sont fondamentalement
diff{\'e}rentes, bien qu'elles sont de m{\^e}me ordre. En effet,
pour r{\'e}soudre l'{\'e}quation de Faraggi et Matone il faut r{\'e}soudre
en premier lieu l'EHJQS pour obtenir l'action $S_0$ qui d{\'e}pend
des solutions de l'{\'e}quation de Schr{\"o}dinger. Ainsi, elle ne peut pas {\^e}tre
r{\'e}solue ind{\'e}pendemment de l'{\'e}quation de Schr{\"o}dinger. Par contre,
notre {\'e}quation peut {\^e}tre r{\'e}solue comme n'importe quelle {\'e}quation
diff{\'e}rentielle ind{\'e}pendemment de l'EHJQS et de l'{\'e}quation de Schr{\"o}dinger.

\bigskip
\noindent
{\it 2.1.2 deuxi{\`e}me m{\'e}thode}
\medskip

Reprenons la forme de l'action r{\'e}duite donn{\'e}e par l'Eq. (3.24)
du Chap. 1
$$
S_0=\hbar \arctan \left \{
{\theta_1+\mu\theta_2  \over \nu\theta_1+\theta_2 }
\right \} \ ,
$$
$\theta_1$ et $\theta_2$ {\'e}tant deux solutions r{\'e}elles
ind{\'e}pendantes de l'{\'e}quation de
Schr{\"o}dinger. En posant [37]
$$
\phi_1=\nu\theta_1+\theta_2 \ ,\ \ \ \ \ \ \  \phi_2=\theta_1+\mu\theta_2 \ ,
\xeqno{2.4}
$$
on peut {\'e}crire
$$
S_0=\hbar \arctan \left(
{ \phi_2 \over \phi_1 }
\right)\ . \xeqno{2.5}
$$
En d{\'e}rivant l'Eq. (2.5) par rapport {\`a} $x$, on obtient
$$
{\partial S_0 \over \partial x}=\hbar
{{\phi_1} {\phi'_2}-
{\phi'_1}  {\phi_2  } \over
 {{\phi_1}^2+{\phi_2}^2 } }\ , \xeqno{2.6}
$$
o{\`u} $\phi'_1$ et $\phi'_2$ sont les premi{\`e}res d{\'e}riv{\'e}s de $\phi_1$ et $\phi_2$
par rapport {\`a} $x$.
En rempla{\c c}ant l'{\'e}quation (1.36) dans l'Eq. (2.6), on peut {\'e}crire
$$
\phi_1 {\phi'_2}-{\phi'_1}  \phi_2
={2 \over \hbar }{E-V\over \dot {x}} \left ({\phi_1}^2+
{\phi_2}^2 \right ) \; . \xeqno{2.7}
$$
Dans le but d'{\'e}liminer les fonctions $\phi_1$ et
$\phi_2$ et leurs d{\'e}riv{\'e}es, d{\'e}rivons
l'Eq. (2.7) par rapport {\`a} $x$, tout en tenant
compte du fait que le premier membre s'annule puisqu'il repr{\'e}sente
le Wronskien, constant, des fonctions $\phi_1$ et  $\phi_2$.
Alors, on peut d{\'e}duire
$$
\phi_1 \phi'_1+\phi_2 \phi'_2 =
{1 \over 2}\left ( {1 \over E-V} {dV \over dx}+
 {\ddot {x} \over \dot{x}^2} \right) \left({\phi_1}^2+
{\phi_2}^2 \right ) \; . \xeqno{2.8}
$$
D{\'e}rivons l'Eq. (2.8) par rapport {\`a} $x$ et tenons compte du fait que
$$
{\phi''_1}=-{2 m \over {\hbar}^2}(E-V) {\phi_1}\; , \ \ \ \ \ \ \ \
{\phi''_2}=-{2 m \over {\hbar}^2} (E-V) {\phi_2} \; ,
$$
on obtient alors [37]
$$
\left({{\phi'_1}}^2+{{\phi'_2}}^2 \right )-
\left ( {1 \over E-V}
 {dV \over dx}+ {\ddot {x} \over
 \dot {x}^2} \right)
\left(\phi_1 \phi'_1 + \phi_2 \phi'_2 \right)
+\left [-{2m \over {\hbar}^2} (E-V)- \right. \ \ \ \  \ \ \ \
\ \ \ \  \ \ \ \ \ \ \ \  \ \ \ \
$$
$$
 \left. \ \ \ \  \ \ \ \ {1 \over 2} {1 \over (E-V)^2} \ \left({dV \over dx}\right)^2
- {1 \over 2(E-V)} {d^2V \over dx^2}
- \ {\dot {\ddot {x}} \over 2 \dot {x}^3}+
 {{\ddot {x}}^2 \over \dot {x}^4} \right ]\
 \left ({\phi_1}^2+{\phi_2}^2 \right )=0 \ . \xeqno{2.9}
$$
Maintenant, en r{\'e}solvant le syst{\`e}me constitu{\'e} par
les Eqs. (2.7) et (2.8), on peut d{\'e}terminer les expressions de
$\phi'_1$ et $\phi'_2$
$$
\phi'_1={1 \over 2}\left({1 \over E-V}{dV \over dx}+{\ddot{x} \over \dot{x}^2}
\right)\phi_1-
{2 \over \hbar}{E-V \over \dot{x}}\phi_2\ ,
$$
$$
\phi'_2={2 \over \hbar}{E-V \over \dot{x}}\phi_1+{1 \over 2}\left({1 \over E-V}{dV \over dx}+{\ddot{x} \over \dot{x}^2}\right)\phi_2\ . \
$$
En {\'e}levant au carr{\'e} et en additionnant membre {\`a} membre ces deux
derni{`e}res {\'e}quations, on aboutit {\`a} [37]
$$
{{\phi'_1}}^2+{{\phi'_2}}^2 = \left [{4\over {\hbar}^2}
{(E-V)^2 \over {\dot{x}}^2}+ {1
\over 4(E-V)^2}
{\left ({dV \over dx} \right)}^2\right. \ \ \ \ \ \ \ \ \ \ \ \ \ \ \ \ \ \ \ \ \ \ \ \
$$
$$
 \left.\ \ \ \ \ \ \ \ \ \ \ \ \ \ \ \ \ \ \ \ \ \ \ \  +{{\ddot{x}}^2 \over 4{\dot{x}}^4}+
 {1 \over 2(E-V)}{dV \over dx}
{\ddot{x} \over {\dot{x}}^2} \right]
\left({\phi_1}^2+{\phi_2}^2 \right) \ .   \xeqno{2.10}
$$
En rempla{\c c}ant les Eqs. (2.8) et (2.10) dans l'Eq. (2.9),
on d{\'e}duit une {\'e}quation dans laquelle tous les termes
sont proportionnels  {\`a} $({\phi_1}^2+{\phi_2}^2)$.
En divisant l'{\'e}quation obtenue par $(\phi_1^2+\phi_2^2)$,
on aboutit exactement {\`a} l'{\'e}quation IPLNQ donn{\'e}e
par l'Eq. (2.3). Donc, avec deux mani{\`e}res
diff{\'e}rentes on arrive {\`a} la m{\^e}me {\'e}quation IPLNQ,
le point de d{\'e}part {\'e}tant {\`a} chaque fois l'{\'e}quation de dispersion (1.36).

\bigskip
\noindent
{\bf 2.2. {\'E}tude du cas de la particule libre }
\medskip

Dans le cas de la particule libre, l'{\'e}quation IPLNQ s'{\'e}crit [37]
$$
E^2-{m{\dot{x}}^2 \over 2} E+{{\hbar}^2\over 8} \left[{3\over 2}
\left({\ddot{x}\over \dot{x}}\right)^2-{\dot{\ddot{x}} \over \dot{x}}
\right]=0 \; , \xeqno{2.11}
$$
et l'Eq. (1.36) se ram{\`e}ne {\`a}
$$
\dot{x}{\partial S_0 \over \partial x}=2E \ . \xeqno{2.12}
$$
L'Eq. (2.11) repr{\'e}sente  l'{\'e}quation dynamique
qui r{\'e}git le mouvement de la particule libre.
Essayons maintenant d'{\'e}tablir l'{\'e}quation des trajectoires.

\bigskip
\noindent
{\it 2.2.1 R{\'e}solution {\`a} partir de l'{\'e}quation IPLNQ }
\medskip

Reprenons l'Eq. (2.11) et faisons le changement de variables
suivant [37]:
$$
U=\sqrt{2mE}\; x \; ,
\ \ \ \ \ \ \ \ q=\sqrt{2E \over m} \; t \; ,  \xeqno {2.13}
$$
o{\`u} $U$ et $q$ sont des variables qui ont les dimensions,
respectivement, d'une action et d'une distance. Avec ces nouvelles
variables, l'Eq. (2.11) se ram{\`e}ne {\`a}
$$
{1\over 2m}\left({dU \over dq} \right)^{2}-E={{\hbar}^2\over 4m}
\left[{3\over 2}\left({dU \over dq}\right)^{-2}
\left({d^2U \over dq^2}\right)^2-{\left({dU \over dq}\right)}^{-1}
\left( {d^3U \over dq^3}\right)\right] \; .\xeqno {2.14}
$$
Cette {\'e}quation a la m{\^e}me forme que l'EHJQS,
{\'e}crite pour un potentiel nul. Ceci nous permet d'utiliser
la solution de cette derni{\`e}re pour r{\'e}soudre l'Eq. (2.14).
Cependant, si on pose
$$
\theta_3=\nu\theta_1+\theta_2\
$$
dans l'Eq. (3.24) du Chap. 1, $S_0$ prend la forme
$$
S_0=\hbar \arctan \left[(1-\mu\nu){\theta_1 \over \theta_3}+\mu \right]\ .
\xeqno{2.15}
$$
Donc, de mani{\`e}re analogue, on peut d{\'e}duire la forme suivante de la solution $U$
de l'Eq. (2.14)
$$
U={\hbar} \arctan{\left[\sigma {{\psi_1}\over{\psi_2}}+\omega \right]}
+ U_0 \; , \xeqno {2.16}
$$
o{\`u} $\sigma$, $\omega$ et $U_0$ sont des constantes
d'int{\'e}gration r{\'e}elles, avec $\sigma\not=0$. Les fonctions
$\psi_1$ et $\psi_2$ sont deux solutions r{\'e}elles ind{\'e}pendantes
de l'{\'e}quation
$$
-{{\hbar}^2\over 2m} {d^2\psi \over dq^2}=E{\psi} \; .
$$
Choisissons
$
{\psi_1}=\sin{\left({\sqrt{2mE}\; q }/ {\hbar}\right)}
$
et
$
 {\psi_2}=\cos{\left({\sqrt{2mE}\; q }/ {\hbar}\right)}.
$
Alors, il s'en suit que
$$
x(t)={\hbar\over\sqrt{2mE}} \arctan{\left[\sigma \tan{\left({2Et
\over\hbar}\right)}+ \omega \right]}+x_0  \; .\xeqno {2.17}
$$
$x$ d{\'e}pend des quatre constantes d'int{\'e}gration $E$,
$\sigma$, $\omega$ et $x_0$. Remarquons que dans le cas
o{\`u} $\sigma=1$ et $\omega=0$, on retrouve l'{\'e}quation horaire classique
$$
x(t)= \sqrt{2E \over m}\;t+x_0 \; .
$$
\medskip
\noindent
{\it 2.2.2 R{\'e}solution {\`a} partir de l'{\'e}quation de dispersion  }
\medskip

 L'{\'e}quation de dispersion (2.12) permet d'{\'e}crire
$$
{\partial S_0 \over \partial x}={2E \over \dot{x}}\ ,
$$
ce qui donne
$$
dS_0=2E\ dt \ .
$$
En int{\'e}grant cette derni{\`e}re, on obtient
$$
S_0=2E(t-t_0)  \ .
\xeqno {2.18}
$$
En utilisant l'Eq. (2.15), on peut {\'e}crire l'action r{\'e}duite de
la particule libre comme suit
$$
S_0=\hbar \arctan \left \{ a \tan\left({\sqrt{2mE}x \over \hbar}\right)+b
\right \} \ .\xeqno {2.19}
$$
En injectant l'Eq. (2.19) dans l'Eq. (2.18), on peut d{\'e}duire
$$
x(t)={\hbar\over\sqrt{2mE}} \arctan \left \{
A \tan \left( {2E (t-t_0)/ \hbar} \right)+B \right \} \ .\xeqno {2.20}
$$
Cette derni{\`e}re {\'e}quation repr{\'e}sente l'{\'e}quation
horaire du mouvement, les constantes $A$ et $B$ {\'e}tant l'ensemble
des variables cach{\'e}es. Pour $A=1$ et $B=0$, l'Eq. (2.20) se
ram{\`e}ne {\`a} l'{\'e}quation horaire classique.
Notons qu'on peut montrer que les Eqs. (2.17) et (2.20) sont
{\'e}quivalentes.

\bigskip
\noindent
{\bf 2.3. Comparaison avec les r{\'e}sultats de Floyd }
\medskip

Pour comparer nos r{\'e}sultats obtenus pour la particule libre avec
ceux de Floyd, utilisons la forme de l'action r{\'e}duite donn{\'e}e par
l'Eq. (2.37) du Chap. 1.

Dans le cadre de notre formulation, d'apr{\`e}s l'Eq. (2.15), on a
$$
2E(t-t_0)_d=\hbar \arctan \left \{
{b \tan \left( {\sqrt{2mE}x / \hbar} \right)+c/2 \over
\sqrt{ab-c^2/4}}
\right \} \ . \xeqno{2.21}
$$
Cependant, Floyd donna une autre {\'e}quation des trajectoires,
{\`a} partir de $\partial S_0 / \partial E=t-t_0$, qui
est (Eq. (2.41), Chap.1)
$$
(t-t_0)_{Fl}= {(ab-c^2/4)^{1 \over 2} \sqrt{2m/E}\ x
\over
a+b+\left[(a-b)^2+c^2 \right]^{1 \over 2} \cos \left \{
2(2mE)^{1 \over 2}{x / \hbar}+\cot^{-1}[(b-a)/c]
\right\}} \ . \xeqno{2.22}
$$
Remarquons qu'en passant {\`a} la limite classique dans l'Eq. (2.21)
$$
\lim_{\hbar \to 0}(t-t_0)_d=
{(ab-c^2/4)^{1 \over 2} \sqrt{2m/E}\ x
\over
a+b+\left[(a-b)^2+c^2 \right]^{1 \over 2} \cos \left \{
2(2mE)^{1 \over 2}{x / \hbar}+\cot^{-1}[(b-a)/c]
\right\}} \ . \xeqno{2.23}
$$
on retrouve exactement le r{\'e}sultat de l'Eq. (2.22).
Ainsi, les trajectoires de Floyd
se pr{\'e}sentent comme une limite classique de nos trajectoires.
Effectivement, dans l'approche de Floyd, c'est uniquement {\`a} la limite
classique qu'on a
$$
\lim_{\hbar \to 0}S_0=2E(t-t_0)\ ,
$$
alors que dans notre approche, l'Eq. (2.18) indique clairement que
$$
S_0=2E(t-t_0)
$$
est valable sans passer {\`a} la limite classique ${\hbar \to 0}$. Ce r{\'e}sultat
est en accord avec le fait que, dans notre
formulation, l'action r{\'e}duite est donn{\'e}e par (Eq. (1.15))
$$
S=S_0-Et=\int_{t_0}^{t} \ L_q \, dt=\int_{t_0}^{t}  E dt=E\; (t-t_0) \ ,
\xeqno{2.24}
$$
{\`a} une constante pr{\`e}s.
Dans ce cas, on pourrait dire que notre formalisme est plus g{\'e}n{\'e}ral
que celui de Floyd.

\bigskip
\headline={
\vbox{
    \xline{ {\craw Chapitre 2.} \hfill {\craw 3.Coordonn{\'e}e quantique et ...} {\bf \ \ \folio}}
    \medskip \hrule

}
}

\bigskip
\noindent
{\bf 3. COORDONN{\'E}E QUANTIQUE ET VERSION QUANTIQUE

DU TH{\'E}OR{\`E}ME DE JACOBI}
\bigskip
Dans la Sec. 3 du Chap. 1, nous avons vu  comment Faraggi et Matone
ont introduit la transformation quantique
$$
x \to \hat{x}
$$
apr{\`e}s laquelle l'EHJQS prend la forme classique. Dans cette section,
nous allons reproduire les r{\'e}sultats des Secs. 1 et 2 de ce chapitre
en faisant appel {\`a} cette transformation.

Reprenons l'EHJQS et {\'e}crivons-la sous la forme
$$
{1 \over 2m}\left({\partial S_0 \over \partial x}\right)^2 \left[
1-{\hbar^2 \over 2} \left( {\partial S_0 \over \partial x} \right)^{-2}
\left \{ S_0 \ , x \right \} \right]+V(x)=E\ .
$$
En d{\'e}finissant
$$
\left({\partial x \over \partial \hat{x} }\right)^2=
\left[
1-{\hbar^2 \over 2}\left( {\partial S_0 \over \partial x} \right)^{-2}
\left \{ S_0 \ , x \right \} \right]=
{2m(E-V) \over ({\partial S_0 / \partial x})^2  }
 \ , \xeqno {3.1}
$$
et en tenant compte de $\hat{V}(\hat{x})=V(x)$, on peut {\'e}crire
$$
{1 \over 2m}\left({\partial \hat{S}_0 \over \partial \hat{x}}\right)^2+
\hat{V}(\hat{x})=E \ . \xeqno {3.2}
$$
Tout d'abord, faisons la remarque que la coordonn{\'e}e quantique $\hat{x}$
est r{\'e}elle dans les r{\'e}gions permises et purement imaginaire
dans les r{\'e}gions interdites.
En comparant l'Eq. (3.1) avec l'Eq. (1.34), on d{\'e}duit
$$
\left({\partial x \over \partial \hat{x} }\right)^2=
{1 \over f(x\ ,E \ ,\mu\ ,\nu)} \ . \xeqno {3.3}
$$
Comme on l'a d{\'e}j{\`a} remarqu{\'e}, $f(x\ ,E \ ,\mu\ ,\nu)$ ne d{\'e}pend que
de la variable $x$ et des constantes d'int{\'e}gration. Il est
de m{\^e}me pour $\partial x / \partial \hat{x}$. On a donc
$$
\left({\partial x \over \partial \hat{x} }\right)=
\left({\partial \hat{x} \over \partial x }\right)^{-1} \ , \xeqno {3.4}
$$
Ce qui permet de r{\'e}{\'e}crire l'Eq. (3.3) sous la forme
$$
\left({\partial \hat{x} \over \partial x }\right)^2=
 f(x\ ,E \ ,\mu\ ,\nu) \ . \xeqno {3.5}
$$
Avant d'aller plus loin, {\'e}non{\c c}ons les hypoth{\`e}ses suivantes:

1)-Sous la transformation quantique $x \to \hat{x}$,
l'action, le potentiel, le Lagrangien et l'Hamiltonien restent invariants
$$
\hat{L_q}(\hat{x})=L_q(x) \ , \xeqno{3.6.a}
$$
$$
\hat{H_q}(\hat{x})=H_q(x) \xeqno{3.6.b}
$$
$$
\hat{S_0}(\hat{x})=S_0(x)\ , \xeqno{3.6.c}
$$
$$
\hat{V}(\hat{x})=V(x) \ , \xeqno{3.6.d}
$$

2)- Le comportement dynamique suivant la coordonn{\'e}e quantique $\hat{x}$
ob{\'e}it aux lois de la m{\'e}canique classique.

\bigskip
\noindent
{\bf 3.1. Lagrangien et Hamiltonien du syst{\`e}me quantique}
\medskip

Reprenons l'EHJQS {\'e}crite suivant $\hat{x}$
$$
{1 \over 2m}\left({\partial \hat{S}_0 \over \partial \hat{x}}\right)^2+
\hat{V}(\hat{x})=E \ .
$$
Comme on l'a dit plus haut, elle a une apparence classique.
En utilisant les hypoth{\`e}ses d'invariance des quantit{\'e}s physiques
(Eqs. (3.6)), et de la
validit{\'e} du formalisme analytique classique par rapport {\`a} $\hat{x}$,
on peut d{\'e}finir une quantit{\'e} de mouvement $\hat{P}$ telle que
$$
\hat{P}={\partial \hat{S}_0 \over \partial \hat{x}} \ , \xeqno{3.7}
$$
et ainsi, d{\'e}finir un Lagrangien et un Hamiltonien quantiques
comme suit
$$
\hat{L}_q(\hat{x}\ , \dot{\hat{x}})=L_q={m \dot{\hat{x}}^2 \over 2}
-\hat{V}(\hat{x})
\ , \xeqno{3.8}
$$
$$
\hat{H}_q(\hat{x}\ ,\hat{P})=H_q={ \hat{P}^2 \over 2m}+\hat{V}(\hat{x})
 \ . \xeqno{3.9}
$$
Notons que
$$
\dot{\hat{x}}={d\hat{x} \over dt}=
\dot{x}  {\partial \hat{x} \over \partial x }
\ , \xeqno{3.10}
$$
$$
\hat{P}={\partial \hat{S}_0 \over \partial \hat{x}}={\partial S_0 \over \partial x}
{\partial x  \over \partial \hat{x}}=P{\partial x  \over \partial \hat{x}}
\ . \xeqno{3.11}
$$
En injectant les Eqs. (3.10) et (3.11) dans les Eqs. (3.8) et (3.9)
respectivement, et en tenant compte de $\hat{V}(\hat{x})=V(x)$,
on aboutit {\`a} [37]
$$
\hat{L}_q=L_q={m \dot{x}^2 \over 2}
\left( {\partial \hat{x} \over \partial x} \right)^2-V(x) \ , \xeqno{3.12}
$$
$$
\hat{H}_q=H_q={ P^2 \over 2m}
\left( {\partial x  \over \partial \hat{x}} \right)^2+V(x) \ . \xeqno{3.13}
$$
Remarquons que $\hat{L}_q$ et $\hat{H}_q$ prennent une
apparence classique. Alors, en utilisant le
\hbox{formal-\hskip -5pt plus 5pt}\break isme
Lagrangien, on d{\'e}finira l'action du syst{\`e}me quantique
$\hat{S}(\hat{x})$ comme suit
$$
\hat{S}(\hat{x})=\int \hat{L}_q(\hat{x}\ , \dot{\hat{x}}) \ dt \ . \xeqno{3.14}
$$
Cette d{\'e}finition est {\'e}quivalente {\`a} la d{\'e}finition
donn{\'e}e par l'Eq. (1.15)
de ce chapitre . Effectivement, puisque
$$
S(x)=\int L_q dt \ ,
$$
et si on tient compte des relations (3.6.a)et (3.6.c),
on arrive directement {\`a}
l'action (3.14).

De la m{\^e}me mani{\`e}re qu'en m{\'e}canique classique, le
principe de moindre action appliqu{\'e} {\`a} l'Eq. (3.14) nous
m{\`e}ne {\`a} l'{\'e}quation de Lagrange
$$
{d \over dt} \left( {\partial \hat{L}  \over \partial \dot{\hat{x}}} \right)-
{\partial \hat{L}  \over \partial \hat{x}} =0 \ , \xeqno{3.15}
$$
menant, {\`a} son tour, {\`a} l'{\'e}quation de mouvement
$$
m \ddot{\hat{x}}=-{d\hat{V} \over d\hat{x}}  \ . \xeqno{3.16}
$$
Remarquons que les Eqs. (3.15) et (3.16) sont {\'e}quivalentes aux Eqs. (1.19)
et (1.22). En effet, montrons en premier lieu que les Eqs. (3.15) et
(1.19) sont {\'e}quivalentes. Notons que
$$
L(x,\dot{x},\Gamma)=\hat{L}[\hat{x}(x,\Gamma),\dot{\hat{x}}(x,\dot{x},\Gamma)] \ .
$$
De cette derni{\`e}re {\'e}quation, on peut {\'e}crire
$$
{\partial L \over \partial x}={\partial \hat{L} \over \partial \hat{x}}
{\partial \hat{x} \over \partial x}+
{\partial \hat{L} \over \partial \dot{\hat{x}}}
{\partial \dot{\hat{x}} \over \partial x} \ , \xeqno{3.17}
$$
et
$$
{\partial L \over \partial \dot{x}}=
{\partial \hat{L} \over \partial \dot{\hat{x}}}
{\partial \dot{\hat{x}} \over \partial \dot{x}} \ . \xeqno{3.18}
$$
En reprenant l'Eq. (1.19) et en tenant compte des Eqs. (3.17) et (3.18), on
peut {\'e}crire
$$
{d \over dt}\left({{\partial \hat{L} \over \partial \dot{\hat{x}}}
{\partial \dot{\hat{x}} \over \partial \dot{x}}}\right)-
{\partial \hat{L} \over \partial \hat{x}}
{\partial \hat{x} \over \partial x}-
{\partial \hat{L} \over \partial \dot{\hat{x}}}
{\partial \dot{\hat{x}} \over \partial x}=0 \ ,
$$
ce qui se ram{\`e}ne {\`a}
$$
{d \over dt}\left({\partial \hat{L} \over \partial \dot{\hat{x}}}\right)
{\partial \dot{\hat{x}} \over \partial \dot{x}}+
{\partial \hat{L} \over \partial \dot{\hat{x}}}{d \over dt}\left(
{\partial \dot{\hat{x}} \over \partial \dot{x}}
\right)-
{\partial \hat{L} \over \partial \hat{x}}
{\partial \hat{x} \over \partial x}-
{\partial \hat{L} \over \partial \dot{\hat{x}}}
{\partial \dot{\hat{x}} \over \partial x}=0 \ . \xeqno{3.19}
$$
En tenant compte de l'Eq. (3.10) et du fait que
$$
{\partial \dot{\hat{x}} \over \partial x}=
{d \over dt}{\partial \hat{x} \over \partial x}\ ,
$$
l'Eq. (1.19) se ram{\`e}ne {\`a} l'Eq. (3.15).

Maintenant, montrons que les Eq. (3.16) et (1.22) sont {\'e}quivalentes.
Pour cela rempla{\c c}ons l'Eq. (3.10)
dans l'Eq. (3.16). Alors, on peut {\'e}crire
$$
m{d \over dt}\left( \dot{x}{\partial \hat{x} \over \partial x} \right)=
-{dV \over dx}{\partial x  \over \partial \hat{x}} \ ,
$$
ce qui se ram{\`e}ne {\`a}
$$
m\ \ddot{x} {\partial \hat{x} \over \partial x}+m\ \dot{x}
{d \over dt}\left( {\partial \hat{x} \over \partial x}\right)+
{dV \over dx}{\partial x  \over \partial \hat{x}}=0\ .
$$
En multipliant par $\partial \hat{x} / \partial x$, on obtient
$$
m\ \ddot{x} \left({\partial \hat{x} \over \partial x}\right)^2+
{m \over 2}\ \dot{x}
{d \over dt}\left( {\partial \hat{x} \over \partial x}\right)^2+
{dV \over dx}=0\ . \xeqno{3.20}
$$
En utilisant la relation (3.5) et le fait que
$$
{df \over dt}=\dot{x}\ {df \over dx}\ ,
$$
on d{\'e}duit que
$$
m  f(x,E,\mu,\nu)\ddot{x}+
{m\dot{x}^2 \over 2}  {df \over dx}+{dV \over dx}=0   \ .
$$
On retrouve ainsi l'Eq. (1.22). Donc, les Eqs. (1.22) et (3.16)
sont {\'e}quivalentes. \hbox{Remar-\hskip -5pt plus 5pt}\break quons qu'en int{\'e}grant
l'Eq. (3.20) et en utilisant l'Eq. (3.1), on arrive {\`a}
l'{\'e}quation de dispersion (1.36). Il est aussi {\`a} noter
qu'{\`a} partir de l'Eq. (3.16), en utilisant
l'expression (3.1) de
$(\partial \hat{x} / \partial x)^2$ et en tenant compte de l'Eq. (1.36),
on aboutit {\`a} l'{\'e}quation IPLNQ.

Tous ces r{\'e}sultats peuvent {\^e}tre reproduits {\`a} partir d'un
formalisme Hamiltonien \hbox{corres-\hskip-2pt plus 2pt}\break pondant {\`a} la coordonn{\'e}e $\hat{x}$.
Effectivement, comme en m{\'e}canique  classique, on peut
{\'e}crire les {\'e}quations canoniques suivantes:
$$
\ {\partial \hat{H} \over \partial \hat{P}}=\dot{\hat{x}} \ , \xeqno{3.21.a}
$$
$$
{\partial \hat{H} \over \partial \hat{x}}=-\dot{\hat{P}} \ . \xeqno{3.21.b}
$$
Remarquons qu'{\`a} partir des Eqs. (3.21), on peut reproduire l'Eq. (3.16).
En effet, en utilisant l'expression de $\hat{H}$ (Eq. (3.9))
dans les Eqs. (3.21), on trouve
$$
\hat{P}=m \dot{\hat{x}}\ ,
$$
et
$$
{d\hat{V} \over d\hat{x}}= - {d\hat{P} \over dt}=-m \ddot{\hat{x}} \ ,
$$
ce qui correspond {\`a} l'Eq. (3.16) qui reproduit l'{\'e}quation IPLNQ.
Donc, il est {\'e}vident que les deux
formulations Lagrangienne et Hamiltonienne suivant la coordonn{\'e}e quantique
$\hat{x}$ sont {\'e}quivalentes. De plus, elles sont
{\'e}quivalentes au formalisme introduit dans les Secs. 1 et 2.

\bigskip
\noindent
{\bf 3.2. Transformations canoniques et th{\'e}or{\`e}me de Jacobi }
\medskip

Maintenant, on peut chercher des transformations
$$
(\hat{x},\hat{P}) \to (\hat{X},\hat{\Phi})
$$
qui gardent les {\'e}quations canoniques invariantes.
Ces transformations sont dites \hbox{canoni-\hskip -5pt plus 5pt}\break ques, et sont d{\'e}finies
par rapport {\`a}
($\hat{x}$,$\hat{P}$) exactement comme en m{\'e}canique classique. Il est ainsi
possible de choisir une transformation pour laquelle le nouveau
Hamiltonien du syst{\`e}me soit nul. Ceci nous permet de reproduire
l'EHJQS par rapport {\`a} la coordonn{\'e}e $\hat{x}$. De m{\^e}me,
on peut {\'e}noncer le th{\'e}or{\`e}me de Jacobi comme suit :

Pour tout syst{\`e}me quantique de coordonn{\'e}e quantique $\hat{x}$,
les {\'e}quations du mouvement seront d{\'e}duites de l'{\'e}quation [37]
$$
\left [{\partial \hat{S}_0 \over \partial E}\right]_{\hat{x}=cte}=t-t_0
\ . \xeqno{3.22}
$$
Maintenant, montrons que ce th{\'e}or{\`e}me nous m{\`e}ne directement aux
{\'e}quations du
\hbox{mouve-\hskip -5pt plus 5pt}\break
ment quantique. Pour cela, reprenons l'Eq. (3.22) puis
d{\'e}rivons par rapport {\`a} $\hat{x}$
$$
{\partial \ \over \partial \hat{x}}
\left(
{\partial  \hat{S}_0 \over \partial E}
\right)={dt \over d\hat{x}}\ ,
$$
ce qui donne
$$
{\partial \; \over \ \partial E}
\left(
{\partial \hat{S}_0 \over \partial \hat{x}}
\right) =
{dt \over d\hat{x}}\ .
$$
En utilisant l'Eq. (3.2), on peut {\'e}crire
$$
{\partial  \over \ \partial E}
[2m (E-\hat{V})]^{1 \over 2}=
{\partial t \over \partial x}{dx \over d\hat{x}}\ ,
$$
ou encore
$$
{1 \over 2}[2m (E-\hat{V}(\hat{x}))]^{-{1 \over 2}}=
{\partial x \over \partial \hat{x}}{1 \over \dot{x}} \ .
$$
En tenant compte du fait que $\hat{V}(\hat{x})=V(x)$, on peut {\'e}crire
$$
{1 \over 2}[2m (E-V(x))]^{-{1 \over 2}}
{\partial \hat{x} \over \partial x }={1 \over \dot{x}} \ .
 \xeqno{3.23}
$$
En utilisant l'Eq. (3.1), on aboutit {\`a}
$$
\dot{x}{\partial S_0 \over  \partial x \ }=2(E-V)\ ,
$$
qui est l'{\'e}quation de dispersion (1.36),
{\`a} partir de laquelle on a d{\'e}duit l'{\'e}quation IPLNQ.
On peut aussi d{\'e}river l'Eq. (3.20) {\`a} partir de l'Eq. (3.23).
Effectivement, en {\'e}levant
au carr{\'e} les deux termes de l'Eq. (3.23), on a
$$
{m \over 2} \dot{x}^2
\left( {\partial \hat{x} \over \partial x } \right)^2=E-V(x)\ .
$$
En d{\'e}rivant cette derni{\`e}re par rapport {\`a} $t$, on aura
$$
m\ \dot{x} \ddot{x}\left( {\partial \hat{x} \over \partial x } \right)^2+
{m \over 2} \dot{x}^2
{d \over dt} \left( {\partial \hat{x} \over \partial x } \right)^2+
{dV \over dx} \dot{x}=0\; ,
$$
ce qui se ram{\`e}ne {\`a} l'Eq. (3.20)
$$
m\ \ddot{x}\left( {\partial \hat{x} \over \partial x } \right)^2+
{m \over 2} \dot{x}
{d \over dt} \left( {\partial \hat{x} \over \partial x } \right)^2+
{dV \over dx} \dot{x}=0\; .
$$
%
%

\bigskip
\noindent
{\bf 3.3 Comparaison avec le th{\'e}or{\`e}me de Jacobi introduit par Floyd}
\medskip

En postulant la validit{\'e} du th{\'e}or{\`e}me de Jacobi
en m{\'e}canique quantique, Floyd utilisa l'{\'e}quation
$$
\left [{\partial S_0 \over \partial E}\right]_{x=cte}=t-t_0
\ . \xeqno{3.24}
$$
qui est une relation classique r{\'e}sultant d'une transformation
canonique particuli{\`e}re \hbox{permet-\hskip -2mm plus 2mm}\break
tant d'aboutir {\`a} l'EHJC qui est du
premier ordre. Ainsi, l'Eq. (3.24) n'est pas
ad{\'e}quate {\`a} l'EHJQS qui est une {\'e}quation diff{\'e}rentielle du
troisi{\`e}me ordre. Dans notre point de vue, le th{\'e}or{\`e}me de Jacobi
doit {\^e}tre appliqu{\'e} quand nous utilisons la coordonn{\'e}e quantique
$\hat{x}$ avec laquelle l'EHJQS prend la forme classique [37].
C'est pour cette raison qu'on utilise la version quantique
de ce th{\'e}or{\`e}me donn{\'e}e par l'Eq. (3.22).
Remarquons qu'en rempla{\c c}ant $\hat{S}_0(\hat{x})$ par $S_0(x)$ dans l'Eq.
(3.22), la d{\'e}rivation par rapport {\`a} $E$ ne laisse pas $\hat{x}$
invariante, comme il est pr{\'e}cis{\'e} dans l'Eq. (3.22) puisque $\hat{x}$
est une fonction de $E$ [37].
De ce fait, en substituant $S_0(x)$ par
$\hat{S}_0(\hat{x})$ dans l'Eq. (3.24), on a

$$
t-t_0=\left[ {\partial  \over \ \ \partial E}S_0(x\, ,\mu,\nu,E) \right]_{x=cte}
$$
$$
\ \ \ \ \ \ \ \ \ \ \ \ \ \ \  =
\left[ {\partial  \over \ \ \partial E}\hat{S}_0[\hat{x}(x\,  ,\mu,\nu,E)\ ,E]
\right]_{x=cte}
$$
$$
\ \ \ \ \ \ \ \ \ \ \ \ \ \ \ \   =
{\partial \hat{S}_0 \over \partial \hat{x}}
\left( {\partial \hat{x}\over \partial E} \right)_{x=cte}+
\left({\partial \hat{S}_0 \over \partial E}\right)_{\hat{x}=cte} \ .
\xeqno{3.25}
$$
Cette derni{\`e}re {\'e}quation nous montre clairement que la version
quantique du th{\'e}or{\`e}me de Jacobi (3.22) telle que nous l'avons utilis{\'e}e
n'est pas compatible avec celle utilis{\'e}e par Floyd et donn{\'e}e
par l'Eq. (3.24). Ainsi, nos trajectoires diff{\`e}rent bien de
celles de Floyd. De m{\^e}me, nos moments conjugu{\'e}s sont diff{\'e}rents
puisque celui de Floyd est obtenu {\`a} partir de l'{\'e}quation (3.24).

Enfin, nous pouvons dire qu'en approchant la m{\'e}canique quantique par
la coordonn{\'e}e quantique $\hat{x}$, nous reproduisons toutes les
{\'e}quations de mouvement d{\'e}j{\`a} {\'e}tablies dans
les Secs. 1 et 2 de ce chapitre. En fait, $\hat{x}$ constitue une coordonn{\'e}e pour
laquelle le formalisme analytique classique est valable,
et reproduit ainsi les {\'e}quations du mouvement quantique.
\bigskip
{\bf Conclusion:}
\medskip

Nous avons {\'e}tablie dans ce chapitre une {\'e}quation IPLNQ (Eq. (2.3))
{\`a} partir d'un formalisme analytique qu'on a con{\c c}u d'apr{\`e}s
nos constations sur la forme de l'action r{\'e}duite tir{\'e}e de
l'EHJQS (Eq. (2.11), Chap. 1). En fait, cette forme
d{\'e}pend de deux solutions r{\'e}elles et
ind{\'e}pendantes de l'{\'e}quation de Schr{\"o}dinger qui d{\'e}pendent de l'{\'e}nergie.
Elle d{\'e}pend aussi de deux constantes
d'int{\'e}gration qui jouent le r{\^o}le de param{\`e}tres cach{\'e}s. Ceci
indique que l'{\'e}quation fondamentale doit {\^e}tre une
{\'e}quation diff{\'e}rentielle du quatri{\`e}me ordre.
Ainsi, nous avons construit
un Lagrangien dont l'expression est donn{\'e}e par l'Eq. (1.16).
A partir de ce Lagrangien, nous avons {\'e}tablie la relation de dispersion (1.36)
qui exprime une certaine dispersion de la partie cin{\'e}tique de l'{\'e}nergie
de la particule entre la quantit{\'e} de mouvement classique $m\dot{x}$
et le moment conjugu{\'e} $\partial S_0 / \partial x$ le long
de la coordonn{\'e}e $x$. A l'aide de cette {\'e}quation nous avons pu {\'e}tablir
l'{\'e}quation IPLNQ. Remarquons que l'{\'e}quation de dispersion
$$
\dot{x}{\partial S_0 \over \partial x}= 2(E-V(x))
$$
contient les deux solution de l'{\'e}quation de Schr{\"o}dinger qui sont pr{\'e}sentes
dans l'expression du moment conjugu{\'e}. De m{\^e}me, en r{\'e}solvant l'{\'e}quation
IPLNQ pour la particule libre, on remarque dans sa solution la
pr{\'e}sence des deux fonction $\sin(\sqrt{2mE}x/\hbar)$ et
 $\cos(\sqrt{2mE}x/\hbar)$ qui sont les solution de l'{\'e}quation de
Schr{\"o}dinger pour la particule libre. Il serait de m{\^e}me
pour les autres cas,
puisque l'{\'e}quation IPLNQ d{\'e}rive  de l'EHJQS (voir Sec. 2.1), laquelle
n'est soluble que si l'{\'e}quation de Schr{\"o}dinger l'est. Donc, il
est {\'e}vident qu'en r{\'e}solvant l'{\'e}quation IPLNQ on doit
reproduire les fonctions d'ondes.
Pour conclure, en m{\'e}canique quantique,
l'{\'e}quation de Schr{\"o}dinger bien qu'elle permet de conna{\^\i}tre
l'{\'e}nergie de la particule,
n'est pas fondamentale puisqu'elle
ne donne aucune description dynamique du comportement de la particule.
Par contre, l'EHJQS et l'{\'e}quation IPLNQ sont
\hbox{fondamenta-\hskip -1mm plus 1mm}\break les.
En effet, en utilisant le th{\'e}or{\`e}me de Jacobi dans le cadre de
 sa version qauntique, l'EHJQS, tout comme l'{\'e}quation IPLNQ nous
permettent de pr{\'e}voir les trajectoires de la particule.

Enfin, pour comparer la m{\'e}canique quantique, pour laquelle
les deux aspects \hbox{corpus-\hskip -2mm plus 2mm}\break
culaire et ondulatoire sont r{\'e}concili{\'e}s,
{\`a} la m{\'e}canique classique, pour laquelle ces
deux aspects sont
s{\'e}par{\'e}s, introduisons le diagramme figurant dans la page suivante.

\vfill\eject

    \line{ \hfill {\bf M{\'e}canique quantique\ \ \ \ } \hfill \hfill
    {\bf Th{\'e}orie classique} \hfill\hfill}
\medskip
    \line{ \hfill 1)- {\'e}quation de Scr{\"o}dinger  \hfill \hfill\hfill
    1)- {\'e}quation
    d'une onde dans \hfill\hfill }
\medskip
     \line{\hfill\hfill \hfill
    un milieu isotrope [2]\hfill
    }

\medskip
    \line
                 { \hfill
    $-(\hbar^2 /2m)(\partial^2{\psi} / \partial x^2)+
    V(x)\psi =i\hbar(\partial \psi /\partial t)$
    \hfill\hfill
    $
     {\partial^2\psi / \partial x^2}
     - {1 \over {\cal V}_0^2}{\partial^2\psi / \partial x^2}=F(x)\psi
    $\hfill
    }

    \medskip
    \line
                 {\hfill

    2)- action quantique \ \ \ \hfill\hfill\hfill  2)- a)- phase de l'onde      \hfill\hfill\hfill}

    \medskip
    \line
                 {\hfill
    $\psi (x,t)=A(x)\{\alpha \exp[i(Et-S_0(x))/\hbar]$
    \hfill\hfill\hfill\hfill\hfill\hfill
    $\psi (x,t)=a(x)\exp[2\pi i(\nu t- \varphi_1(x))]$\hfill\hfill\hfill
    }

    \medskip
    \line
                {\hfill
    $+\beta \exp[i(-Et+S_0(x))/\hbar] \}$\hfill\hfill\hfill
    $\varphi=\nu t-\varphi_1(x,t)$\hfill\hfill}

    \medskip
    \line
                 {\hfill
    $S=S_0(x,E,\mu,\nu)-Et$
    \hfill\hfill b)- action classique \hfill}

    \medskip
    \line
                 {\hfill $S_0=\hbar\arctan \{$
    $\sigma(\theta / \phi)+\omega \}$
    \hfill\hfill\hfill $S=S_0(x,E)-Et$ \hfill\hfill}

    \medskip
    \line
                 {\hfill\hfill\hfill
    3)- EHJQS\hfill\hfill\hfill \hfill \hfill \hfill \hfill \hfill \hfill
    \hfill \hfill
    \hfill\hfill
    \hfill\hfill3)- a) {\'e}quation de l'optique g{\'e}om{\'e}trique  }

    \medskip
    \line
                  {
     $(1 / 2m)(\partial S_0 / \partial x)^2-
    (\hbar^2 / 4m) \{ S_0\, ,x \}$
    \hfill\hfill\hfill\hfill
    $\left({\partial \varphi_1 / \partial x}\right)^2=
    {1 / \lambda (x) ^2}$
    \hfill\hfill}
    \medskip
    \line
                 {\hfill
    $+V(x)=E$
    \hfill\hfill\hfill\hfill}

    \medskip
    \line
                 {\hfill\hfill\hfill\hfill\hfill
    b)- EHJC\ \ \
    \hfill\hfill}

    \medskip
    \line
                 {\hfill\hfill
     $(1 / 2m)(\partial S_0 / \partial x)^2+V(x)=E$
    }

    \medskip
    \line {\hfill
    4)- La loi de Newton quantique \hfill\hfill 4)- la loi de Newton
    classique\hfill}

    \medskip
    \line
                {\hfill
    voir l'Eq. (2.3) de ce chapitre \hfill\hfill
    $m{\ddot{x}}=-{dV / dx}$ \hfill}

    \medskip
    \line
                {\hfill
    $(1/2)\dot{x}(\partial S_0 / \partial x)+V=E$
    \hfill\hfill\hfill $E=m\dot{x}^2/2+V(x)$
    \hfill}

\vfill\eject

\headline={}

\centerline{\ftitle CHAPITRE 3}\bigskip

\bigskip
\bigskip
\centerline{\ftitle GENERALISATION DE L'EHJQS A TROIS  }
\bigskip
\centerline{\ftitle  DIMENSIONS DANS LE CAS}
\bigskip
\centerline{\ftitle D'UN POTENTIEL A SYM{\'E}TRIE SPH{\'E}RIQUE}
\bigskip
\bigskip
\bigskip

\bigskip
\bigskip
\bigskip

La construction d'une {\'e}quation de Hamilton-Jacobi {\`a} une dimension,
dans le cadre de la m{\'e}canique quantique, est un pas tr{\`e}s important
pour l'{\'e}laboration d'une th{\'e}orie quantique
d{\'e}terministe qui r{\'e}tablirait
la r{\'e}alit{\'e} causale des ph{\'e}nom{\`e}nes quantiques.
\hbox{N{\'e}an-\hskip -5pt plus 5pt}\break
moins, une telle {\'e}quation
n'est pas suffisante pour une description convenable de la r{\'e}alit{\'e},
car en fait, les ph{\'e}nom{\`e}nes physiques se r{\'e}alisent
souvent dans un espace r{\'e}el {\`a}
trois dimensions. Pour cette raison, une g{\'e}n{\'e}ralisation {\`a} trois
dimensions de l'EHJQS s'av{\`e}re indispensable. C'est dans ce sens qu'on
introduira dans ce chapitre une tentative de g{\'e}n{\'e}ralisation
qui sera sans doute plus ou moins cons{\'e}quente pour la nouvelle
approche que nous avons pr{\'e}sent{\'e} dans le chapitre pr{\'e}c{\'e}dent.
 Pour cela, nous allons {\'e}tudier le cas d'un syst{\`e}me quantique
plong{\'e} dans un potentiel {\`a} sym{\'e}trie sph{\'e}rique.

\bigskip
\noindent
{\bf 1. L'EHJQS A TROIS DIMENSIONS}
\bigskip
 On commencera tout d'abord dans cette section par l'exposition de
l'{\'e}quation de Schr{\"o}dinger {\`a} trois dimensions pour un potentiel {\`a}
sym{\'e}trie sph{\'e}rique.

\vfill\eject
\noindent
{\bf  1.1. Equation de Schr{\"o}dinger {\`a} trois dimensions

 pour un potentiel {\`a} sym{\'e}trie sph{\'e}rique}
\medskip
{\'E}crivons l'{\'e}quation de Schr{\"o}dinger en coordonn{\'e}es
sph{\'e}riques ($r$,$\vartheta$,$\varphi$) [38]
$$
{\partial ^2 \Psi \over \partial r^2}+
{2 \over r}{\partial  \Psi \over \partial r}+
{1 \over r^2}{\partial ^2 \Psi \over \partial \vartheta^2}
+{\cot \vartheta \over r^2}\,{\partial  \Psi \over \partial \vartheta}+
{1 \over r^2 \sin^{2} \vartheta }\,{\partial ^2 \Psi \over \partial \varphi^2}+
{2m \over \hbar^2}[E-V(r)] \Psi=0 \ . \xeqno {1.1}
$$
$\Psi (r$,$\vartheta$,$\varphi)$ est la fonction d'onde
du syst{\`e}me quantique.
En s{\'e}parant les variables dans la fonction $\Psi$ telle que
$$
\Psi (r\ ,\vartheta\ ,\varphi)=R(r)\, T(\vartheta) F(\varphi)\ , \xeqno {1.2}
$$
l'Eq. (1.1) se s{\'e}pare aux trois {\'e}quations [38]
\headline={
\vbox{
    \xline{{\craw Chapitre 3.} \hfill  {\craw 1. Les EHJQS {\`a} trois dimensions}\ {\bf \ \  \folio}}
    \medskip \hrule
}
}
$$
{r^2 \over R}{d^2 R \over dr^2}+{2r \over R}{dR \over dr}+
{2m r^2 \over \hbar^2} (E-V(r))=\lambda \ , \xeqno {1.3}
$$
$$
{d^2T \over d\vartheta^2}+\cot \vartheta {dT \over d\vartheta}+
\left( \lambda -{m_{\ell}^2 \over \sin^2 \vartheta} \right) T=0 \ , \ \ \ \
\xeqno {1.4}
$$
$$
{1 \over F} \ {d^2F \over d\varphi^2} =-m_{\ell}^2\ . \ \ \ \ \ \ \ \ \ \ \ \ \ \ \ \
 \ \ \ \ \ \ \ \ \ \ \ \ \ \ \ \ \ \ \  \xeqno{1.5}
$$
$\lambda$ est une constante telle que
$$
\lambda=\ell (\ell+1) \ , \xeqno{1.6}
$$
$\ell$ {\'e}tant un nombre entier positif ou nul. $m_{\ell}$ est
un nombre entier qui v{\'e}rifie
$$
-\ell \leq m_{\ell}\leq \ell \ . \xeqno{1.7}
$$
$\ell (\ell+1)$ est la valeur propre du carr{\'e} de l'op{\'e}rateur
moment cin{\'e}tique $L^2$, et
$ m_{\ell}$ est la valeur propre de l'op{\'e}rateur
projection $L_z$ du moment cin{\'e}tique suivant la direction $z$.
Les solutions des Eqs. (1.3), (1.4) et (1.5) sont pr{\'e}sent{\'e}es dans
la r{\'e}f{\'e}rence [38]. Les solutions de l'Eq. (1.4) peuvent s'exprimer
en fonction des polyn{\^o}mes de Legendre, alors que
les solutions de l'Eq. (1.5) sont des fonctions p{\'e}riodiques de forme
$e^{im_{\ell}\varphi}$. Pour les fonctions radiales, on distingue deux cas.
Le premier cas est celui de la liaison ($E<0$), pour lequel
l'Eq. (1.3) n'a de solutions que pour des valeurs particuli{\`e}res de $E$.
Dans ce cas, des nombres entiers interviennent dans l'expression
de la fonction $R(r)$. Par contre, le deuxi{\`e}me cas est celui
de la diffusion ($E>0)$, pour lequel l'Eq. (1.3) poss{\`e}de des
solutions pour toutes les valeurs r{\'e}elles de $E$.

Maintenant, essayons d'{\'e}crire les Eqs. (1.3), (1.4) et (1.5)
sous la forme de l'{\'e}quation de Schr{\"o}dinger {\`a} une dimension. Pour cela,
commen{\c c}ons d'abord par l'Eq. (1.3) et introduisons la fonction
${\cal X}$ d{\'e}finie par
$$
R={{\cal X} \over r} \ . \xeqno{1.9}
$$
En d{\'e}rivant deux fois successivement l'Eq. (1.9)
par rapport {\`a} $r$, on obtient
$$
{dR \over dr}={1 \over r}{d{\cal X} \over dr}-{{\cal X} \over r^2}\ ,
\xeqno{1.10}
$$
$$
{d^2R \over dr^2}={1 \over r}{d^2{\cal X} \over dr^2}-
2{d{\cal X}/dr \over r^2 }+2{{\cal X} \over r^3}\ . \xeqno{1.11}
$$
En rempla{\c c}ant ces derni{\`e}res {\'e}quations dans
l'Eq. (1.3), on peut {\'e}crire [39]
$$
{-\hbar^2 \over 2m}  {d^2{\cal X} \over dr^2}+\left[ V(r)+
{\lambda \hbar^2 \over 2mr^2}\right]{\cal X}=E{\cal X} \ . \xeqno{1.12}
$$
Cette {\'e}quation ressemble {\`a} l'{\'e}quation de Schr{\"o}dinger
unidimensionnelle avec un potentiel fictif
$$
V^{'}(r)=V(r)+{\lambda \hbar^2 \over 2mr^2}\ . \xeqno{1.13}
$$

De m{\^e}me, on peut ramener la forme de l'Eq. (1.4) {\`a}  celle de
l'{\'e}quation de Schr{\"o}dinger unidimensionnelle. Effectivement,
reprenons l'Eq. (1.4) et introduisons la fonction
${\cal T}(\vartheta)$ d{\'e}finie par
$$
{\cal T}(\vartheta)=\sin^{1 \over 2}\vartheta \; T(\vartheta) \ .  \xeqno{1.14}
$$
En d{\'e}rivant deux fois successivement l'Eq. (1.14) par
rapport {\`a} $\vartheta$, on aboutit {\`a}
$$
{dT \over d\vartheta}={1 \over \sin^{1 \over 2}\vartheta }
{d{\cal T} \over d\vartheta}-{1 \over 2}
{\cos\vartheta\over \sin^{3  \over 2}\vartheta }{\cal T}\ , \xeqno{1.15}
$$
$$
{d^2T \over d\vartheta^2}={1 \over \sin^{1 \over 2}\vartheta }
{d^2{\cal T} \over d\vartheta^2}-
{\cos\vartheta \over \sin^{3  \over 2}\vartheta }
{d{\cal T} \over d\vartheta}+{3 \over 4}
{\cos^{2}\vartheta \over \sin^{3\over 2}\vartheta}
+{1 \over 2}{{\cal T} \over \sin^{1 \over 2}\vartheta }  \ . \xeqno{1.16}
$$
En rempla{\c c}ant les Eqs. (1.15) et (1.16) dans l'Eq. (1.4), on obtient [39]
$$
{d^2{\cal T} \over d\vartheta^2}+(\lambda+{1 \over 4}){\cal T}+
{(1/4-m_{\ell}^{2}) \over \sin^{2}\vartheta } {\cal T}=0 \ . \xeqno{1.17}
$$
On remarque que l'Eq. (1.17) se ram{\`e}ne {\`a}
l'{\'e}quation de Schr{\"o}dinger unidimensionnelle
d'un syst{\`e}me de potentiel et d'{\'e}nergie dont les formes sont
$$
V(\vartheta)= {(m_{\ell}^{2}-1/4) \over \sin^{2}\vartheta }{\hbar^2 \over 2m}
\ , \xeqno{1.18}
$$
et
$$
E_{\vartheta}=(\lambda+{1 \over 4}){\hbar^2 \over 2m}\ . \xeqno{1.19}
$$
On remarque aussi que l'Eq. (1.5) est de la forme de l'{\'e}quation
de Schr{\"o}dinger \hbox{unidimension-\hskip -5pt plus 5pt}\break nelle
d'un syst{\`e}me libre d'{\'e}nergie {\'e}gale {\`a} ${m_{\ell}^2 \hbar^2 / 2m}$.

En fin de compte, les Eqs. (1.3), (1.4) et (1.5) se
ram{\`e}nent toutes {\`a}  l'{\'e}quation de Schr{\"o}dinger
unidimensionnelle d'{\'e}nergie et de potentiel
particuliers. Ce fait sera utilis{\'e} dans les sections suivantes
pour l'{\'e}tablissement et la r{\'e}solution des EHJQS.

\bigskip
\noindent
{\bf 1.2. L'EHJQS radiale}
\medskip

En reprenant l'Eq. (1.12), et en se basant sur l'{\'e}criture
de la fonction d'onde {\`a} une dimension, donn{\'e}e par
l'Eq. (3.25) du Chap.1, {\'e}crivons que [39]
$$
{\cal X}(r)=A(r)\left [ \alpha e^{{i \over \hbar} Z(r)}+
\beta e^{-{i \over \hbar} Z(r)} \right ]\ . \xeqno {1.20}
$$
De la m{\^e}me mani{\`e}re que dans le cas unidimensionnel [29], en injectant
la relation (1.20) dans l'Eq. (1.12),
on aboutit aux deux {\'e}quations suivantes:
$$
{1 \over 2m}\left({dZ \over dr} \right)^2 -{\hbar^2 \over 2m}{1 \over A}{d^2 A \over dr^2}
+V(r)+{\lambda\hbar^2 \over 2m\; r^2}=E\ , \xeqno {1.21}
$$
$$
{d^2Z \over dr^2}+{2 \over A} {dA \over dr}{dZ \over dr}=0\ .
\xeqno {1.22}
$$
En int{\'e}grant cette derni{\`e}re {\'e}quation, on obtient
$$
A(r)=k \left({dZ \over dr}\right)^{-{1 \over 2}}\ , \xeqno {1.23}
$$
$k$ {\'e}tant une constante d'int{\'e}gration.
En injectant l'Eq. (1.23) dans l'Eq. (1.21), on peut {\'e}crire [39]
$$
{1 \over 2m} \left( {dZ \over dr} \right)^2-
{\hbar ^2 \over 4m} \left \{ Z,r\right \}+V(r)+
{\lambda \hbar^2 \over 2m r^2}=E\ , \xeqno{1.24}
$$
o{\`u} la quantit{\'e} $\left \{ Z,r\right \}$ repr{\'e}sente
le Schwarzien de $Z(r)$ (voir Chap. 1). L'Eq. (1.24)
repr{\'e}sente l'EHJQS radiale. La fonction $Z(r)$ est l'{\it action r{\'e}duite radiale}

\vfill\eject
\bigskip
\noindent
{\bf 1.3. L'EHJQS angulaire suivant $\vartheta$ }
\medskip

Maintenant, reprenons l'Eq. (1.17) et {\'e}crivons ${\cal T}(\vartheta)$
sous la forme [39]
$$
{\cal T}(\vartheta)=\xi(\vartheta)  \left[
\gamma e^{{i \over \hbar}L(\vartheta)}+
\varepsilon e^{-{i \over \hbar}L(\vartheta)}
\right] \ . \xeqno{1.25}
$$
Apr{\`e}s avoir inject{\'e} cette derni{\`e}re {\'e}quation
dans l'Eq. (1.17), de la m{\^e}me mani{\`e}re que dans le cas unidimensionnel [29],
on arrive aux {\'e}quations suivantes:
$$
\left({dL \over d\vartheta} \right)^2-
{\hbar^2 \over \xi}{d^2 \xi \over d\vartheta^2}-
{(1/4-m_{\ell}^{2}) \over \sin^{2}\vartheta } \hbar^2=
(\lambda+{1 \over 4})\hbar^2
\ , \xeqno{1.26}
$$
$$
2 {d\xi \over d\vartheta}{dL \over d\vartheta}+
\xi {d^2L \over d\vartheta^2}=0 \ . \xeqno{1.27}
$$
L'Eq. (1.27) peut se mettre sous la forme
$$
\xi(\vartheta)=k^{'}\left( dL \over d\vartheta \right)^{-{1\over 2}}
 \ , \xeqno{1.28}
$$
$k^{'}$ {\'e}tant une constante d'int{\'e}gration.
En rempla{\c c}ant cette derni{\`e}re {\'e}quation dans l'Eq. (1.26),
on aboutit {\`a} [39]
$$
\left({dL \over d\vartheta}\right)^2-{\hbar^2 \over 2}\left \{L,\vartheta
\right \}+
{(m_{\ell}^2-1/4) \over \sin^2\vartheta}\hbar^2=
(\lambda +{1 \over 4})\hbar^2 \ .
\xeqno{1.29}
$$
L'Eq. (1.29) repr{\'e}sente l'EHJQS angulaire suivant la direction $\vartheta$.
La fonction $L(\vartheta)$ sera appel{\'e}e l'{\it action r{\'e}duite angulaire
suivant $\vartheta$.}

\bigskip
\noindent
{\bf 1.4. L'EHJQS
angulaire suivant $\varphi$ }
\medskip
Reprenons l'Eq. (1.5) et {\'e}crivons la fonction $F(\varphi)$ sous la forme [39]
$$
F(\varphi)=\eta(\varphi)\left [ \sigma e^{{i \over \hbar} M(\varphi)}+
\omega e^{-{i \over \hbar} M(\varphi)} \right ]\ . \xeqno {1.30}
$$
En rempla{\c c}ant cette derni{\`e}re {\'e}quation dans l'Eq. (1.5),
de la m{\^e}me fa{\c c}on que dans le cas unidimensionnel,
on aboutit aux {\'e}quations
$$
\left({dM \over d\varphi}\right)^2-
{\hbar^2 \over \eta(\varphi)}{d^2\eta(\varphi) \over d\varphi^2}-
m_{\ell}^2\hbar^2=0 \ , \xeqno{1.31}
$$
$$
{\eta(\varphi) \over 2}{d^2M(\varphi) \over d\varphi^2}+
{d\eta(\varphi) \over d\varphi}\left({dM \over d\varphi}\right)=0 \ .
\xeqno{1.32}
$$
La solution  de cette {\'e}quation est
$$
\eta(\varphi)=k^{''} \left({dM \over d\varphi}\right)^{-{1 \over 2}} \ ,
\xeqno{1.33}
$$
$k^{''}$ {\'e}tant une constante d'int{\'e}gration.
En injectant la relation (1.33) dans l'Eq. (1.31), on arrive {\`a} [39]
$$
\left({dM \over d\varphi}\right)^2-{\hbar^2 \over 2}
\left \{ M,\varphi \right \}=m_{\ell}^2 \hbar ^2 \ . \xeqno{1.34}
$$
L'Eq. (1.35) repr{\'e}sente l'EHJQS suivant la direction angulaire $\varphi$.
Ainsi, la fonction $M(\varphi)$ repr{\'e}sente
l'{\it action r{\'e}duite angulaire suivant $\varphi$}.

\headline={
\vbox{
    \xline{\craw Chapitre 3. \hfill      {\craw 2. R{\'e}solution des EHJQS {\`a} trois dimensions}{\bf \ \ \folio}}
    \medskip \hrule
}
}
\bigskip
\bigskip
\noindent
{\bf 2. R{\'E}SOLUTION DES EHJQS A TROIS DIMENSIONS}
\medskip

Dans cette section on va r{\'e}soudre les Eqs. (1.24), (1.29) et (1.34)
pour conna{\^\i}tre la forme des fonctions $Z(r)$, $L(\vartheta)$ et $M(\varphi)$
que nous avons introduit dans la Sec.1.

\medskip
\noindent
{\bf 2.1. La forme de l'action r{\'e}duite radiale}
\medskip

En tenant compte de l'Eq. (1.13), l'Eq. (1.24) peut se mettre sous la forme
$$
{1 \over 2m}\left({dZ \over dr} \right)^2-
{\hbar^2 \over 4m}\left \{Z,r\right \}+V^{'}(r)-E=0 \ , \xeqno{2.1}
$$
qui a une forme analogue {\`a} l'EHJQS unidimensionnelle (Eq. (2.11) du Chap.1).
Du fait de cette analogie, on d{\'e}duit la solution suivante de l'Eq. (2.1) [39]
$$
Z(r)=\hbar \arctan \left \{  {{\cal X}_1+\mu{\cal X}_2 \over
{\nu\cal X}_1+{\cal X}_2} \right \} \ , \xeqno{2.2}
$$
o{\`u} ${\cal X}_1$ et ${\cal X}_2$ sont deux solutions r{\'e}elles et ind{\'e}pendantes
de l'Eq. (1.12), et $\mu$ et $\nu$ sont des constantes d'int{\'e}gration.
La forme de l'action r{\'e}duite $Z(r)$ ressemble
bel et bien {\`a} la forme de l'action r{\'e}duite unidimensionnelle,
donn{\'e}e par l'Eq. (3.24) du Chap.1. Toutefois, les fonctions
${\cal X}_1$ et ${\cal X}_2$ sont diff{\'e}rentes de $\theta$ et $\phi$.
${\cal X}_1$ est donn{\'e}e par  l'Eq. (1.9)
$$
{\cal X}_1=r\ R_1 \ ,
$$
alors que ${\cal X}_2$ peut {\^e}tre calcul{\'e} {\`a}
partir du Wronskien ${\cal W}$
de ${\cal X}_1$ et ${\cal X}_2$
$$
{\cal W} =K={\cal X}_1 {d{\cal X}_2 \over dr}-
{d{\cal X}_1 \over dr}{\cal X}_2 \ , \xeqno{2.3}
$$                                                                             %
o{\`u} $K$ est une constante. De l'Eq. (2.3), on peut {\'e}crire
$$
{\cal X}_2(r)=K{\cal X}_1(r) \int {dr \over {\cal X}_1^2(r) } \ .\xeqno{2.4}
$$
En rempla{\c c}ant cette derni{\`e}re relation dans l'Eq. (2.2), on
obtient
$$
Z(r)=\hbar \arctan \left \{
{K\mu \int {dr \over {\cal X}_1^2 }+1 \over
\nu+K  \int {dr \over {\cal X}_1^2 }}\right \} \ .
\xeqno{2.5}
$$
En adoptant les notations de Floyd, donc en utilisant les constantes $a$,
$b$ et $c$, on aura
$$
Z(r)=\hbar \arctan \left \{
{Kb\ \int {dr \over {\cal X}_1^2 }+c/2 \over
\sqrt{ab-c^2/4}}\right \} \ .
\xeqno{2.6}
$$
%
%

\noindent
{\bf 2.2. La forme de l'action r{\'e}duite angulaire suivant $\vartheta$}      \medskip

Maintenant, essayons de r{\'e}soudre l'Eq. (1.29)
$$
\left({dL \over d\vartheta}\right)^2-{\hbar^2 \over 2}\left \{L,\vartheta
\right \}+
{(m_{\ell}^2-1/4) \over \sin^2\vartheta}\hbar^2=(\lambda +{1 \over 4})\hbar^2 \ .
$$
Puisque cette derni{\`e}re {\'e}quation peut {\^e}tre vue comme une EHJQS
unidimensionnelle, avec un potentiel et une {\'e}nergie donn{\'e}s
par les Eqs. (1.18) et (1.19), sa solution peut {\^e}tre
{\'e}crite sous la forme [39]
$$
L(\vartheta)=\hbar \arctan \left \{
{{\cal T}_1(\vartheta)+\delta {\cal T}_2(\vartheta) \over
\varrho{\cal T}_1(\vartheta)+ {\cal T}_2(\vartheta)
}\right \}\ ,
\xeqno{2.7}
$$
o{\`u} $\varrho$ et $\delta$ sont des constantes d'int{\'e}gration.
On peut voir aussi que le Wronskien de deux solutions r{\'e}elles
et ind{\'e}pendantes ${\cal T}_1$ et ${\cal T}_2$ de l'Eq. (1.17)
est une constante qu'on notera $k$. Alors, en explicitant
l'expression du Wronskien $W$ et
en int{\'e}grant, on obtient
$$
{\cal T}_2=k\, {\cal T}_1\int {dx \over {\cal T}_1 ^2 }
\ . \xeqno{2.8}
$$
De cette {\'e}quation et de l'Eq. (1.14), on peut d{\'e}duire
$$
{{\cal T}_2 \over {\cal T}_1}={T_1 \over T_2}=k\int {dx \over {\cal T}_1 ^2 }
\ , \xeqno{2.9}
$$
o{\`u} $T_1$ et $T_2$ sont deux solutions r{\'e}elles et ind{\'e}pendantes
de l'Eq. (1.4)
correspondant {\`a} ${\cal T}_1$ et ${\cal T}_2$ respectivement.
Donc, l'Eq. (2.7) peut se mettre sous la forme
$$
L(\vartheta)=\hbar \arctan \left \{
{\mu k\int {dx \over {\cal T}_1 ^2 }+1  \over
\nu + k\int {dx \over {\cal T}_1 ^2 }
}\right \}\ .
\xeqno{2.10}
$$
ou encore
$$
L(\vartheta)=\hbar \arctan \left \{
{T_1 + \mu T_2  \over \nu T_1 + T_2 }
\right \} \ .
\xeqno{2.11}
$$
Pour s'assurer que les expressions donn{\'e}es dans les Eqs. (2.7),
(2.10) et (2.11) sont des solutions de l'Eq. (1.29), {\'e}crivons l'Eq. (2.11)
{\`a} l'aide des notations de Floyd
$$
L(\vartheta)=\hbar \arctan \left \{
{b \ (T_1 / T_2) + c/2 \over \sqrt{ab-c^2/4}}
\right \}
$$
En injectant l'Eq. (2.11) dans l'Eq. (1.29), on obtient
$$
{\hbar^2 \alpha^2 (ab-c^2/4)  \over
\sin^2 \vartheta (b T^2_1+aT^2_2+cT_1T_{2})^2}-
\hbar^2 \left \{ -{1 \over 4}cot^2\vartheta -{1 \over 2}+ \right.
$$
$$
\left.{(ab-c^2/4)\alpha^2 \over \sin^2 \vartheta (b T^2_1+aT^2_2+cT_1T_2)^2}+
\left(bT_2+{c \over 2}T_1\right)
{[(d^2 T_2/ d\vartheta^2)+cot\theta (dT_2 /d\vartheta)] \over
(b T^2_1+aT^2_2+cT_1T_2)}+ \right.
$$
$$
\left.
\left(aT_1+{c \over 2}T_2\right)
{[(d^2 T_1/ d\vartheta^2)+cot\vartheta (dT_1 /d\vartheta)] \over
(b T^2_1+aT^2_2+cT_1T_2)} \right \}-\lambda\hbar^2-{\hbar^2 \over 4}+
{m_{\ell}^2\hbar^2 \over \sin^2 \vartheta }-{\hbar^2 \over 4 \sin^2 \vartheta}=0 \ .  $$

\noindent
Ceci permet d'{\'e}crire l'{\'e}quation suivante
%
%
$$
-\left(bT_2+{c \over 2}T_1\right)
\left[{d^2 T_2 \over d\vartheta^2}+cot\vartheta
{dT_2 \over d\vartheta}\right]-\left(aT_1+{c \over 2}T_2\right)
\left[{d^2 T_1 \over d\vartheta^2}+cot\vartheta
{dT_1 \over d\vartheta}\right]+
$$
$$
\left( {m_{\ell}^2 \over \sin^2 \vartheta}- \lambda\right)\hbar^2
(b T^2_1+aT^2_2+cT_1T_2)=0 \ ,
$$
ce qui se ram{\`e}ne {\`a}
$$
\left(bT_2+{c \over 2}T_1\right)
\left[{d^2 T_2 \over d\vartheta^2}+cot\vartheta {dT_2 \over d\vartheta}+
\left(\lambda- {m_{\ell}^2 \over \sin^2 \vartheta} \right)T_2 \right]+
$$
$$
\left(aT_1+{c \over 2}T_2\right)
\left[{d^2 T_1 \over d\vartheta^2}+cot\vartheta {dT_1 \over d\vartheta}+
\left(\lambda- {m_{\ell}^2 \over \sin^2 \vartheta} \right)T_1\right]=0 \ .
\xeqno{2.12}
$$

Puisque $T_1$ et $T_2$ sont des solutions de l'{\'e}quation
de Schr{\"o}dinger angulaire suivant $\vartheta$ (Eq. (1.4)),
alors l'Eq. (2.12) est satisfaite.
Donc, les relations (2.7), (2.10) et (2.11) sont bien
les solutions de l'EHJQS angulaire suivant $\vartheta$.

\bigskip
\noindent
{\bf 2.3. La forme de l'action r{\'e}duite angulaire suivant $\varphi$ }              \medskip

Reprenons l'Eq. (1.34)
$$
\left({dM \over d\varphi}\right)^2-{\hbar^2 \over 2}
\left \{ M,\varphi \right \}-m_{\ell}^2 \hbar ^2=0 \ .
$$
Comme on l'a dit plus haut, cette {\'e}quation s'identifie
avec l'EHJQS {\`a} une dimension pour un potentiel nul et une {\'e}nergie
${m_{\ell}^2 \hbar^2 / 2m}$. Donc sa solution doit {\^e}tre de la forme
$$
M(\varphi)=\hbar \arctan \left \{ {F_1+\epsilon F_2) \over \tau F_1+F_2}
\right \} \ ,
$$
o{\`u} $\epsilon$ et $\tau$ sont des constantes d'int{\'e}gration et
$F_1$ et $F_2$ sont deux solutions r{\'e}elles et ind{\'e}pendantes
de l'Eq. (1.5). En choisissant ces deux solutions de la fa{\c c}on
suivante
$$
F_1=\cos (m_{\ell}\varphi) \ ,
$$
$$
F_2=\sin (m_{\ell}\varphi) \ ,
$$
la solution de l'Eq. (1.34) est [39]
$$
M(\varphi)=\hbar \arctan \left \{
{\ cos (m_{\ell}\varphi)+\epsilon\sin (m_{\ell}\varphi ) \over
\tau \cos (m_{\ell}\varphi)
+\sin (m_{\ell}\varphi)}
\right \} \ .
\xeqno{2.13}
$$
Cette derni{\`e}re {\'e}quation peut {\^e}tre {\'e}crite, suivant
les notations de Floyd sous la forme
$$
M(\varphi)=\hbar \arctan \left \{
b\ \tan(m_{\ell}\varphi)+c/2 \over \sqrt{ab-c^2/4}
\right \} \ .\xeqno{2.14}
$$

Notons que les relations donn{\'e}es par (1.24), (1.29) et (1.34)
sont de formes g{\'e}n{\'e}rales identiques.
Cette forme est celle de l'action r{\'e}duite
dans le cas du probl{\`e}me {\`a} une dimension.
Tout ceci semble indiquer que les mouvements dans toutes
les directions \hbox{appart-\hskip -1pt plus 2pt}\break iennent {\`a} une m{\^e}me
classe d{\'e}finie par la forme g{\'e}n{\'e}rale de l'action,
bien que les arguments des arcs tangentes sont diff{\'e}rents.

\medskip
\noindent
{\it Remarque: il est {\'e}vident que les EHJQS angulaires ainsi que
leurs solutions g{\'e}n{\'e}rales
ne d{\'e}pendent pas du potentiel. Par contre, la solution de
l'EHJQS radiale
est {\'e}troitement li{\'e}e {\`a} la forme du potentiel,
puisque ce dernier d{\'e}termine la forme des fonctions ${\cal X}_1$
et ${\cal X}_2$}.

\bigskip
\bigskip
\noindent
{\bf 3. L'EHJQS {\'E}CRITE POUR L'ACTION R{\'E}DUITE TOTALE}
\bigskip

Reprenons les trois EHJQS [39]
$$
{1 \over 2m} \left( {dZ \over dr} \right)^2-
{\hbar ^2 \over 4m} \left \{ Z,r\right \}+V(r)+
{\lambda \hbar^2 \over 2m r^2}=E \ ,
$$
$$
\left({dL \over d\vartheta}\right)^2-{\hbar^2 \over 2}\left \{L,\vartheta
\right \}+{(m_{\ell}^2-1/4) \over \sin^2\vartheta}\hbar ^2=
(\lambda +{1 \over 4})\hbar^2 \ ,
$$
$$
\left({dM \over d\varphi}\right)^2-{\hbar^2 \over 2}
\left \{ M,\varphi \right \}=m_{\ell}^2 \hbar ^2 \ .
$$
En tirant la valeur de $\lambda$ de l'{\'e}quation radiale et celle de
$m_{\ell}^2$ {\`a} partir
de l'{\'e}quation suivant $\varphi$, et en rempla{\c c}ant dans
l'{\'e}quation suivant $\vartheta$, on obtient l'{\'e}quation suivante:
$$
{1 \over 2m} \left[ \left( {dZ \over dr} \right)^2+
{1 \over r^2}\left({dL \over d\vartheta}\right)^2+
{1 \over r^2 \sin^2\vartheta} \left({dM \over d\varphi}\right)^2 \right]-
{\hbar ^2 \over 4m}\left[ \left \{ Z,r\right \}+{1 \over r^2}
\left \{L,\vartheta \right \}+ \right.
$$
$$
\left. {1 \over r^2 \sin^2\vartheta}\left \{ M,\varphi \right \} \right]+
V(r)-E- {\hbar ^2 \over 8m\ r^2}-   {\hbar ^2 \over 8m\ r^2\ \sin^2\vartheta}   =0 \ .\xeqno{3.1}    $$
Maintenant, d{\'e}finissons l'action r{\'e}duite totale du syst{\`e}me quantique
comme {\'e}tant la somme des actions $Z(r)$, $L(\vartheta)$, $M(\varphi)$
$$
S_0(r,\vartheta,\varphi)= Z(r)+L(\vartheta)+M(\varphi)\ . \xeqno{3.2}
$$
En utilisant cette d{\'e}finition, l'Eq. (3.1) peut s'{\'e}crire comme suit [39]
$$
{1 \over 2m} \left(\vec{\nabla}_{r,\vartheta,\varphi} S_0 \right)^2-
{\hbar ^2 \over 4m}\left[ \left \{ S_0,r\right \}+{1 \over r^2}
\left \{S_0,\vartheta \right \}+
{1 \over r^2 \sin^2\vartheta}\left \{ S_0,\varphi \right \} \right]+
$$
$$
V(r)-E-{\hbar ^2 \over 8m\ r^2}-   {\hbar ^2 \over 8m\ r^2\ \sin^2\vartheta}
=0\; . \xeqno{3.3}
$$
L'Eq. (3.3) peut {\^e}tre consid{\'e}r{\'e}e comme une g{\'e}n{\'e}ralisation
{\`a} trois dimensions de l'EHJQS pour le cas d'un potentiel
{\`a} sym{\'e}trie sph{\'e}rique.
\bigskip
\noindent
{\bf 4. CONCLUSION}
\medskip
La g{\'e}n{\'e}ralisation {\`a} trois dimensions constitue un
objectif tr{\`e}s important pour notre nouvelle approche,
et l'{\'e}tablissement d'EHJQS {\`a} trois dimensions dans le
cas d'un potentiel {\`a} sym{\'e}trie sph{\'e}rique constitue la
premi{\`e}re {\'e}tape d'une telle g{\'e}n{\'e}ralisation. En effet,
la construction des EHJQS {\`a} trois dimensions de m{\^e}me
forme que celle de l'EHJQS unidimensionnelle, et le fait
d'aboutir {\`a} des solutions de formes g{\'e}n{\'e}rales {\'e}quivalentes
indique qu'il est possible de construire une approche dynamique
qui devra {\^e}tre une \hbox{g{\'e}n{\'e}rali-\hskip -5pt plus 5pt}\break
sation {\`a} trois dimensions
de l'approche que nous avons d{\'e}velopp{\'e} au chapitre pr{\'e}c{\'e}dent.
Une telle construction {\`a} trois dimensions pourrait constituer
un argument tr{\`e}s puissant en faveur de cette approche.

\vfill\eject
\headline={
\vbox{
    \xline{   {\bf  \folio}}}

}
\footline={}

\bigskip
\bigskip
\bigskip
\bigskip

\centerline{\ftitle CONCLUSION}
\bigskip
\bigskip
\bigskip

\bigskip
\bigskip
\bigskip

\bigskip
\bigskip
\bigskip

Cette th{\`e}se est pr{\'e}sent{\'e}e dans le cadre d'une
approche d{\'e}terministe de la m{\'e}canique quantique,
qui consid{\`e}re les objets quantiques comme {\'e}tant
des corpuscules localis{\'e}s dans l'espace de mani{\`e}res permanente,
et ayant des trajectoires bien d{\'e}finies.
Nous avons ainsi pr{\'e}sent{\'e} dans ce travail une nouvelle approche bas{\'e}e sur
un formalisme analytique qui s'apparente {\`a} celui de la m{\'e}canique
classique. Par la suite, nous avons obtenu de deux mani{\`e}res
une {\'e}quation qui repr{\'e}sente l'{\'e}quivalent en m{\'e}canique
quantique de la premi{\`e}re int{\'e}grale de la loi de Newton (IPLNQ, Eq. (2.3)
du Chap. 2). La premi{\`e}re m{\'e}thode consiste {\`a} introduire
la fonction $f(x,E,\mu,\nu)$ dans la partie cin{\'e}tique
du Lagrangien et de l'Hamiltonien du syst{\`e}me quantique.
Les param{\`e}tres $\mu$ et $\nu$ jouent le r{\^o}le de variables cach{\'e}es.
Cette fonction d{\'e}crit les effets quantiques et tend vers 1 {\`a} la limite
classique $\hbar \to 0$.
La pr{\'e}sence des param{\`e}tres $\mu$ et $\nu$ est d{\^u} au fait que
l'{\'e}quation fondamentale qui d{\'e}crit le mouvement quantique est
du quatri{\`e}me ordre en $x$. La deuxi{\`e}me m{\'e}thode est bas{\'e}e sur
la coordonn{\'e}e quantique $\hat{x}$ introduite par Faraggi et Matone [36].
Ainsi, la coordonn{\'e}e $\hat{x}$ ram{\`e}ne l'EHJQS qui est du troisi{\`e}me
ordre {\`a} une {\'e}quation du premier ordre, et nous permet alors d'appliquer
correctement le th{\'e}or{\`e}me de Jacobi qui nous conduit exactement {\`a}
l'{\'e}quation IPLNQ d{\'e}j{\`a} obtenue par la premi{\`e}re m{\'e}thode. De m{\^e}me,
par les deux m{\'e}thodes, on obtient une {\'e}quation dynamique
qui lie le produit de la vitesse $\dot{x}$ et du moment conjugu{\'e}
$\partial S_0/\partial x$ {\`a} l'{\'e}nergie cin{\'e}tique du syst{\`e}me quantique.

L'{\'e}quation IPLNQ est du troisi{\`e}me ordre en $x$
et contient la premi{\`e}re et la deuxi{\`e}me d{\'e}riv{\'e}e de $V$
par rapport {\`a} $x$. Les termes contenant les d{\'e}riv{\'e}es d'ordres deux et      trois de $x$ sont
proportionnels {\`a} $\hbar^2$, et s'annulent donc {\`a} la limite classique.
Ainsi, {\`a} la limite classique l'{\'e}quation IPLNQ se ram{\`e}ne {\`a} l'{\'e}quation
classique de conservation de l'{\'e}nergie.

Le fait que la loi quantique de Newton est une {\'e}quation
diff{\'e}rentielle du quatri{\`e}me ordre implique l'existence de
deux constantes d'int{\'e}gration non classiques, $\mu$ et $\nu$,
dans l'expression de $x(t,E,\mu,\nu,x_0)$. Ces constantes sont
{\`a} l'origine de notre ignorance de l'{\'e}tat de mouvement de la particule
{\`a} l'{\'e}chelle quantique. Pour d{\'e}terminer cet {\'e}tat de mouvement, on doit
conna{\^\i}tre quatre conditions initiales, alors qu'en m{\'e}canique
classique, deux seulement sont suffisantes. Ces constantes sont aussi
{\`a} l'origine de l'existence de micro-{\'e}tats dans l'EHJQS non
d{\'e}tect{\'e}s par l'{\'e}quation de Schr{\"o}dinger.

Ainsi, les {\'e}quations du mouvement
quantique sont tr{\`e}s particuli{\`e}res bien qu'elles se ram{\`e}nent vers
les {\'e}quations classique pour $\hbar \to 0$. On note aussi que
l'{\'e}nergie cin{\'e}tique
de la particule est dispers{\'e}e entre deux sortes
d'impulsions, la quantit{\'e}
de mouvement classique $m \dot{x}$ et le moment conjugu{\'e}
$\partial S_0/ \partial x$.
Alors, on ne pourra plus lier l'{\'e}nergie s{\'e}par{\'e}ment {\`a} ces deux
sortes d'impulsions, comme c'est le cas en m{\'e}canique classique
$$
{m \dot{x}^2 \over 2}=E-V
$$
ou
$$
{1 \over 2m}\left( {\partial S_0 \over \partial x}\right)^2 =E-V
$$
En fait, la relation de dispersion est propre {\`a} la m{\'e}canique quantique,
 et elle est donn{\'e}e par

$$
\dot{x}{\partial S_0^{cla} \over \partial x} =2(E-V)\ .
$$

Bien que la g{\'e}n{\'e}ralisation {\`a} trois dimensions des {\'e}quations
de mouvement , {\'e}tablies au Chap.2, se heurte {\`a} d'{\'e}norme             difficult{\'e}s,
l'{\'e}tablissement de l'actions r{\'e}duite constitue la premi{\`e}re {\'e}tape  pour
la construction d'un formalisme bas{\'e} sur une vue dynamique {\`a} trois dimensions
de la m{\'e}canique quantique.

Notre approche, dans le cadre de sa g{\'e}n{\'e}ralisation
{\`a} trois dimensions, peut {\^e}tre d'un apport
consid{\'e}rable pour la physique. premi{\`e}rement, elle r{\'e}tablirait
une r{\'e}alit{\'e} objective de la nature, ce qui aurait des implication
philosophiques importantes. Deuxi{\`e}mement, elle pr{\'e}sente un espoir
pour fonder une th{\'e}orie quantique de la gravitation, objectif
\hbox{indispen-\hskip -2mm plus 2mm}\break sable, pour achever l'unification des quatre
interactions fondamentales.

\vfill\eject

\noindent
{\bf REFERENCES}
\medskip

\item{[1]}
L.~de Broglie, {\it Sur le parall{\'e}llisme entre la dynamique du point
mat{\'e}riel et l'optique g{\'e}om{\'e}trique}, {\it Journal de physique},
Serie VI, t. VII, n1, 1926, p. 1-6.

\item{[2]}
L.~de Broglie, {\it Les incertitudes de'Heisenberg et l'interpr{\'e}tation
probabiliste de la m{\'e}canique ondulatoire} , (Gauthier-Villars, 1982), Chap. I.

\item{[3]}
L.~de Broglie, Encyclop{\'e}die Fran{\c c}aise, tome II, La physique, p. 2-26-2 {\`a} 2-28-2.

\item{[4]}
L.~de Broglie, Encyclop{\'e}die Fran{\c c}aise, tome II, La physique, p.   2-60-3 {\`a} 2-60-7.

\item{[5]}
 L.~de Broglie, Encyclop{\'e}die Fran{\c c}aise, tome II, La physique, p.   2-48-15 {\`a} 2-50-4.

\item{[6]}
 L.~de Broglie, Encyclop{\'e}die Fran{\c c}aise, tome II, La physique, p.   2-50-5 {\`a} 2-62-5.

\item{[7]}
C.~J.~ Davisson, L.~H. Germer, {\it Phys. Rev.} {\bf 30}, 705 (1927).

\noindent
I. Estermann, O. Stern, {\it Zeit f\"ur Physik}.  {\bf 61}, 95 (1930).

\noindent
H. B{\"o}rsch, {\it Naturwiss} {\bf 28}, 709 (1940).

\item{[8]}
L.~de Broglie, {\it Les incertitudes de'Heisenberg et l'interpr{\'e}tation
probabiliste de la m{\'e}canique ondulatoire} , (Gauthier-Villars, 1982), Chap. XII.

\noindent
L.~de Broglie, {\it  Comp. rend.}  {\bf 183} , 447 (1926); {\bf 184} , 273 (1927);
{\bf 185} , 380 (1927);

\item{[9]}
L.~de Broglie, {\it Les incertitudes de'Heisenberg et l'interpr{\'e}tation
probabiliste de la m{\'e}canique ondulatoire} , (Gauthier-Villars, 1982),
Pr{\'e}face de L.~de Broglie.

\item{[10]}
L.~de Broglie, {\it Les incertitudes de'Heisenberg et l'interpr{\'e}tation
probabiliste de la m{\'e}canique ondulatoire} , (Gauthier-Villars, 1982), Chap. II.

\item{[11]}
A.~Einstein, N.~Rosen, and B.~Podolsky, {\it Phys. Rev}. {\bf 47}, 777 (1935).

\item{[12]}
D.~Bohm, {\it Phys. Rev}. {\bf 85}, 166 (1952);\ \  {\bf 85}, 180 (1952);\ \
D.~Bohm and J.~P.~Vigier, {\it Phys. Rev}. {\bf 96}, 208 (1954).

\item{[13]}
E.~Madelung, {\it Z. Physik.} {\bf 40}, 332 (1926).

\item{[14]}
T.~Tabakayasi, {\it progr. theoret. Phys.}  (Japan) {\bf  8}, 143 (1952);
{\bf  9}, 187 (1953) .

\noindent
Finkelstein, Lelevier and Ruderman, {\it Phys. Rev.} {\bf 83}, 326 (1951).

\item{[15]}
A.~Aspect, P.~Garnier and G.~Roger, {\it Phys. Rev. Lett.}  {\bf 47} , 460 (1982);
{\bf 49} , 91 (1982);

\noindent
A.~Aspect, J.~Dalibard,  G.~Roger, {\it Phys. Rev. Lett.}  {\bf 47} , 460 (1982).

\item{[16]}
J.~S.~Bell, {\it Physics}, Vol. {\bf 1}, N 3, 195 (1964).

\item{[17]}
P.~Roussel, {\it Journal de Physique} {\bf 47}, 393 (1986).

\item{[18]}
E.~R.~Floyd, arXiv:quant-ph/0009070.

\item{[19]}
E.~R.~Floyd, {\it Phys. Rev }. {\bf D 26}, 1339 (1982).

\item{[20]}
E.~R.~Floyd, {\it Phys. Rev }. {\bf D 29}, 1842 (1984).

\item{[21]}
E.~R.~Floyd, {\it Phys. Rev}. {\bf D 34}, 3246 (1986).

\item{[22]}
E.~R.~Floyd, {\it Found. Phys. Lett. }  {\bf  9}, 489  (1996);
arXiv:quant-ph/9707092.

\item{[23]}
E.~R.~Floyd, {\it Int. J. Mod. Phys}. {\bf A 15}, 1363 (2000),\ \
arXiv:quant-ph/9907092.

\item{[24]}
A.~E.~Faraggi and M.~Matone, {\it  Int. J. Mod. Phys}.  A15, 1869 (2000);
arXiv:hep-th/9809127.

\item{[25] }
A.~E.~Faraggi and M.~Matone, {\it Phys. Lett}. {\bf B 437},
369 (1998), arXiv:hep-th/9711028.

\item{[26] }
A.~E.~Faraggi and M.~Matone, {\it Phys. Lett}. {\bf B 450},
34 (1999), arXiv:hep-th/9705108.

\item{[27] }
A.~E.~Faraggi and M.~Matone, {\it Phys. Lett}. {\bf B 445},
357 (1998), arXiv:hep-th/9809126.

\item{[28] }
G.~Bertoldi, A.~E.~Faraggi and M.~Matone, {\it Class. Quant. Grav.}  {\bf 17},
3965 (2000), arXiv:hep-th/9909201.

\item{[29] }
A.~Bouda, {\it Found. Phys. Lett.} 14, 17 (2001), arXiv:quant-ph/0004044.

\item{[30] }
E.~Chpolski, Physique Atomique (Mir, 1978), tome II, p. 74-76.

\item{[31] }
E.~R.~Floyd, {\it Found. Phys. Lett. }  {\bf  14}, 17  (2001);
arXiv:quant-ph/9708007.

\item{[32] }
E.~R.~Floyd, {\it Phys. Rev.}  {\bf  D 25}, 1547 (1982).

\item{[33]}
A.~Messiah, {\it Quantum Mechanics} (Wiley, New York, 1961), Vol. I, p. 88.

\item{[34] }
E.~Chpolski, Physique Atomique (Mir, 1978), tome II, p. 168-174.

\item{[35] }
H.~B.~Dwith, {\it Table of Integrals and other Mathematical Data} 4th ed.
(McMillan, New York, 1961) ¶858.524; ¶858.534

\item{[36] }
A.~E.~Faraggi and M.~Matone, {\it Phys. Lett}. {\bf A 249}, 180 (1998),
arXiv:hep-th/9801033.

\item{[37] }
A.~Bouda and T.~Djama, arXiv:quant-ph/0103071.

\item{[38] }
C.~Cohen-Tannoudji, B.~Diu et  F.~Lalo\"e, {\it M{\'e}canique quantique}
(Hermann, 1977), tome 1.

\item{[39] }
T.~Djama, arXiv:quant-ph/0111142

\bye